\documentclass[12pt]{iopart}
\usepackage{color}
\usepackage{graphicx}
\usepackage{epsf}
\usepackage{graphicx,epsfig}
\usepackage{tabularx}
\usepackage{epstopdf}
\def\cs2{c_{s}^{2}}

 \def\be   {\begin{equation}}   \def\ee   {\end{equation}}
 \def\ba   {\begin{array}}      \def\ea   {\end{array}}
 \def\bea  {\begin{eqnarray}}   \def\eea  {\end{eqnarray}}
 \def\bean {\begin{eqnarray*}}  \def\eean {\end{eqnarray*}}

\begin{document}

\title{Large non-Gaussianities in the Effective  Field Theory  Approach to Single-Field Inflation:\\ the Bispectrum}

\author{Nicola Bartolo$^{1,2}$, Matteo Fasiello$^{3}$, Sabino Matarrese$^{1,2}$ and 
Antonio Riotto$^{2,4}$}
\vspace{0.4cm}
\address{$^1$ Dipartimento di Fisica ``G. Galilei'', Universit\`{a} degli Studi di 
Padova,  via Marzolo 8, I-35131 Padova, Italy} 
\address{$^2$ INFN, Sezione di Padova, via Marzolo 8, I-35131 Padova, Italy}
\address{$^3$ Dipartimento di Fisica ``G. Occhialini'', Universit\`{a} degli Studi di Milano Bicocca and INFN, Sezione di Milano Bicocca, Piazza della Scienza 3, I-20126 Milano, Italy}
\address{$^4$ CERN, Theory Division, CH-1211 Geneva 23, Switzerland\\
\vskip 0.5cm
DFPD-2010-A-05\,\, CERN-PH-TH/2010-076}
\vskip 0.5cm
\eads{\mailto{nicola.bartolo@pd.infn.it}, \mailto{matteo.fasiello@mib.infn.it}, 
\mailto{sabino.matarrese@pd.infn.it} and \mailto{riotto@mail.cern.ch}}

\begin{abstract}
The methods of effective field theory are used to study generic theories of
inflation with a single inflaton field and to perform a general analysis
of the associated non-Gaussianities. We investigate the 
amplitudes and shapes of  the various generic three-point correlators, the   bispectra,   
which may be generated by different classes of single-field inflationary models.  Besides the well-known results for the
DBI-like models and the ghost inflationary theories, we point out   
that  curvature-related interactions may give rise to large non-Gaussianities  in the form of 
bispectra characterized by a flat shape which, quite interestingly, is independently produced by several interaction terms. In a subsequent work, we will perform a similar
general analysis for the non-Gaussianities generated by the generic four-point correlator,  the  trispectrum.
\end{abstract}
\newpage

\tableofcontents

\section{Introduction}
The inflationary paradigm is central in modern cosmology as it naturally provides an explanation for  long standing issues as the flatness and horizon problems. The simplest standard, single-field slow-roll models of inflation already nicely account for the scale invariant primordial power 
spectrum of the Cosmic Microwave Background (CMB) anisotropies \cite{smoot92,bennett96,gorski96,wmap3,wmap5,kom}. On the other hand, in order to get a deeper understanding of the inflaton dynamics, one needs to consider observables sensitive to deviations from Gaussianity, which 
might be studied in the form of the three-point correlator of cosmological perturbations, the so-called bispectrum \cite{bisp,bispectrum,BabichCZ,chen-bis}, 
the connected four-point correlator, the so-called trispectrum \cite{trispectrum}, loop corrections to the power spectrum \cite{loop} and so on. In recent years, a lot of progress has been done in this direction resulting in the characterization of a wide variety of inflationary models through their higher order correlation functions~\cite{reviewk}.

Complementing this theoretical effort, new experiments provide a better sensitivity to  deviations from Gaussian statistics (see, e.g.~\cite{Liguori1,VerdeM}). 
The very recent launch of the Planck satellite~\cite{Pl,Mand} (and the continued analysis of WMAP data~\cite{kom}) presents us with the exciting opportunity to actually test this zoo of possibilities and makes more urgent the classification and refinement of the different predictions at the level of bispectrum and  trispectrum of curvature perturbations 
(see, e.g.,~\cite{koyamarev,ourrev,chenrev} for comprehensive and updated reviews).

In this context, the effective theory approach proposed in \cite{luty,eft08} and further analyzed in \cite{w-e} represents a powerful tool. Indeed, the effective Lagrangian approach provides an efficient way to  generically study large classes of  inflationary models. Furthermore, this formalism  provides a  clear-cut dictionary between deviations from the standard scenario 
(almost Gaussian perturbations) and higher-order invariant operators in the action; it sheds new light on effects due to symmetries in the action (notably, 
a reduced speed of sound often automatically results in an enhanced non-Gaussianity); it allows some general and immediate calculational advantages, by exploiting the
so called decoupling regime; finally, 
one can classify almost any specific inflationary model switching on some particular operators in the Lagrangian, thus allowing a more unifying approach.
The effective field theory formalism naturally comprises Lagrangians with higher order terms. These terms call for an UV completion of the underlying theory. 
In this perspective one looks with particular interest at string theory inspired models, such as DBI infaltion \cite{DBI}, and possible explanations in this context for k-inflation models \cite{mukh1,mukh2}.

 They are  retrievable in the effective theory  approach as the effective field theory that lives in a particular region of the parameters space.

In the spirit of effective theory, one should also include all possible curvature-generated terms ($\delta R^\mu_{\,\,\,\,\mu}{}^2$,...) 
in writing the most generic effective theory  Lagrangian  of inflation up to the desired 
order in perturbation theory. In this paper we perform a thorough analysis of the
amplitude and shapes of the three-point correlators which arise from the various classes of inflationary models driven by a single
scalar field described within the generic effective theory approach. Besides reproducing the known results exisiting in the literature, 
that is to say that large non-Gaussianities may be generated if the sound speed is smaller than unity, 
we also show that single-field models of inflation may produce large non-Gaussianities in the form of three-point correlators whose amplitude
is peaked in the so-called flat configuration, and precisely corresponding to a shape of the bispectrum where two of the momenta are roughly one half of the third one $k_1 =2 k_2=2 k_3$. 
Such bispectra emerge from curvature-generated interactions that have not been discussed so far in the literature. Let us recall that a shape of this type, peaked on flat triangles, was first found in \cite{ssz05} where it was obtained as the result of a linear combination of two operators with proper weights \footnote{It is important to note that these weights are not fixed, indeed there is an order one interval in \cite{ssz05} that produces shapes compatible with the flat template.}. We stress here that our results further expand the space of operators that generates this kind of shape and allow one to easily obtain it from different, independent operators. Before the results of \cite{ssz05}, the only scenario compatible with such a flat shape 
(also called folded or squashed) for the primordial bispectrum was that of models of inflation which relax the assumption of a regular Bunch-Davies vacuum, 
as studied in~\cite{chen-bis,HT,Pier}.  In fact, up to \cite{ssz05}, it was standard procedure to associate the primordial bispectrum evaluated in the Bunch-Davies vacuum  to the so-called equilateral or local shapes~\cite{BabichCZ}.\footnote{We refer to~\cite{FShellard,Liguori2} for a detailed analysis of possible shapes of the primordial bispectra and to the discussion in Ref.~\cite{ssz05} for an approach that includes the flat shape. (see also Ref.~\cite{chen-bis}).} 
Here  we highlight that such a flat configuration for the primordial non-Gaussianities might be generated in single-field models of inflation in a more general context than previously thought. 
It is worth pointing out another aspect in which our results differ from previous ones. 
We are able to provide a solution for the equations of motion of the primordial perturbations that can 
interpolate between known cases, like DBI or Ghost inflation. We are able to study in  full generality 
these intermediate situations also when interaction terms are included.  
In a follow-up paper \cite{future}, we will perform a similar analysis for the trispectrum.

The  paper is organized as follows. 
In section 2 we give a brief account on how to write the Lagrangian we employ in the following sections, following Ref.~\cite{eft08}. The reader who is familiar with this approach might wish  to skip this part and jump  directly to the third-order Eq.~(\ref{action}). We also comment on the inflationary models one can span within this approach 
and on the claim of its generality. The calculational algorithm used in this section can be extended to higher orders as well.
In section 3 we derive the equation of motion for the scalar field which in the effective theory approach is used to describe the cosmological perturbations \cite{tilted} and we provide an 
analytical solution to this general equation.
In section 4 we present the calculation of the amplitude of the bispectrum  for each interaction term in six different configurations capturing the simplest single field slow roll model, DBI inflation, ghost inflation and intermediate cases. We focus on  regions in the parameter space where 
non-Gaussianity is  large. 
In section 5 we study  the shape of the bispectra arising from the  individual interaction terms identifying, among other things, interesting shapes from the  curvature-related contributions. We then summarize our findings and comment on further work in section 6.

\section{The inflationary action up to third order in the effective theory approach}
Our initial goal is to write down the complete theory of single-field models of inflation   up to third order in perturbations. 
We will follow the  effective theory approach first introduced in  Ref.  \cite{eft08}.  Let us give a brief summary of the procedure. 
The scalar field $\phi$ responsible for inflation is splitted as usual as an unperturbed part plus a fluctuating one:
\be
\phi(\vec x, t)=\phi_0 (t)+ \delta \phi(\vec x, t).
\ee 
One then chooses to work in the the comoving (or unitary) gauge for which $\delta \phi =0$  \cite{luty} . The 
 Lagrangian will no more be invariant under full spacetime diffeomorphisms (diffs) but only under the spatial reparametrizations.
 This is the starting point to write the most general  space diffs invariant Lagrangian at the desidered order in perturbation theory in an effective theory approach.
In \cite{eft08} the authors prove that, once an approximate shift-simmetry is operated, their second and third order action is the most general one  (see also  \cite{w-e} for an interesting perspective on the most general effective Lagrangian for inflation). 
One can then use the Stueckelberg procedure  to restore full spacetime invariance. 
As a by-product of this procedure, the degree of freedom hidden in the metric shows up again as a scalar field. 

The most general space diffs-invariant action in unitary gauge is  schematically written as \cite{eft08}:
\be
S = \int d^4 x \sqrt{-g} \; F(R_{\mu\nu\rho\sigma}, g^{00}, K_{\mu\nu},\nabla_\mu,t),\label{unper}
\ee
where $K_{\mu\nu}$ is the extrinsic curvature on which we will elaborate more later and the indices on $g^{00}$ are free indices.  Considering fluctuations around a FRW background amounts to studying the following action
\begin{eqnarray}
S &=& \int d^4x \sqrt{-g} \; \Big[\frac12 M_{\rm Pl}^2 R + M_{\rm Pl}^2  \dot H  g^{00} - M_{\rm Pl}^2 \Big(3 H^2 +\dot H\Big) + \nonumber \\  && 
\sum_{n\geq 2} F^{(n)}(g^{00}+1,\delta K_{\mu\nu}, \delta R_{\mu\nu\rho\sigma};\nabla_\mu;t)\Big],
\label{genspacei}
\end{eqnarray}
where the  $F^{(n)}$ functions contain fluctuations which are at least quadratic.
The next step is to   restore full spacetime reparametrization invariance. To see how it works, we  borrow a simple example from \cite{eft08} and consider the following sample action terms
\be
\int d^4x\;  \sqrt{-g} \left[A(t)+B(t)g^{00}(x)\right] \ \label{ex1}.
\ee
We are interested in time reparametrization $t\rightarrow t+\xi_{0}(\vec x,t); \quad \vec x \rightarrow \vec x $, under which the above action (after a simple variable redefinition) reads
\be
\fl \int d^4x\;  \sqrt{- g(x)} \left[A( t-\xi_0(x))+B(t-\xi_0(x)) \frac{\partial (t-\xi_0(x))}{\partial x^\mu}\frac{\partial (t-\xi_0(x))}{\partial x^\nu} g^{\mu\nu}(x)\right].\label{ex2}
\ee
Upon promoting $\xi_0$ to a field, $\xi_0(x)=-\pi(x)$ and requiring the following gauge transformation rule $\pi(x)\rightarrow \pi(x)-\xi_0(x)$ on $\pi$,  the above action is invariant under full spacetime diffeomorphisms. The  scalar degree of freedom $\pi$ makes its 
appearance in  the time dependence of the $A,B$ coefficients and in the transformed metric.
Furthermore, under time reparametrization the metric  $g^{\mu\nu}$ transforms as follows:
\begin{eqnarray}
 \widetilde g^{\alpha \beta} = \frac{\partial \widetilde g^{\alpha}}{\partial x^\mu} \frac{\partial \widetilde g^{\beta}}{\partial x^\nu} g^{\mu\nu}\, , 
\end{eqnarray}
which implies
\begin{eqnarray}
g^{ij}\rightarrow g^{ij}; \quad g^{0i}\rightarrow (1+\dot\pi)g^{0i} +g^{ij}\partial_j \pi; \label{metric} 
\\ \nonumber
g^{00}\rightarrow (1+\dot\pi)^2 g^{00} +2(1+\dot\pi)g^{0i}\partial_i \pi+ g^{ij}\partial_i \pi \partial_j \pi \label{ex3}.
\end{eqnarray}
This procedure has been borrowed, conceptually unchanged, from standard gauge theory: a Goldstone boson which transforms non linearly under the gauge transformation provides the longitudinal component of  a massive   gauge boson. At sufficiently high energy 
such  Goldstone boson  becomes the only relevant 
 degree of freedom. This is the so-called equivalence theorem. The same is true for our case: for sufficiently high energy
 the mixing with gravity becomes irrelevant and the scalar $\pi$ becomes the only relevant mode in the dynamics. This is the
 so-called decoupling regime. 
Let us clarify this concept with a simple example. Consider the following contribution, taken from  Eq.~(\ref{genspacei})
\be
M_{\rm Pl}^{2} \dot H g^{00} \rightarrow M_{\rm Pl}^{2} \dot H ((1+\dot\pi)^2 g^{00} +2(1+\dot\pi)g^{0i}\partial_i \pi+ g^{ij}\partial_i \pi \partial_j \pi).
\ee
We focus on  the quadratic part of the first term in the above equation. Upon canonical normalization, $\pi_{c}=M_{\rm Pl}{\dot H}^{1/2}\pi$ and $ g^{00}_{c}=M_{\rm Pl}g^{00}$, one gets

\be
M_{\rm Pl}^{2} \dot H(  g^{00} + 2 \dot \pi  g^{00} + {\dot\pi}^2 g^{00} )=   {\dot\pi_{c}}^2 + 2 {\dot H}^{1/2}\dot\pi_{c} g^{00}_{c}  + \dot H g^{00}_{c}. \label{dec}
\ee
Consider the second term of Eq.~(\ref{dec}) which mixes gravity with the scalar. Since $\dot \pi_{c}\sim E \pi_c$, at energies higher than $\sim {\dot H}^{1/2}$ the term  ${\dot\pi_c}^2$ dominates the dynamics. This turns out to be true in general: the number of derivatives (which in Fourier mode would basically give an energy-dependent coefficient in front of $\pi$) is higher in  terms containing only $\pi$'s  than in the mixed terms and therefore there exists an energy threshold above which the scalar decouples from gravity. Since in explicitating the $F^{(2)}$ 
term in Eq.~(\ref{genspacei}) there can be, in principle, other quadratic terms that go like ${\dot \pi}^2$, one has to consider which one is the leading kinetic term and determine the canonically normalized field $\pi_c$ and the energy threshold accordingly. To take the safe route, one might well take the energy threshold, $E_{\rm mix}$, to be the highest one of this set. Since one is concerned with correlators just after horizon crossing, 
one concludes that the decoupling procedure works as long as the decoupling energy is smaller than
the Hubble rate $H$.  More precisely, we can anticipate  that the kinetic terms in $F^{(2)}$ which are going to matter in our discussion 
come with coefficients  $M_2^4$ and $M_{\rm Pl}^2 \epsilon H^2$. The condition   $E_{\rm mix} < H$ is then 
 satisfied if $M_{\rm Pl}^2 \epsilon H^2 > M_2^4$, where $\epsilon=-\dot{H}/H^2$ is a slow-roll parameter; if this is not the case we need to assume $M_2^4< M_{\rm Pl}^2 H^2 $.

From now on we will work in the decoupling regime. In considering  the terms of Eq.~(\ref{genspacei}), we will therefore use only the unperturbed entries of the metric tensor.
To write  the effective Lagrangian up to third order, we  start from  Eq.~(\ref{genspacei}) and follow the algorithm given in \cite{eft08}. Fluctuations are encoded in the $F^{(n)}$ terms. In order to be as general as possible, we also include all possible 
contributions up to third order   coming from extrinsic curvature $K_{\mu\nu}$ terms.  In fact, it is instructive at this stage to step back and consider the action in Eq. (\ref{unper}). Given a theory which is space diffs-invariant, one can always identify a slicing of spacetime, described by a timelike function $\widetilde t(x)$,  which realizes time diffeomorphism: on surfaces of constant $\widetilde t$ the time symmetry breaking scalar is also constant. Before selecting a gauge, there is still the freedom to make a choice on $\widetilde t(x)$ and working in the unitary gauge amounts to requiring $\widetilde t=t$. In order to describe the geometry of this preferred slicing, one employs the extrinsic curvature tensor. In writing
 down such a  tensor, one needs  two ingredients: the  unit normal vector $n_\mu$, perpendicular to the constant $\widetilde t$ surfaces and the induced metric $h_{\mu\nu}$. These are defined as
\be
n_\mu=\frac{\partial_\mu \widetilde t}{\sqrt{-g^{\mu\nu}\partial_\mu \widetilde t \partial_\nu \widetilde t}}\label{n-h} \rightarrow \frac{\delta_\mu^0}{\sqrt{-g^{00}}}; \quad h_{\mu\nu}=g_{\mu\nu}+n_\mu n_\nu,
\ee
which allows us to write
\be
\fl K_{\mu \nu} \equiv h^{\sigma}_{\mu} \nabla_{\sigma}n_{\nu}\,\,=\,\, \frac{\delta^{0}_{\nu} \partial_{\mu} g^{00}}{2(-g^{00})^{3/2}} + \frac{\delta^{0}_{\nu} \delta^{0}_{\mu} g^{0\sigma} \partial_{\sigma} g^{00}}{2(-g^{00})^{5/2}}- \frac{g^{0\epsilon}(\partial_{\mu}g_{\epsilon \nu} + \partial_{\nu}g_{\epsilon \mu}-\partial_{\epsilon}g_{\mu \nu})}{2(-g^{00})^{1/2}}.
\ee
The above expressions can be used to  write explicitly the most generic third  order action for the fluctuations around the FRW background 
within the effective theory approach  of Ref.  \cite{eft08}:
\begin{eqnarray}
\label{eq:actiontad}
\fl S_{3}& = &\int  d^4 x \; \sqrt{- g} \Big[ \frac12 M_{\rm Pl}^2 R + M_{\rm Pl}^2 \dot H
g^{00} - M_{\rm Pl}^2 (3 H^2 + \dot H) + \frac{1}{2!}M_2(t)^4(g^{00}+1)^2 \nonumber \\
\fl &+&\frac{1}{3!}M_3(t)^4 (g^{00}+1)^3- \frac{\bar M_1(t)^3}{2} (g^{00}+1)\delta K^\mu {}_\mu
-\frac{\bar M_2(t)^2}{2} \delta K^\mu {}_\mu {}^2  \nonumber \\
\fl &-&\frac{\bar M_3(t)^2}{2} \delta K^\mu {}_\nu \delta K^\nu {}_\mu  
 -\frac{\bar M_4(t)^3}{3!} (g^{00}+1)^2 \delta K^\mu {}_\mu -\frac{\bar M_5(t)^2}{3!} (g^{00}+1) \delta K^\mu {}_\mu {}^2  \nonumber \\
\fl &-&\frac{\bar M_6(t)^2}{3!}(g^{00}+1) \delta K^\mu {}_\nu \delta K^\nu {}_\mu  -\frac{\bar M_7(t)}{3!}  \delta K^\mu {}_\mu {}^3-\frac{\bar M_8(t)}{3!}  \delta K^\mu {}_\mu {}  \delta K^\nu {}_\rho \delta K^\rho {}_\nu \nonumber \\
\fl  &-&\frac{\bar M_9(t)}{3!}   \delta K^\mu {}_\nu {}  \delta K^\nu {}_\rho \delta K^\rho {}_\mu
\Big] \; .\label{genv}
\end{eqnarray}
 The coefficients $M_{2},M_{3}$ and ${\bar M_1},\cdots,{\bar M_9}$ are to be considered generic. The $\bar{M}$ coefficients multiply curvature-generated interactions. A given particular set of them 
 will specify a given inflationary theory. The action (\ref{genv}) is not yet  invariant under full diffeormophisms though. One needs to follow exactly the steps illustrated in Eqs~ (\ref{ex1}), (\ref{ex2}) and (\ref{ex3}) and promote $\xi_0$ to a field $\pi$ with the proper gauge transformation.

In the decoupling limit we find:

\begin{eqnarray}
\fl  S_3&=&\int d^4 x \sqrt{-g}\left[ M_{\rm Pl}^{2}\dot H (\partial_{\mu} \pi)^2 
+ M_2(t)^4\left(2{\dot\pi}^2 -2\dot\pi \frac{(\partial_i \pi)^2}{a^2}\right) -\frac{4}{3}M_3(t)^4{\dot\pi}^3 
\right.\nonumber\\
\fl &-& \frac{\bar M_1(t)^3}{2}\left(\frac{-2 H (\partial_i \pi)^2 }{a^2} +\frac{(\partial_i \pi)^2 \partial_j^2 \pi}{a^4}  \right) -\frac{\bar M_2(t)^2}{2} \left( \frac{(\partial_i^2\pi)(\partial_j^2\pi)   +H (\partial_i^2\pi) (\partial_j \pi)^2  +2\dot\pi \partial_{i}^2 \partial_j \pi \partial_j \pi}{a^4} \right) 
\nonumber\\
\fl &-&\frac{\bar M_3(t)^2}{2} \left( \frac{ (\partial_i^2\pi)(\partial_j^2\pi) +2H(\partial_i \pi)^2\partial_i^2 \pi + 2 \dot\pi \partial_{i  j}^{2}\pi \partial_j \pi  }{a^4} \right)   -\frac{2}{3}\bar M_4(t)^3 \frac{1}{a^2}{\dot\pi}^2 \partial_i^2\pi +  \frac{\bar M_5(t)^2}{3}  \frac{\dot\pi}{a^4}(\partial_i^2\pi)^2 
\nonumber\\
\fl & +&\left.\frac{\bar M_6(t)^2}{3}  \frac{\dot\pi}{a^4}(\partial_{ij}\pi)^2 -\frac{\bar M_7(t)}{3!} \frac{(\partial_i^2 \pi)^3}{a^6}-\frac{\bar M_8(t)}{3!}  \frac{\partial_i^2 \pi}{a^6}(\partial_{jk}\pi)^2-\frac{\bar M_9(t)}{3!}\frac{1}{a^6}\partial_{ij} \pi\partial_{jk} \pi\partial_{ki}\pi
\right].\label{action}
\end{eqnarray}
A few clarifying comments are in order:
\begin{itemize}
\item If we consider  terms only up to second-order, for $M_{2}=\bar M_{1,2,3}=0$ one recovers the usual quadratic Lagrangian for the fluctuations, with  sound speed  $c_s^2 =1$ and the standard  solution to the equations of motion. Switching on $M_{2}$ amounts to allowing models with sound speed smaller than unity, $1/c_s^2 =1- 2M_2^4/(M_{\rm Pl}^2 \dot H)$, which are often linked to a high level of primordial non-Gaussianity \cite{chen-bis,eft08}. Furthermore,  turning on $\bar{M}_{2,3}$ in the de Sitter limit,  one recovers  Ghost inflation \cite{ghost}. On the same lines, keeping all the $\bar M$'s vanishing,  but going to third  and higher order with the $M$'s,   one can retrieve the interactions that describe DBI inflation \cite{DBI,chen-bis, chen-tris}. The list of correspondences continues with K-inflation theories and others, thus showing how the effective action approach provides a unifying perspective on inflationary models \cite{eft08}.
\item The action in Eq.~(\ref{action}) has already been  written with  large non-Gaussianities in mind.
This means  that, at every order in fluctuations and for  each $M$ and $ \bar M$ coefficients,  we have  selected  those 
leading terms which will eventually generate large three-point correlators. To clarify this point,  we provide   a simple example. Let us consider the terms up to second order in Eq.~(\ref{action}) and set conveniently   $\bar M_{1,2,3}=0$. The properly normalized solution to the equation of motion will be the usual $\pi_k(\tau) \propto i H e^{-i k c_s \tau}(1+i k c_s \tau)$. It is straightforward to verify that, at the horizon crossing , 
 $\dot \pi \sim H \pi$ and $\nabla \pi \sim H/c_s\,\, \pi$. Therefore, among the $\pi$ terms with the same number of derivatives, the ones with the highest number of space derivatives dominate in the $c_s\ll 1$ limit. Generalizing these estimates for the classical solution (which we will  describe below) 
obtained from the equation of motion of our complete action, one selects  the terms in Eq.~(\ref{action}).
\item 
there is also the  comparison between same perturbative order but different $M$ terms to be made.
In the literature, all non zero coefficients  in front of the various operators are generically assumed to be of the same order (see for example the discussion 
concerning the orthogonal configuration  in Ref. \cite{ssz05} for an interesting perspective). 
We shall not restrict ourselves to this situation. Note that,  were the coefficients to be all of the 
same order, one could already identify the dominant
operators.  For example, consider the third order contributions $\bar M_1^3 (\partial_i \pi)^2 \partial_j^2 \pi/a^4\sim M^3 (H^4/c_s^4) {\pi}^3$; for $c_s\ll 1$ this will be a leading contribution with respect to, say, $\bar M_2(t)^2 H (\partial_i^2\pi) (\partial_j \pi)^2\sim M^2 H (H^4/c_s^4){\pi}^3$. This is due to the fact that in the effective Lagrangian  every additional derivative comes with a $H/M\ll 1$ factor attached: one is basically doing an  $H/M$ expansion where M is roughly the energy range of the underlying theory.  In the last example we have intentionally picked terms with the same power of $c_s$ at the denominator. Let us  now look at the $\bar M_7(t) (\partial_i^2 \pi)^3 \sim M (H^6/c_s^6){\pi}^3 $ term though; comparing this contribution with the $\bar M_1$ term amounts to comparing $M^3$ with $M H^2/c_s^2$. We see that for a very small speed of sound the $\bar M_7$ contribution may still be relevant. 
These examples justify our strategy of   including all the terms in Eq.~(\ref{action}) compatible with  $c_s\ll 1$ (we will make more comments on this point in Sec.~\ref{ampl}).
\end{itemize}
Having written the action, we now proceed to elaborate about  the solution to the equation of motion for the scalar degree of freedom $\pi$.  
\section{Solution to the equation of motion for the scalar degree of freedom}
To solve for the equation of motion for the scalar degree of freedom $\pi$,  we write the Lagrangian to second order in the perturbations. Note that in the equation below, as opposed to what was done in writing the third order action in Eq.~(\ref{action}), we include all possible terms, even the ones which are subleading and are therefore not expected to 
give large non-Gaussianities.
\bea 
\fl  \mathcal{L}_2&=&a^3\Bigg[ M_{\rm Pl}^{2}\dot H (\partial_{\mu} \pi)^2 
+ 2 M_2^4{\dot\pi}^2 - \bar M_1^3 H \left(3 {\dot \pi}^2 -   \frac{(\partial_i \pi)^2 }{2 a^2} \right)
  -\frac{\bar M_2^2}{2}\Big( 9H^2 {\dot \pi}^2 \nonumber \\  \fl &-&3H^2 \frac{(\partial_i \pi)^2}{a^2} +\frac{1}{a^4} (\partial_i^2\pi)^2\Big) -\frac{\bar M_3^2}{2}\left(3H^2 {\dot \pi}^2   -H^2 \frac{(\partial_i \pi)^2}{a^2}+\frac{1}{a^4} (\partial_j^2\pi)^2\right) \Bigg].\label{lagr2}
\eea
Following a common procedure, we have suppressed the time dependence in the $M$ coefficients since we choose to work at leading order in the generalized slow roll approximation. 
After the usual change of variable,  $\pi(\vec k, t(\tau))=u(\vec k, \tau)/a(\tau)$, the equation of motion reads
\be
u'' -\frac{2}{\tau^2} u + \alpha_0 k^2 u + \beta_0 k^4 \tau^2 u=0, \label{eom}
\ee
where  $\alpha_0, \beta_0$ are time independent (again, at leading order in slow roll) dimensionless coefficients:
\bea
 \alpha_0 = \frac{-M_{\rm Pl}^{2}\dot H - \bar M_{1}^{3}H/2-3/2 \, \bar M_2^2 H^2-\bar M_3^2 H^2/2}{-M_{\rm Pl}^{2}\dot H +2M_{2}^{4}-3 \bar M_{1}^{3}H -9/2\, \bar M_2^2 H^2 -3/2\, \bar M_3^2 H^2}\, ,\nonumber \\ \beta_{0}=\frac{(\bar M_{2}^{2} +\bar M_3^2) H^2}{2(-M_{\rm Pl}^{2}\dot H +2M_{2}^{4} -3 \bar M_{1}^{3}H -9/2\, \bar M_2^2 H^2 -3/2\, \bar M_3^2 H^2)}\, .
\eea
Now, for the sake of simplification and in order to make sharper and more clear the correspondence between switching on the coefficients and recovering any specific inflationary models, we make the following choice: we take $\bar M_3^2 =-3 \bar M_2^2$ and introduce $\bar M_0^2 =2/3\, \bar M_3^2 $. As a consequence, the second order Lagrangian now reads:
\be 
\fl \mathcal{L}_2=a^3\left(  M_{\rm Pl}^{2}\dot H (\partial_{\mu} \pi)^2 
+ 2 M_2^4{\dot\pi}^2 - \bar M_1^3 H \left(3 {\dot \pi}^2 -   \frac{(\partial_i \pi)^2 }{2a^2} \right) -\frac{\bar M_0^2}{2}\left(\frac{1}{a^4} (\partial_i^2\pi)^2 \right) \right)\label{lagr0}.
\ee
Since in the Lagrangian we lost one coefficient in going from $\bar M_{2,3}$ to $\bar{M}_0$ one might well count one less degree of freedom at our disposal in the calculations that follow. On the other hand, as one can easily check, the operators with the  $\bar M_2, \bar M_3$ coefficients give rise to almost the same amplitude and shape for non-Gaussianities. Thus, the price to pay is not that high. 
The  $\alpha_0$  and $\beta_0$ parameters generalize the sound speed $c_s^2$ and the $ H^2 /M^2$ term (first introduced in Ref. \cite{ghost}), respectively. Indeed, considering from now on Eq.~(\ref{lagr0}) as our starting point, the complete expression for these parameters reads: 
\be
\fl \alpha_0 = \frac{-M_{\rm Pl}^{2}\dot H - \bar M_{1}^{3}H/2}{-M_{\rm Pl}^{2}\dot H +2M_{2}^{4}-3 \bar M_{1}^{3}H },  \qquad \beta_{0}=\frac{\bar M_{0}^{2} H^2}{2(-M_{\rm Pl}^{2}\dot H +2M_{2}^{4} -3 \bar M_{1}^{3}H)}, \label{ab}
\ee
so that, for instance,  one recovers the sound speed  $1/c_s^2 =1- 2M_2^4/(M_{\rm Pl}^2 \dot H)$ for $\bar M_{1}=0$. The  solution for standard inflation is recovered 
if $\beta_0=\bar M_1=0$; the one for ghost inflation is obtained  in the de Sitter limit if    $\alpha_0=0=\bar M_1$.
Notice that, as evident from Eq.~(\ref{ab}), there is the possibility of a negative value for $\alpha_0$, the generalized speed of sound. By looking at Eq.~(\ref{eom}) one sees this would immediately result in a dispersion relation which, for $\alpha_0 k^2$ dominating over $\beta_0 k^4 \tau^2$, gives an exponential solution for the modes. As explained in \cite{ssz05}, this fact does not necessarily exclude models with negative $\alpha_0$. In fact, as far as the $\beta_0 k^4 \tau^2$ term dominates soon enough in the UV before the cutoff scale of the theory is reached,  these models are still in principle acceptable as they eventually freeze at the Hubble scale. On the other hand, having an exponential growth, even for a limited energy window, can generate large non-Gaussianities which are ruled out by observations \cite{ssz05}. We plan to perform a detailed analysis of the issues associated with the $\alpha_0$-negative region of the parameters space in \cite{p.s.}.

We can readily  make some educated guesses on the behaviour of the wavefunction before solving the equation of motion. First of all, the typical oscillatory behaviour deep inside the horizon is to be expected: in this regime both $\alpha_0 k^2$ and $\beta_0 k^4 \tau^2$ cause wave-like behaviour on the wavefunction while the $(-2/\tau^2)$ contribution is negligible. This tells us the main contribution to correlation functions will be coming, as usual, from the horizon crossing region and the outside. Note here that, if $\beta_0 \ne 0$, the `ghost inflation' term will eventually lead the oscillation if one goes deep enough inside the horizon.
On the other hand, in the $\tau \rightarrow 0$ limit, the $\tau^{-2}$ term leads  the dynamics and we recover the usual frozen modes. As done for the  DBI case, it is convenient to introduce the notion of an 
effective horizon crossing, placing it where the oscillatory behaviour stops being dominant
\be
 \alpha_0 k^2 + \beta_0 k^4 \tau_{*}^2 \sim  \frac{2}{\tau_*^2} \quad \Rightarrow \quad \tau_* = -\frac{2}{k \sqrt{\alpha_0+ \sqrt{\alpha_0^2 + 8 \beta_0}}}.
\ee
For $\beta_0=0$ and $\alpha_0\sim 1 $ one recovers the usual $k^2 \tau_*^2 \sim 1$  horizon crossing.

Let us now be more quantitative. The solution to Eq.~(\ref{eom}), being of second order, will come with two momentum-dependent integration constants. We have determined their values by requiring to obtain the known standard and ghost solutions in the corresponding limits. The general wavefunction reads 

\begin{eqnarray}
\fl u(k,\tau)&=&\frac{1}{2^{1/4} \tau }i e^{\frac{1}{2} i \sqrt{\beta_0} k^2 \tau ^2} \Bigg[{\rm HyperG} \left(-\frac{1}{4}-\frac{i \alpha_0}{4 \sqrt{\beta_0}},-\frac{1}{2},-i \sqrt{\beta_0} k^2 \tau ^2 \right) C_1(k) \nonumber \\
\fl &+& {\rm LagL}\left(\frac{1}{4}+\frac{i \alpha_0}{4 \sqrt{\beta_0}},-\frac{3}{2},-i \sqrt{\beta_0} k^2 \tau ^2\right) C_2(k)\Bigg],
\end{eqnarray}
where HyperG is the confluent hypergeometric function and LagL stands for the generalized Laguerre polynomial. Upon fixing 
\[
\fl C_1(k,\alpha_0, \beta_0)= \frac{\left(\alpha_0+\sqrt{\beta_0}\right)^{-3/4}\Gamma\left[\frac{5}{4}-\frac{i \alpha_0}{4 \sqrt{\beta_0}}\right]k^{-3/2}}{ 2^{1/4} \sqrt{M_{\rm Pl}^2 \epsilon H^2 +2 M_2^4 -3 \bar M_{1}^{3}H}\,\,\, \Gamma\left[3/2-\frac{ \sqrt{\beta_0}}{4 \left(i \alpha_0+ \sqrt{\beta_0}\right)}\right]}, \qquad C_2(k)=0,
\]
one can recover the normalized wavefunctions of standard inflation \cite{b&d} and ghost inflation \cite{ghost}. We can  rewrite the solution as

\be
\fl \pi(k,\tau)=-\frac{H i e^{\frac{1}{2} i \sqrt{\beta_0} k^2 \tau ^2}\,\, \Gamma\left[\frac{5}{4}-\frac{i \alpha_0}{4 \sqrt{\beta_0}}\right] {\rm HyperG} \left(-\frac{1}{4}-\frac{i \alpha_0}{4 \sqrt{\beta_0}},-\frac{1}{2},-i \sqrt{\beta_0} k^2 \tau ^2\right)}{ \sqrt{2(M_{\rm Pl}^2 \epsilon H^2 +2 M_2^4-3 \bar M_{1}^{3}H)}      \left(\alpha_0+\sqrt{\beta_0}\right)^{3/4} k^{3/2}\,\, \Gamma\left[\frac{5}{4}+\frac{\alpha_0}{4 \alpha_0 -4 i \sqrt{\beta_0}}\right]}\label{pi}, 
\ee
whose asymptotic expression for $\tau \rightarrow 0 $ reads
\be
\fl \pi(k,\tau\simeq 0)\simeq \frac{i H \sqrt{\pi/8}}{\sqrt{M_{\rm Pl}^2 \epsilon H^2 +2 M_2^4-3 \bar M_{1}^{3}H} \,\left(\alpha_0+\sqrt{\beta_0}\right)^{3/4} k^{3/2}\,\,  \Gamma\left[\frac{3}{2}-\frac{\sqrt{\beta_0}}{4 \left(i \alpha_0+\sqrt{\beta_0}\right)}\right]}.\label{asy}
\ee
 Note here the normalization factor in the denominator of Eq.~(\ref{asy}):
\[
\sqrt{M_{\rm Pl}^2 \epsilon H^2 +2 M_2^4-3 \bar M_{1}^{3}H}\equiv {\cal N}.
\]
By looking at Eq.~(\ref{action}) one understands that  this factor originates from  canonically normalizing the  kinetic term, which leads to
to  $\pi_c = {\cal N} \pi$ .
Let us also mention that the expected behaviour of the wavefunction at horizon crossing  we commented above, 
 $\dot \pi \sim H \pi$ and $\nabla \pi \sim H/c_s\,\, \pi$,  are now generalized to 
\be
\dot\pi \sim H \pi, \qquad \nabla \pi \sim    \frac{\sqrt{2}H}{\sqrt{\alpha_0 + \sqrt{\alpha_0^2 + 8 \beta_0}}}\,\, \pi \label{est} 
\ee
It is  then clear  how one can in principle obtain high level of non-Gaussianity by means of  terms in the Lagrangian with space derivatives (requiring $\alpha_0, \beta_0 \ll 1$ in the process) rather than time-like ones. 
A more detailed analysis of the general solution to the equation of motion and of the resulting power spectrum is to be found in \cite{p.s.}.

\section{Amplitude of the primordial non-Gaussianity}
\label{ampl}
In this section we wish to perform a general analysis of the   amplitude of the bispectra stemming from the 
general  third-order interaction terms. The shape analysis will be done in the following section. In the calculations that follow we employ the so-called {\rm in-in} 
formalism \cite{in-in1, in-in3, in-in2, w-qccc}. To compute the amplitude of the non-Gaussianity, indicated by   $f_{\rm NL}$, we proceed as traditionally done in the literature and 
evaluate  the three-point correlator in the so-called  equilateral configuration where all momenta are taken to be equal: $k_1 = k_2 = k_3$. In other words one can write the bispectra of the gauge-invariant 
curvature perturbation $\zeta$ generated by each interaction term (I) as  
\begin{eqnarray}
\langle \zeta(\textbf{k}_1)\zeta(\textbf{k}_2)\zeta(\textbf{k}_3)\rangle_{I}
=(2\pi)^3\delta^3(\textbf{k}_1+\textbf{k}_2+\textbf{k}_3) B_I(k_1,k_2,k_3)\, ,
\end{eqnarray} 
with an amplitude $f^I_{\rm NL}$ defined so that with all three-momenta equal $f^I_{\rm NL}=(6/5) B_I(k,k,k)/P_{\zeta}(k)^2$, where $P_{\zeta}(k)$ is the power spectrum of the curvature perturbation. 
For large non-Gaussianities the linear relation  $\zeta=-H \pi$ suffices  for the bispectrum calculations since quadratic corrections give a negligible contributions. 
Notice also that the values of the integrals appearing in the in-in computations have been specified at horizon crossing 
and the contribution of the integral function at $\tau=-\infty$  has been put to zero mimicking  the effect of the slight rotation of the  $\tau$-axis into the imaginary plane.

Given the broad number of possibilities,  we choose to compute numerically the amplitude of the various bispectra  identifying  six  benchmark points in the $(\alpha_0, \beta_0)$  plane and numerically integrating the exact wavefunctions. Since one is interested in probing models with large non-Gaussianity, the values of $(\alpha_0, \beta_0)$ are taken much smaller than unity as described in the following Table:

\begin{center}
\begin{tabular}{| l || l | l | l | l | l| l | }
\hline			
       Benchmarks   & 1 & 2 &3 &4 &5 &6 \\ \hline 
  $\alpha_0$ & $10^{-2}$  & 0 & $0.5\cdot 10^{-2}$ & $2 \cdot 10^{-7}$ &$10^{-4}$  &$10^{-6}$ \\ \hline
  $\beta_0$  &  0 & $0.5\cdot 10^{-4} $ & $0.25\cdot 10^{-4} $ & $5\cdot 10^{-5}$ & 0&0  \\
\hline  
\end{tabular}
\\
\vspace{5mm}
\end{center}
 Upon using Eq.~(\ref{ab}) one can check that the first four benchmark points   correspond to the same choice of the 
 effective horizon. 
 This choice has been made to suitably perform a comparison between the various cases and in particular against the values of  $f_{\rm NL}$ of purely DBI and  Ghost models which correspond to benchmarks 
1 and 2. 
 The last two configurations (which are pure DBI) probe the space of extremely small $\alpha_0$ and, in interactions with at least two space derivatives, are expected to give a larger amplitude than the first four points, at least if the  interaction terms are regulated by unconstrained masses. To get the feeling for the figures involved, if we restrict ourselves  to the case of theories for which $\alpha_0$ reduces to  the usual sound speed $c_s^2$, typical values of $c_s$ are between  $10^{-3}$ and $10^{-2}$. Note also that the definition of $c_s$ varies according to which operators are switched on in the action (one can well be 
in the de Sitter limit where the only spatial quadratic term has four derivatives; this leads 
to a different $c_s$, see also \cite{ssz05}). Therefore we choose to specify all the amplitudes as a function of the various $M$ and $\bar M$ masses.

 It is convenient at this stage to clearly point out which $M$ coefficients in Eq.~(\ref{action}) are free and underline the relations among the constrained ones. From Eq.~(\ref{ab}) one can see that,
despite fixing $\alpha_0$, as we did in the Table, there is still  the freedom to pick any reasonable value for either $ M_2^4$ or $\bar M_1^3$. Similarly, fixing $\beta_0$ does not completely specify $\bar M_0$. In other words, both $\bar M_1^{3}$ and  $\bar M_0^{2}$ are constrained by our choice of the $(\alpha_0, \beta_0)$ parameters; all the other coefficients are unconstrained. As the  $M$'s are expected to set the   energy  scale of the various underlying theories, they should be  larger than the Hubble rate $H$, and can go up to  $M_{\rm Pl}$.
As elucidated in Ref. \cite{w-e} though , for the action (\ref{action}) to be as general as possible,  one might want to require the $M$'s to be smaller than the Planck mass. That said, some useful inequalities that the mass coefficients must respect can now be reminded. Due to the fact we are working in the decoupling regime, we must require $M_2^4 < M_{\rm Pl}^2 H^2$. Also, the fact that we are probing the 
$(\alpha_0, \beta_0) \ll 1 $ space, imposes  bounds on some masses.  Consider the parameter $\alpha_0$ in Eq.~(\ref{ab}). There are two ways this coefficient can be much smaller than unity. The first and perhaps most natural way, is to ask $M_2^4 \gg {\rm Max}\,(-M_{\rm Pl}^2 \dot H, -\bar M_1^3 H)$ which, due to decoupling inequalities on $M_2$ puts a bound on $\bar M_1$,  $\bar M_1^3 \ll M_{\rm Pl}^2 H$. The other possibility requires a partial cancellation in the numerator of $\alpha_0$, $-M_{\rm Pl}^2 \dot H \sim \bar M_1^3 H/2$ which is certainly possible but it implies  we are neither in the DBI, nor in the ghost regime, both of which have $\bar M_1=0$. Looking at $\beta_0$ we see it is  enough to require $M_0^2 H^2 \ll M_2^4$ or $M_0^2 H^2 \ll -\bar M_1^3 H$ and again, the first condition seems more natural. Let us stress here that, upon requiring the masses to be all of the same order, $M$,  and using that $H/M\ll 1$ in the effective theory,  one can easily obtain small $\alpha_0, \beta_0$ coefficient. 
However, when employing a single mass scale $M$  in the whole Lagrangian, working with tiny values for $\alpha_0$ and $\beta_0$ would put a bound on $M$ and necessarily influence the magnitude of all the interaction terms.
In our analysis we let the $M,\bar M$'s coefficients  be not all of the same order (with some important caveats upon which we expand at the end of this section).

 Below we present the results for each interaction term. All the amplitudes can be written as a dimensionless coefficient, $\gamma_n$, times an ($\alpha_0,\beta_0$)-dependent numerical coefficient.
The terms described in the first subsection  are interactions that have already been discussed in the literature. The novelty here is represented by the fact we are able to study also  interpolating configurations through the third and fourth benchmark points. In the second subsection we report on the amplitudes of the contributions from some curvature-generated terms that have never been discussed in the literature.
 
\subsection{Amplitudes from DBI-like interactions and first two curvature-generated terms}
The amplitudes from DBI-like interactions and first two curvature-generated terms are the following: 
\vskip 0.5cm

 $\bullet$ $ {\cal O}_1=-2 M_2^4 \dot\pi (\partial_i \pi)^2/a^2$

\begin{center}
\begin{tabular}{ |l || l | l | l | l | l| l | }
\hline			
benchmarks   & 1 & 2 &3 &4 &5 &6 \\ \hline 
$f^{M_2}_{\rm NL}$   &     $\,\,\, 10^{2} \gamma_1 \,\,\, $     &    $\,\,\, 8 \cdot  10^{1}\gamma_1\,\,\, $       &     $\,\,\, 6\cdot 10^{1}\gamma_1\,\,\,$     &      $\,\,\, 4\cdot  10^{1}\gamma_1\,\,\,$      &      $\,\,\,  10^{4}\gamma_1\,\,\,$    &    $\,\,\,  10^{6}\gamma_1\,\,\,$  \\ 
  
\hline  
\end{tabular}
\end{center}
 where 
\be 
\gamma_1= \frac{M_2^4}{2M_2^4 +M_{\rm Pl}^2 \epsilon H^2-3 \bar M_{1}^{3}H}.
\ee
We see the parameter $\gamma_1$ can in principle be of order unity. Indeed, if one assumes $M_2^4$ is the largest  term in the denominator (in DBI this would correspond to a very small speed of sound), $\gamma_1$ is roughly $1/2$.\\

$\bullet$ ${\cal O}_2=-4/3 \,\,  M_3^4 {\dot\pi}^3 $\\

\begin{center}
\begin{tabular}{| l || l | l | l | l | l| l | }
\hline			
benchmarks   & 1 & 2 &3 &4 &5 &6 \\ \hline 
$f^{M_3}_{\rm NL}$   &     $\,\,\, 1/2  \gamma_2 \,\,\, $     &    $\,\,\,  10^{-2}\gamma_2\,\,\, $       &     $\,\,\,5 \cdot  10^{-2}\gamma_2\,\,\,$     &      $\,\,\, 10^{-2}\gamma_2\,\,\,$      &      $\,\,\,  1/2 \gamma_2\,\,\,$    &    $\,\,\,  1/2\gamma_2\,\,\,$  \\ 
  
\hline  
\end{tabular}
\end{center}

\noindent where 
\be \gamma_2= \frac{M_3^4}{2M_2^4 +M_{\rm Pl}^2 \epsilon H^2-3 \bar M_{1}^{3}H}.
\ee
The parameter $\gamma_2$ can be even  larger than unity if, for instance,   $M_3$ is larger than  $M_2$. We can see, already at this stage, the effect of small values of $\alpha_0$ and  $\beta_0$ at work: the numerical factor of a spatial derivative-free interaction is much smaller than that of a third order term like the $M_2$ one calculated above which has two spatial derivatives. \\

$\bullet$  \textbf{$ {\cal O}_3=-1/2\,\,  \bar M_1^3 (\partial_i \pi)^2 \partial_j^2 \pi/a^4  $}\\

\begin{center}
\begin{tabular}{| l || l | l | l | l | l| l | }
\hline			
benchmarks   & 1 & 2 &3 &4 &5 &6 \\ \hline 
$f^{\bar M_1}_{\rm NL}$   &     $\,\,\,  10^{5} \gamma_3 \,\,\, $     &    $\,\,\,  10^{3}\gamma_3\,\,\, $       &     $\,\,\,  4 \cdot 10^{4}\gamma_3\,\,\,$     &      $\,\,\, 1.5\cdot  10^{3}\gamma_3\,\,\,$      &      $\,\,\, 10^9 \gamma_3\,\,\,$    &    $\,\,\,  10^{13}\gamma_3\,\,\,$  \\ 
  
\hline  
\end{tabular}
\end{center}
\vspace{5mm}
Due to the fact that $\bar M_1^3$ is a constrained parameter we find that the amplitude equals an $(\alpha_0, \beta_0)$-dependent number times the paramter
\be
\gamma_3 = \frac{-M_{\rm Pl}^2 \epsilon H^2 (1-\alpha_0)+ 2 \alpha_0 M_2^4}{M_{\rm Pl}^2 \epsilon H^2 + 2  M_2^4-3\bar M_1^3 H}.
\ee
One should substitute the various values of $\alpha_0$ in the Table in the $\gamma_3$ expression given above. Barring cancellation between different mass terms, $\gamma_3$ is generally smaller than one and can be as small as $\alpha_0$ itself.\\

$\bullet$  \textbf{$ {\cal O}_4=-1/2 \,\, \bar M_0^2/4\,\,  \left(5H (\partial_i^2\pi) (\partial_j \pi)^2  +4\dot\pi \partial_{i}^2 \partial_j \pi \partial_j \pi \right)/ a^4 $}\\

\begin{center}
\begin{tabular}{| l || l | l | l | l | l| l | }
\hline			
benchmarks   & 1 & 2 &3 &4 &5 &6 \\ \hline 
$f^{\bar M_0}_{\rm NL}$   &     $\,\,\, 3\cdot  10^{5}\,  \gamma_4 \,\,\,$     &    $\,\,\,  10^{5}\, \gamma_4\,\,\, $       &     $\,\,\, 1.3 \cdot 10^{4}\, \gamma_4\,\,\,$     &      $\,\,\,  10^{5}\, \gamma_4\,\,\,$      &      $\,\,\, 3\cdot 10^9 \, \gamma_4\,\,\,$    &    $\,\,\, 3\cdot  10^{13}\, \gamma_4\,\,\,$  \\ 
\hline  
\end{tabular}
\end{center}
\vspace{5mm}\
with 
\be
 \gamma_4 = (\bar M_0^2 H^2 )/(M_{\rm Pl}^2 \epsilon H^2 + 2 M_2^4-3\bar M_1^3 H).
\ee
We see that large numerical coefficients appear in such a case, nevertheless    $\gamma_4=\beta_0\ll 1$.




\subsection{ Amplitudes from curvature-generated novel interaction terms}
We come to the curvature-generated  interaction terms that generate novel bispectra. Their amplitudes are given by
\vskip 0.5cm

$\bullet$  {\bf$ {\cal O}_5=-2/3\,\,  \bar M_4^3 {\dot\pi}^2 \partial_i^2\pi \,\,/ a^2 $}\\

\begin{center}
\begin{tabular}{| l || l | l | l | l | l| l | }
\hline			
benchmarks   & 1 & 2 &3 &4 &5 &6 \\ \hline 
$f^{\bar M_4}_{\rm NL}$   &     $\,\,\, 7 \cdot 10^{2}\, \gamma_5 \,\,\, $     &    $\,\,\,  10^{2} \,\gamma_5\,\,\, $       &     $\,\,\,2 \cdot 10^{2}\, \gamma_5\,\,\,$     &      $\,\,\,  10^{2}\, \gamma_5\,\,\,$      &      $\,\,\, 7\cdot 10^4\, \gamma_5\,\,\,$    &    $\,\,\,   7\cdot 10^6 \,\gamma_5\,\,\,$  \\ 
\hline  
\end{tabular}
\end{center}
\vspace{5mm}

where 
\be 
\gamma_5 = (\bar M_4^3 H )/(M_{\rm Pl}^2 \epsilon H^2 + 2 M_2^4-3\bar M_1^3 H).
\ee 
The coefficient $\gamma_5$ can be larger than unity. Barring cancellations in the denominator, a $\gamma_5\gg1 $ larger than unity  imposes  $\bar M_4 \gg M_2, \bar M_1$. \\

$\bullet$ {\bf $ {\cal O}_6=1/3\,\,  \bar M_5^2 \dot\pi (\partial_i^2\pi)^2 \,\,/ a^4 $}\\

\begin{center}
\begin{tabular}{ |l || l | l | l | l | l| l | }
\hline			
benchmarks   & 1 & 2 &3 &4 &5 &6 \\ \hline 
$f^{\bar M_5}_{\rm NL}$   &     $\,\,\, 5\cdot  10^{4}\, \gamma_6 \,\,\, $     &    $\,\,\, 1.6 \cdot  10^{4}\,\gamma_6\,\,\, $       &     $\,\,\,2 \cdot 10^{4}\, \gamma_6\,\,\,$     &      $\,\,\, 1.6 \cdot  10^{4}\,\gamma_6\,\,\,$      &      $\,\,\, 4\cdot 10^8\, \gamma_6\,\,\,$    &    $\,\,\, 4 \cdot 10^{12}\gamma_6\,\,\,$  \\ 
\hline  
\end{tabular}
\end{center}
\vspace{5mm}
 where 
\be
\gamma_6 = (\bar M_5^2 H^2 )/(M_{\rm Pl}^2 \epsilon H^2 + 2 M_2^4 -3\bar M_1^3 H).
\ee
To get $\gamma_6\gg 1$, one needs to impose a less natural condition $ \bar M_5^2 H^2\gg M_2^4, \bar M_1^3 H $. \\

$\bullet$  {\bf $ {\cal O}_7=1/3\,\,  \bar M_6^2 \dot\pi (\partial_{ij}\pi)^2 \,\,/ a^4 $}\\

\begin{center}
\begin{tabular}{| l || l | l | l | l | l| l | }
\hline			
benchmarks   & 1 & 2 &3 &4 &5 &6 \\ \hline 
$f^{\bar M_6}_{\rm NL}$   &     $\,\,\,  10^{4}\, \gamma_7 \,\,\, $     &    $\,\,\, 4\cdot  10^{3}\,\gamma_7\,\,\, $       &     $\,\,\,5\cdot 10^{3}\,\gamma_7\,\,\,$     &      $\,\,\, 4\cdot 10^{3}\,\gamma_7\,\,\,$      &      $\,\,\, 10^8\,\gamma_7\,\,\,$    &    $\,\,\, 10^{12}\,\gamma_7\,\,\,$  \\ 
\hline  
\end{tabular}
\end{center}
\vspace{5mm}

where
\be
\gamma_7 = (\bar M_6^2 H^2 )/(M_{\rm Pl}^2 \epsilon H^2 + 2 M_2^4 -3\bar M_1^3 H).
\ee 
The same consideration as for the case of $\bar M_5$ apply here. Note again that the numerical values, especially in the fifth and sixth benchmark points tend to be much larger for the $M$ coefficients with the most spatial derivatives, thus conferming our expectations.\\

$\bullet$  {\bf ${\cal O}_8 = -1/6\,\,  \bar M_7 (\partial_i^2 \pi)^3 \,\,/ a^6 $}\\

\begin{center}
\begin{tabular}{| l || l | l | l | l | l| l | }
\hline			
benchmarks   & 1 & 2 &3 &4 &5 &6 \\ \hline 
$f^{\bar{M}_7}_{\rm NL}$   &     $\,\, 8 \cdot 10^6\,\gamma_8 \,\, $     &    $\,\, 2.6 \cdot 10^{6}\,\gamma_8\,\,$       &     $\,\,3.5 \cdot 10^{6}\,\gamma_8\,\,$     &      $\,\,  2.6 \cdot 10^{6}\,\gamma_8\,\,$      &      $\,\, 8\cdot 10^12 \gamma_8\,\,$    &    $\,\, 8\cdot 10^{18}\gamma_8\,\,$  \\ 
\hline  
\end{tabular}
\end{center}
\vspace{5mm}

with 
\be
\gamma_8 = (\bar M_7 H^3 )/(M_{\rm Pl}^2 \epsilon H^2 + 2 M_2^4  -3\bar M_1^3 H).
\ee 
Notice that in this case , and in the following ones, the $\gamma$ coefficients are naturally expected to be smaller than unity. 
For this interaction, as well as the two following ones, the numerical factor coming from the integration can be quite large, especially in the fifth and sixth benchmark points. 
This is clearly due to the six space derivatives that characterize these interactions. \\

$\bullet$  {\bf $ {\cal O}_9=-1/6\,\,  \bar M_8 \, \partial_i^2 \pi (\partial_{jk} \pi)^2 \,\,/ a^6 $}\\

\begin{center}
\begin{tabular}{| l || l | l | l | l | l| l | }
\hline			
benchmarks   & 1 & 2 &3 &4 &5 &6 \\ \hline 
$f^{\bar{M}_8}_{\rm NL}$   &     $\,\, 2\cdot 10^6 \gamma_9 \,\, $     &    $\,\,  6 \cdot 10^5 \gamma_9\,\, $       &     $\,\, 8\cdot  10^{5}\gamma_9\,\,$     &      $\,\, 6\cdot  10^{5}\gamma_9\,\,$      &      $\,\, 2\cdot 10^12 \gamma_9\,\,$    &    $\,\,  2 \cdot  10^{18}\gamma_9\,\,$  \\ 
\hline  
\end{tabular}
\end{center}
\vspace{5mm}

with 
\be 
\gamma_9 = (\bar{M}_8 H^3 )/(M_{\rm Pl}^2 \epsilon H^2 + 2 M_2^4 -3\bar M_1^3 H).
\ee
Let us stress that one could guess the amplitude of $\bar M_{8,9}$ terms by simply looking at the results obtained for the interaction term tuned by $\bar M_7$ because, althought these terms might produce a different shape for non-Gaussianities, they have essentially the same structure as far as the integration is concerned.\\

$\bullet$  {\bf $ {\cal O}_{10}=-1/6\,\,  \bar M_9 \, \partial_{ij} \pi \partial_{jk} \pi  \partial_{ki} \pi \,\,/ a^6 $}\\

\begin{center}
\begin{tabular}{ |l || l | l | l | l | l| l | }
\hline			
benchmarks   & 1 & 2 &3 &4 &5 &6 \\ \hline 
$f^{\bar M_9}_{\rm NL}$   &     $\,\,10^6 \gamma_{10} \,\, $     &    $\,\,  3\cdot 10^{5}\gamma_{10}\,\, $       &     $\,\,  4\cdot  10^{5} \gamma_{10}\,\,$     &      $\,\, 3\cdot  10^{5}\gamma_{10}\,\,$      &      $\,\, 10^{12} \gamma_{10}\,\,$    &    $\,\, 10^{18}\gamma_{10}\,\,$  \\ 
\hline  
\end{tabular}
\end{center}
\vspace{5mm}

where
\be
\gamma_{10} = (\bar M_9 H^3 )/(M_{\rm Pl}^2 \epsilon H^2 + 2 M_2^4 -3\bar M_1^3 H).
\ee 
One could again just read off the maximum value from $f^{\bar M_7}_{\rm NL}$, they differ by just a factor $1/8$. 
\subsection{Some general considerations on the amplitudes of non-Gaussianities}
It might be worth now to pause and   comment on our findings in relation to earlier results in the literature. As we have stressed, the non-Gaussianity generated from operators proportional 
to the  $\bar M$'s masses arises from curvature terms that are often neglected.  From general considerations, one expects that  the curvature terms, 
especially the ones coming from ${\delta K}^3$-type of contributions,  can often be neglected. This is not generically true though. The general structure of the amplitudes we have found for the bispectra can be schematically written as
\be
{\cal A}_n \sim \frac{1}{\left(\sqrt{\alpha_0 +\sqrt{\alpha_0^2 +8 \beta_0}}\right)^q} \frac{\widetilde{M}_n^{4-q} H^q}{2M_2^4 + M_{\rm Pl}^2 \epsilon H^2  -3\bar M_1^3 H}\, ,
\ee
where $\tilde M_n$ stands for a generic mass coefficient of the type $M, \bar M$.\\ If $\widetilde{M}_n \gg H/(\alpha_0 +(\alpha_0^2+8 \beta_0)^{1/2})^{1/2}$, and the masses are all of the same order, then it is natural to expect the $\widetilde{M}_n$ with the largest exponent to dominate. In this case the dominant terms are those associated 
with $M_2$ and $M_3$, corresponding to DBI inflation.  
Furthemore, we may also recall that large masses appear also in the definition of $\alpha_0, \beta_0$ and, at least in DBI, a large $M_2$ is needed to have a small speed of sound. 

 On the other hand, one can employ the freedom for all the $M$'s not to be of the same order in magnitude. The $\widetilde{M}_n$'s have a natural upper bound that must be smaller than the Planck mass and in the general theory with both DBI and ghost (and more in general curvature terms) operators switched on, one can allow for a very small speed of sound without assuming much on the unconstrained masses. In such cases, the generalized sound speed can be so small that $H/(\alpha_0 +(\alpha_0^2 +8\beta_0)^{1/2})^{1/2}$ is actually larger than $\widetilde M$. Consequently, 
  the amplitudes of the masses with a smaller exponent, typically the barred $\bar M$'s, are not negligible any longer.  We conclude
  that large non-Gaussianities may be induced by theories parametrized by  suitable values of the $\bar M$ masses.\\

It is important at this stage to offer some additional comments on the general structure of the Lagrangian and of the resulting amplitudes.
As clear from Eq.~(\ref{est}), each space derivative acting on the scalar $\pi$ can be schematically written in Fourier space as a term $H \pi$ divided by a number which is much smaller than unity in the cases of interest. Now, it is clear from dimensional analysis that other terms in the action, specifically the ones with less derivatives, will have their own $ H/(\alpha_0+(\alpha_0^2 +8\beta_0)^{1/2})^{1/2}$ factors coming from derivatives but, most importantly, will also  need some generic $M$ factor to make the action properly dimensionless. This is exactly what happens in our theory (see Eq.~(\ref{action})) as well: the exponent of $\bar M_1$ is bigger than the one of $\bar M_2$ and so on. We have employed in the analysis of the amplitudes some freedom on these $\bar M_n$ parameters to show that even terms with higher spatial derivatives can give non negligible contributions to the amplitude. Of course, if one wants to have a reliable effective theory the $\bar M_n$ coefficients must eventually prevail over $H/(\alpha_0+(\alpha_0^2 +8\beta_0)^{1/2})^{1/2}$ (in a Lorentz invariant theory it would suffice to ask for $\bar M_n\gg H$, here we need more) so that it makes sense to consider higher derivatives up to some given finite order, but not further.

\section{The shapes of non-Gaussianities}
In this section we wish to analyze the shapes of the bispectra generated by the   various operators analyzed above.
In calculating the amplitudes on the bispectra in the previous section we have chosen to perform, albeit numerically, calculations with the exact wavefunctions. We decided to counterbalance the loss of information in not having $\alpha_0, \beta_0$ explicit in the result by running the same procedure for six different  benchmark values of the parameters.  In analaysing the shapes we do not enjoy the possibility to use the exact wavefunction any longer as we  must approximate the wavefunctions inside the conformal time $\tau$ integral(s) calculated as prescribed by the in-in formalism. Of course, exact results are always available when the classical solution reduces to the usual, Hankel function $H_{(3/2)}(\tau)$ .
In computing terms like

\be
\int^{\tau}_{-\infty} d \widetilde\tau \frac{1}{H^4 \widetilde\tau^{4}} \left[ \dot\pi_{k1}(t(\widetilde\tau)) \dot\pi_{k2}(t(\widetilde\tau)) \dot\pi_{k3}(t(\widetilde\tau))         \right]\, ,
\ee
we have  choosen  to expand in series each $k$-mode inside the integrand within a region that starts from slightly inside its effective horizon (setting to zero the function in the rest of the interval). This enables us to keep both, parameters and external momenta, arbitrary. This approximation  is justified by the fact that, due to the oscillatory behaviour of the wavefunctions\footnote{Of course, an oscillatory behaviour is not, by itself, enough to provide a cancellation all over the -inside the horizon- region. Indeed in some cases for $\tau \rightarrow - \infty$ the amplitude of the wavefunctions increases and one certainly does not expect this to give zero contribution. On the other hand, much like in the simplest single field slow-roll calculations, one is expected to slightly rotate $\tau$ to the imaginary plane to match the vacuum thus basically putting to zero the contribution at $- \infty$.}, the main contribution to the integral comes only  from the region where  all the wavefunctions are not oscillating anymore (to be safe, 
we actually choose to include the region where the $k$-mode with the latest effective horizon is still within its horizon). Furthermore, it is reasonable to assume that, even if some non-negligible contribution is being left over because of this approximation procedure, it might have a systematic effect on all $k$-modes and would therefore not change the  shapes of  the bispectrum of perturbations. This procedure has been calibrated with the calculations of Ref. \cite{chen-bis} which were performed exactly: we are able to reproduce  the very same shapes for the bispectrum.

Following the scheme used for the amplitudes, we now report below the shape of non-Gaussianity for each interaction term. Within a single interaction term, we consider four configurations in the $(\alpha_0, \beta_0)$-plane. We preliminary found that the shape are not particularly sensible to their absolute values. This is expected as these coefficients,  being the same for each $k$-mode, enter mainly  in the amplitudes, not in  the shapes. Therefore,  we limited our attention to the $\alpha_0/\beta_0$ ratios. We probed the following cases: $\alpha_0\not= 0, \beta_0 \simeq 0$  (DBI);  $\alpha_0\simeq 0, \beta_0\not= 0$ (ghost); $\alpha_0\not= 0 \not= \beta_0$ (with the solution interpolating between DBI and ghost inflation and $\alpha_0$ playing the dominating role in determining where is the effective horizon, we call this configuration $A$) and  $\alpha_0\not= 0 \not= \beta_0$ (with the effect of $\beta_0$ leading in the expression for the horizon, we call this configuration $B$). 

Let us briefly add some general preliminary comments.  We need to point out that when we use the name DBI might generate  some confusion. Indeed,  we use it also to  describe shapes due to interaction terms coming from curvature perturbations, such as $\bar M_{1,2,3....}$. What we mean here  is that we can employ in the integrations the usual wavefunction $\pi_k \propto e^{-i \sqrt{\alpha_0}k \tau }(1+i \sqrt{\alpha_0}k \tau)$ without resorting to any approximation. This wavefunction can be used as far as it is the solution to the equation of motion. This is a condition concerning only second order perturbations:  we need only require $\beta_0=0\Leftrightarrow 0= M_0$ to employ it. We can, at the same time, have extrinsic curvature-driven interaction terms, the $\bar M$'s, and yet find exact results.
 The shapes corresponding to what we denominate DBI configuration have been thoroughly investigated in a number of papers \cite{bispectrum}, but what has been generally left out is the contribution coming from the extrinsic curvature terms: it is indeed possible to have the usual Hankel, $H_{3/2}(\tau)$, wavefunction as a solution to the classical equations and, at the same time, switch on curvature operators like $\bar M_{4,5,6,7,8,9}$. All the shapes obtained in the DBI configuration are generated through exact analytical methods. We will call them as ``exact-DBI configurations''. Should we find, as we will, a shape which is not equilateral in the first configuration, that shape suffers none of any possible limitations the approximated method might introduce. The ghost configuration has been analyzed in depth in many articles, among which \cite{tilted,ssz05}, and again, not all the terms coming from extrinsic curvature have been taken into account. It is important to note though that in  \cite{eft08,tilted,ssz05} the $M$ coefficients multiplying curvature terms have all been chosen of the same order thus resulting in only a couple of leading curvature (they correspond to the $\bar M_1, \bar M_0$  contributions).
 Finally, the shapes in configuration \textit{$A$, $B$} have never been analyzed before.

The shapes reported below have been  obtained by employing the  shape function  $B(1,x_2,x_3)$,  where $x_i=k_i/k_1$,  which posseses  the same $k_{1,2,3}$-dependence as the three point function. What is plotted exactly is $x_2^2 x_3^2 B(1,x_2,x_3)/B(1,1,1)$ in the region satisfying $x_2 \geq x_3 \geq 1-x_2$~\cite{BabichCZ}. 
Let us remind the reader that a shape is called local when it peaks for a small value of, say,  $x_2$, with $x_1=x_3=1$; 
equilateral when it  peaks in the equilateral configuration $x_1=1, x_2=1=x_3$ and is called flat for squashed triangles with $1=x_2+x_3$. In particular, we find, as 
detailed below,  that some novel curvature-generated terms produce a flat bispectrum which specifically peaks for $x_2=x_3=1/2$.   
In presenting the shapes we follow the same order and organization we employed for the amplitudes:

\subsection{ Shapes from DBI-like interactions and first two curvature-generated terms.} 
\vskip 0.5cm

$\bullet$  ${\cal O}_1=-2 M_2^4 \dot\pi (\partial_i \pi)^2/a^2$

\begin{figure}[h]
\includegraphics[scale=0.50]{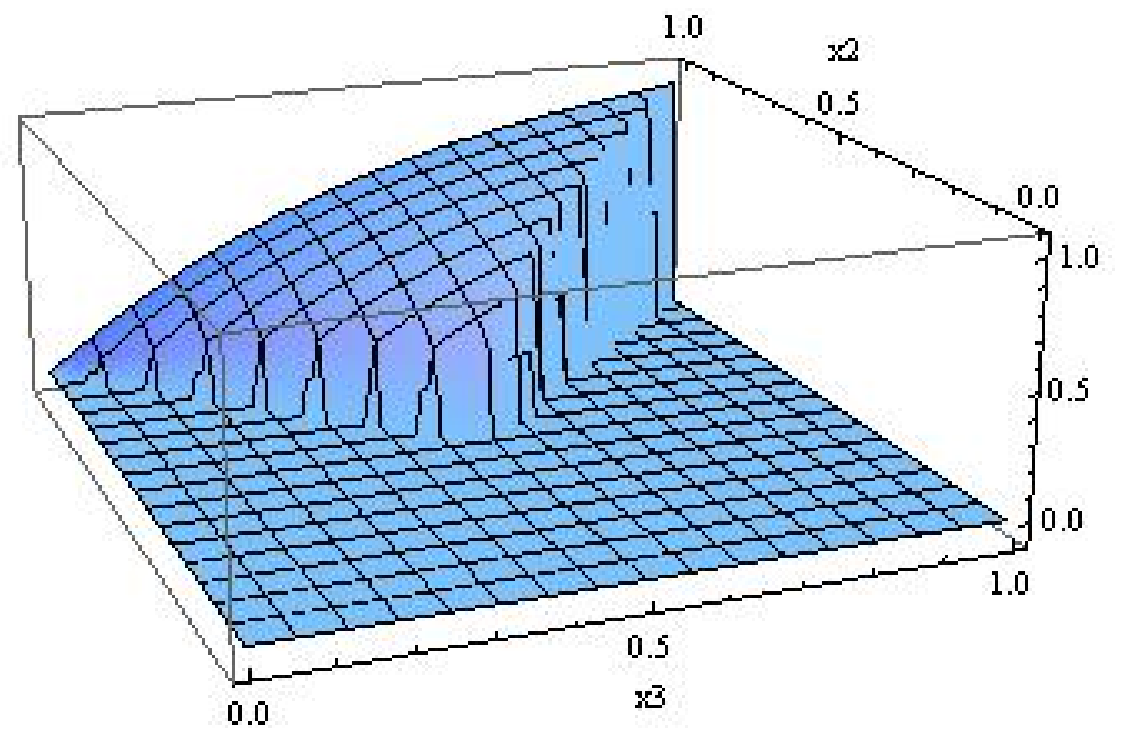}
\hspace{10mm}
\includegraphics[scale=0.43]{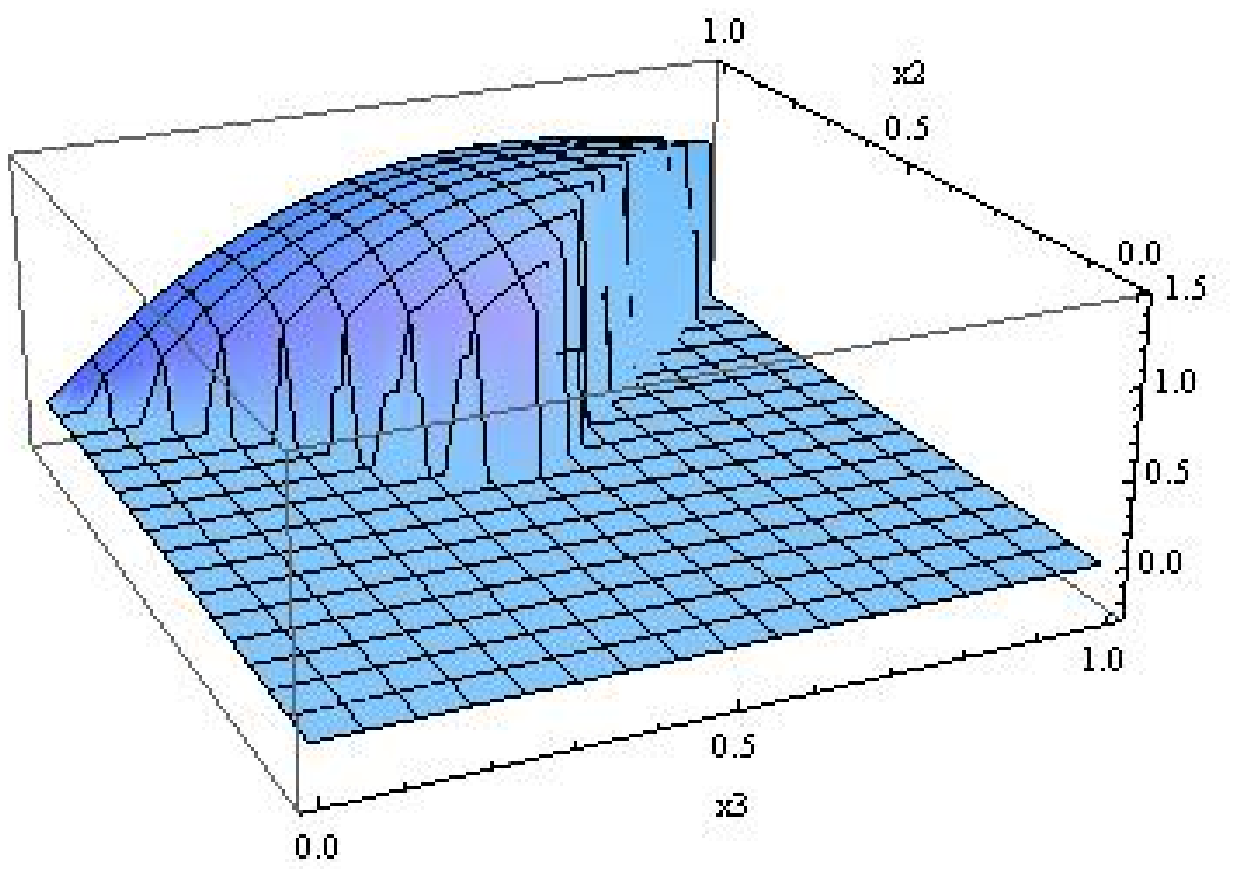}
	\label{fig:M2-1} 
	\caption{DBI configuration on the left, obtained using exact methods; the approximated ghost shape on the right.}
\end{figure}

\begin{figure}[h]
	\includegraphics[scale=0.50]{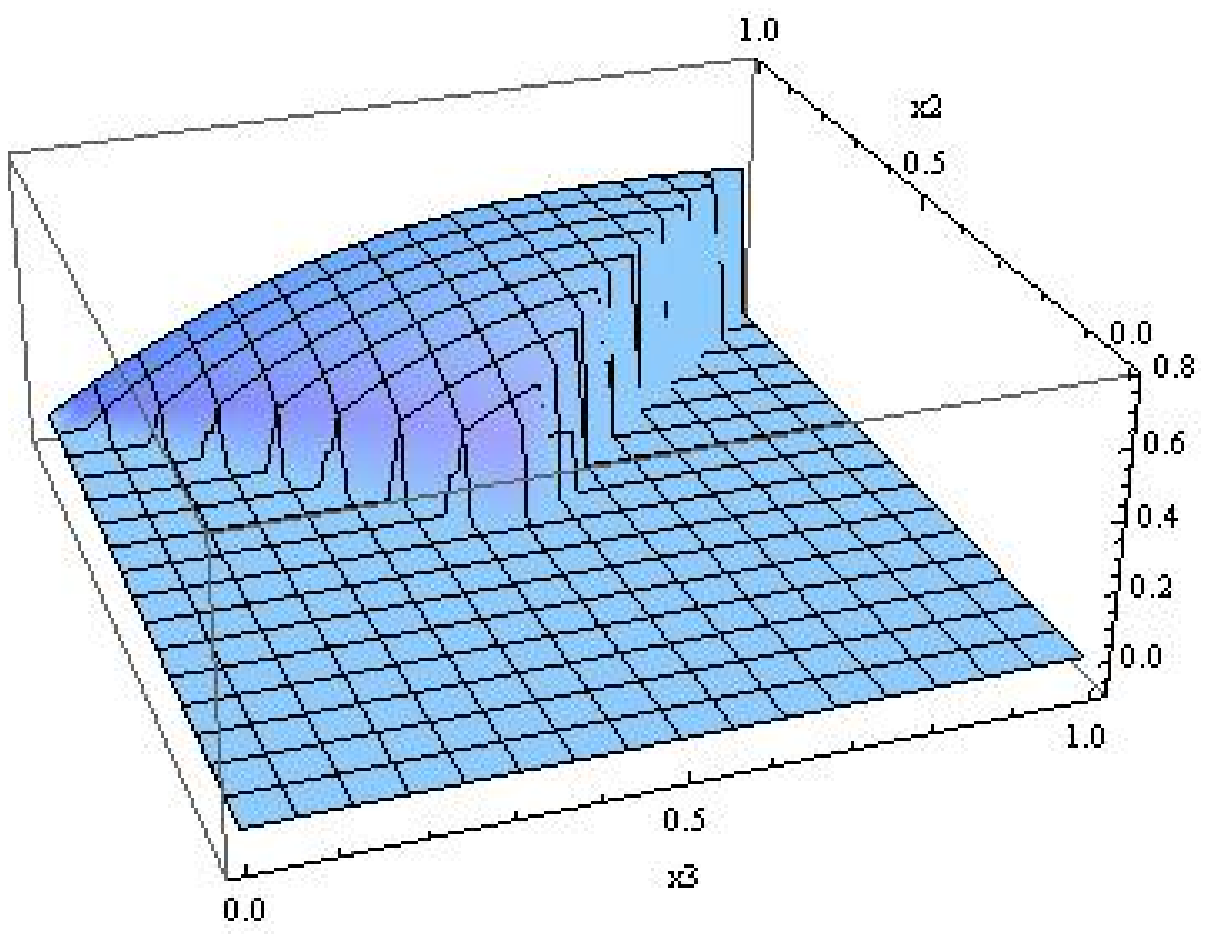}
\hspace{15mm}
	\includegraphics[scale=0.55]{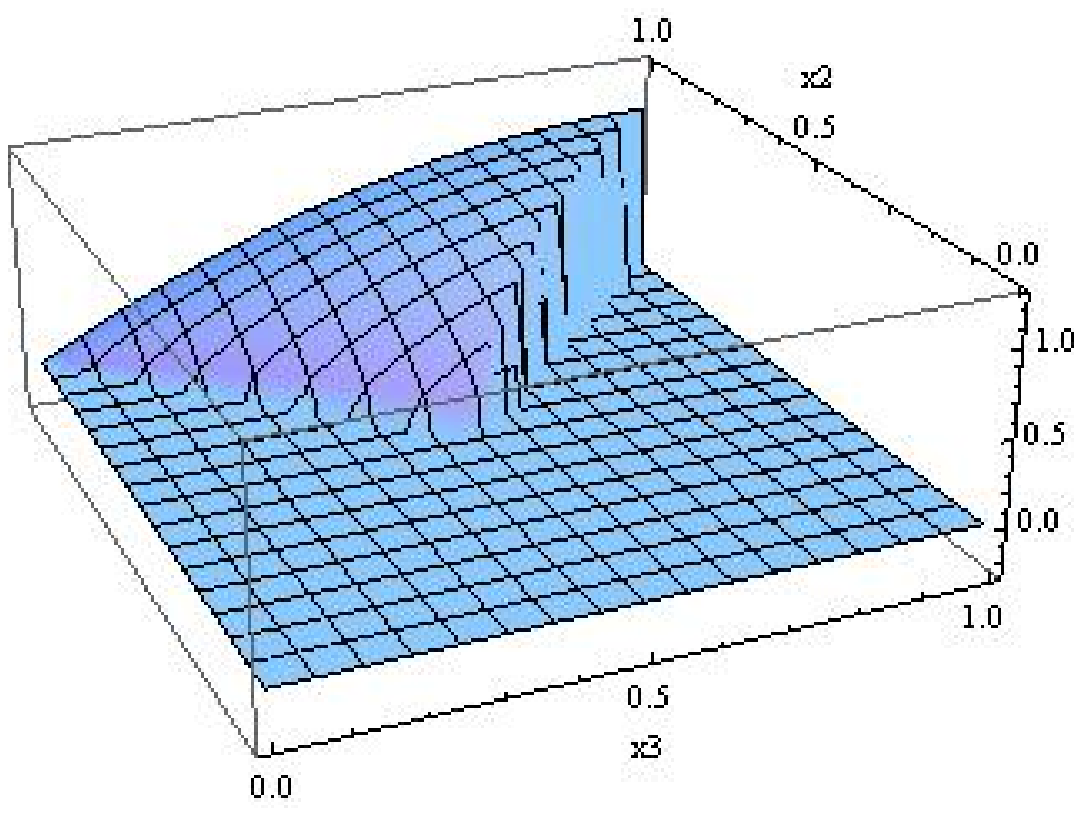}
	\label{fig:M2-2} 
	\caption{$A$ configuration on the left, $B$ configuration on the right.}
\end{figure}
Note that, although more or less sharply, all the four plots are peaked in the equilateral configuration.\\
\newpage 
$\bullet$  ${\cal O}_2=4/3 \,\,  M_3^4 {\dot\pi}^3 $\\

\begin{figure}[h]
	\includegraphics[scale=0.50]{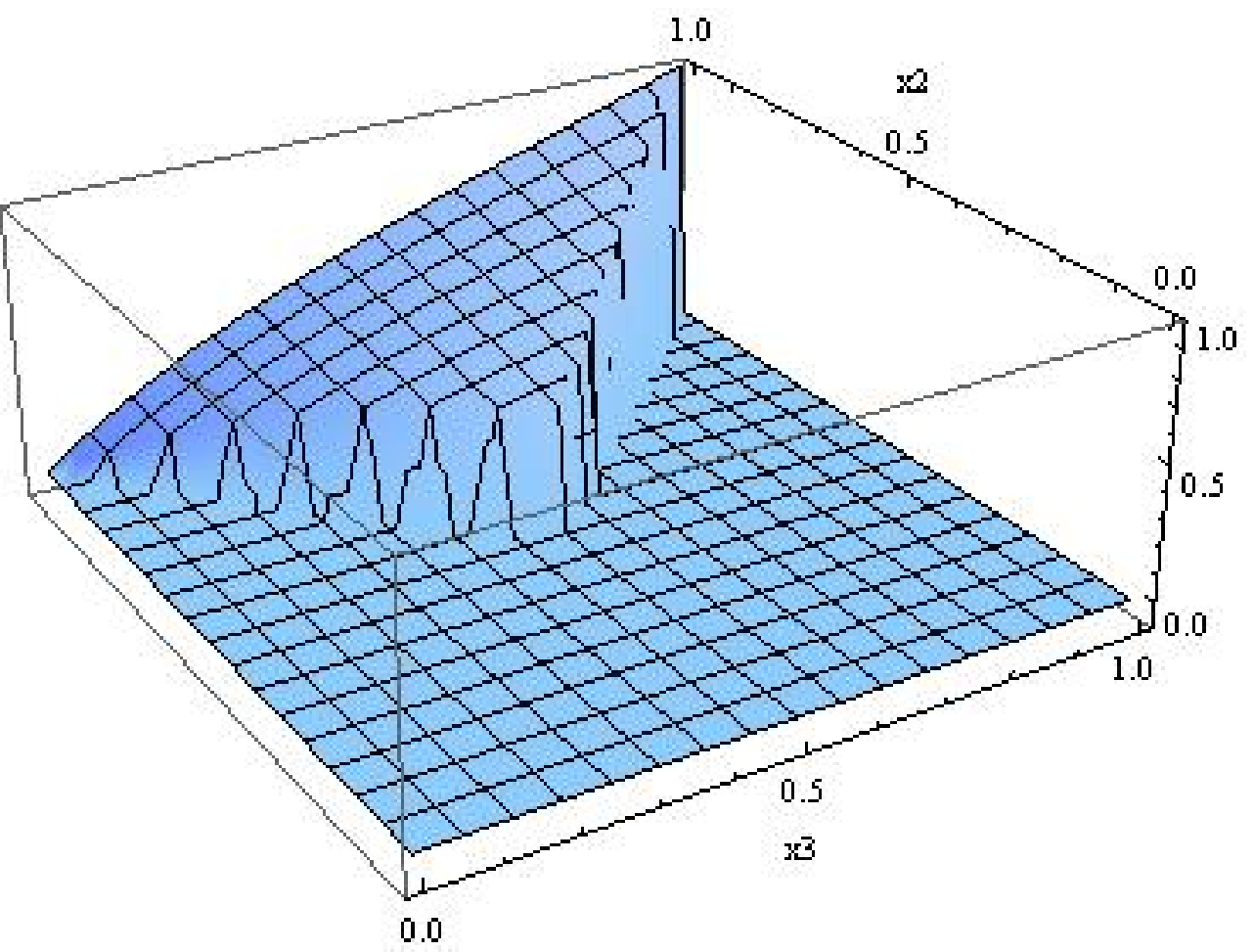}
\hspace{10mm}
	\includegraphics[scale=0.60]{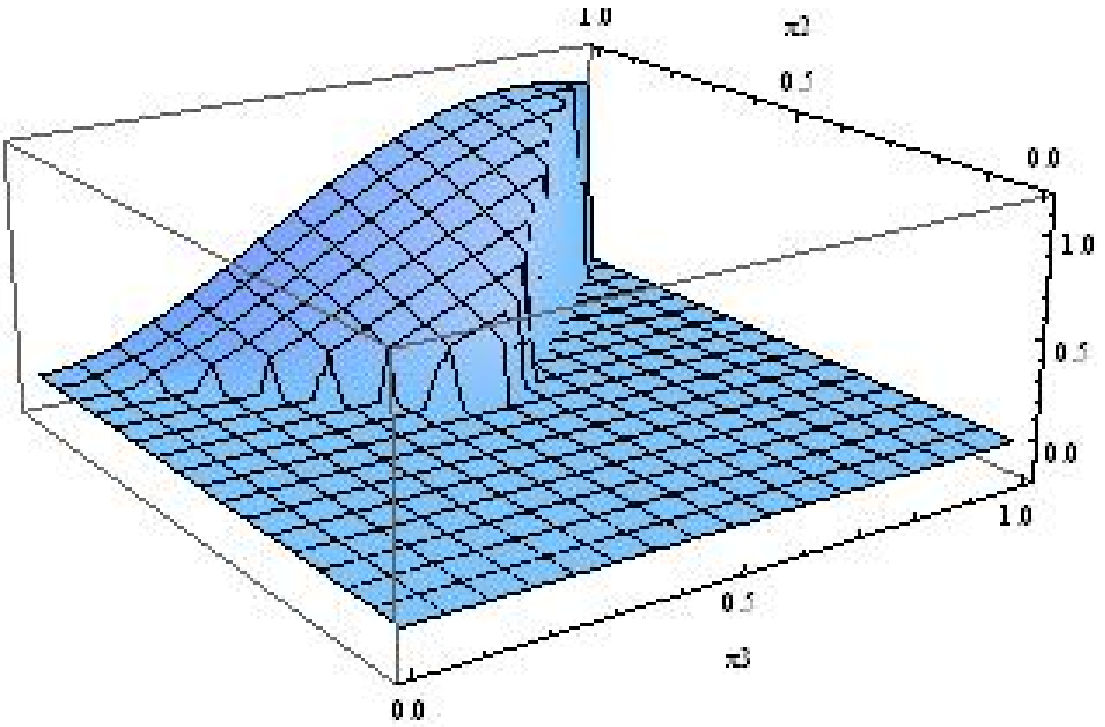}
	\label{fig:M3-1} 
	\caption{exact DBI configuration on the left; approximated ghost shape on the right.}
\end{figure}
\begin{figure}[h]
	\includegraphics[scale=0.65]{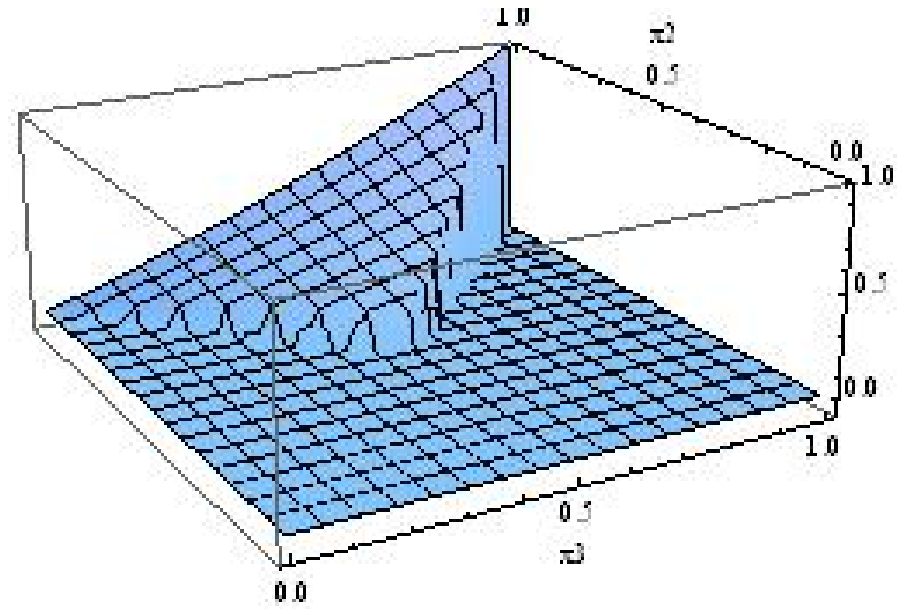}
\hspace{15mm}
	\includegraphics[scale=0.55]{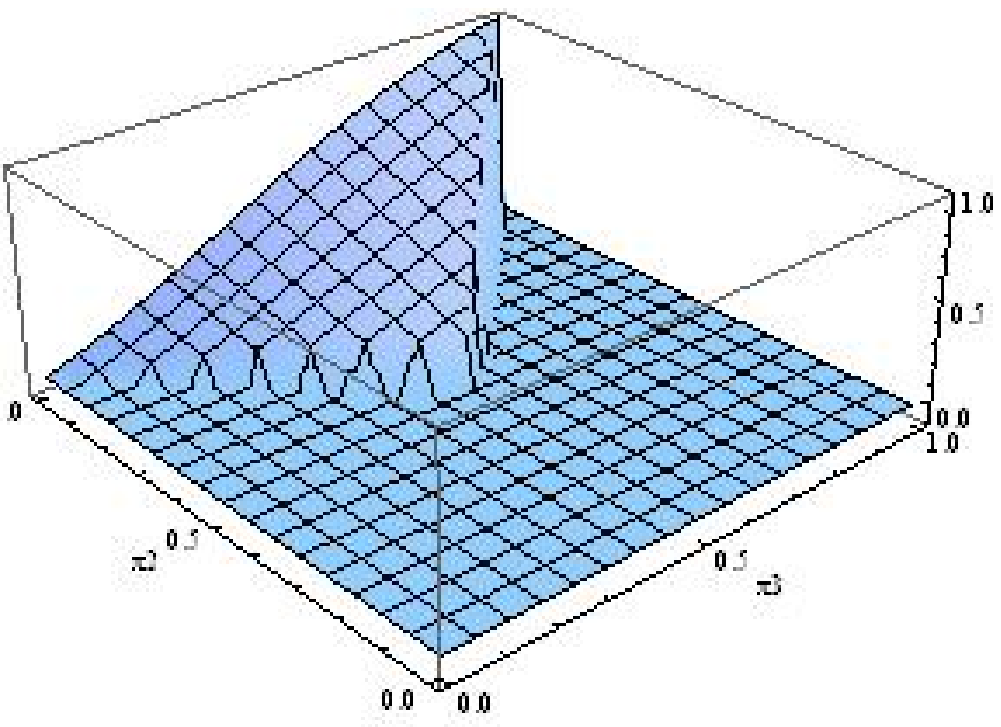}
	\label{fig:M3-2} 
	\caption{$A$ on the left, $B$ configuration on the right for the $M_3$-driven interaction term.}
\end{figure}
We obtain equilateral shapes in all four cases. It is somewhat expected that the general, interpolating 
solution employed in configurations \textit{$A$} and \textit{$B$}, will give qualitatively the same plot, we have verified it  it in these first two rounds of shapes.\\
\newpage 
$\bullet$  ${\cal O}_3= -1/2\,\,  \bar M_1^3 (\partial_i \pi)^2 \partial_j^2 \pi/a^4  $\\

\begin{figure}[h]
	\includegraphics[scale=0.55]{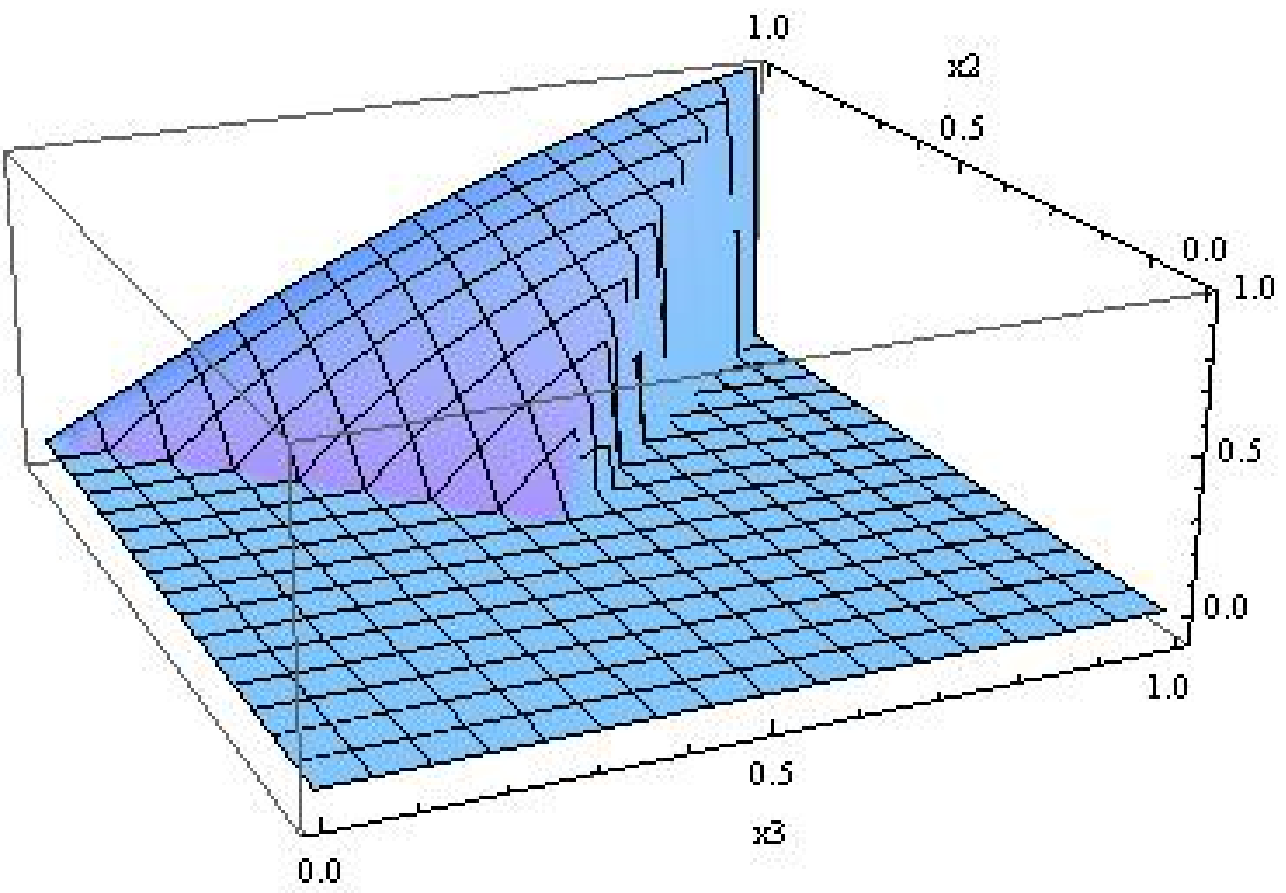}
\hspace{10mm}
	\includegraphics[scale=0.55]{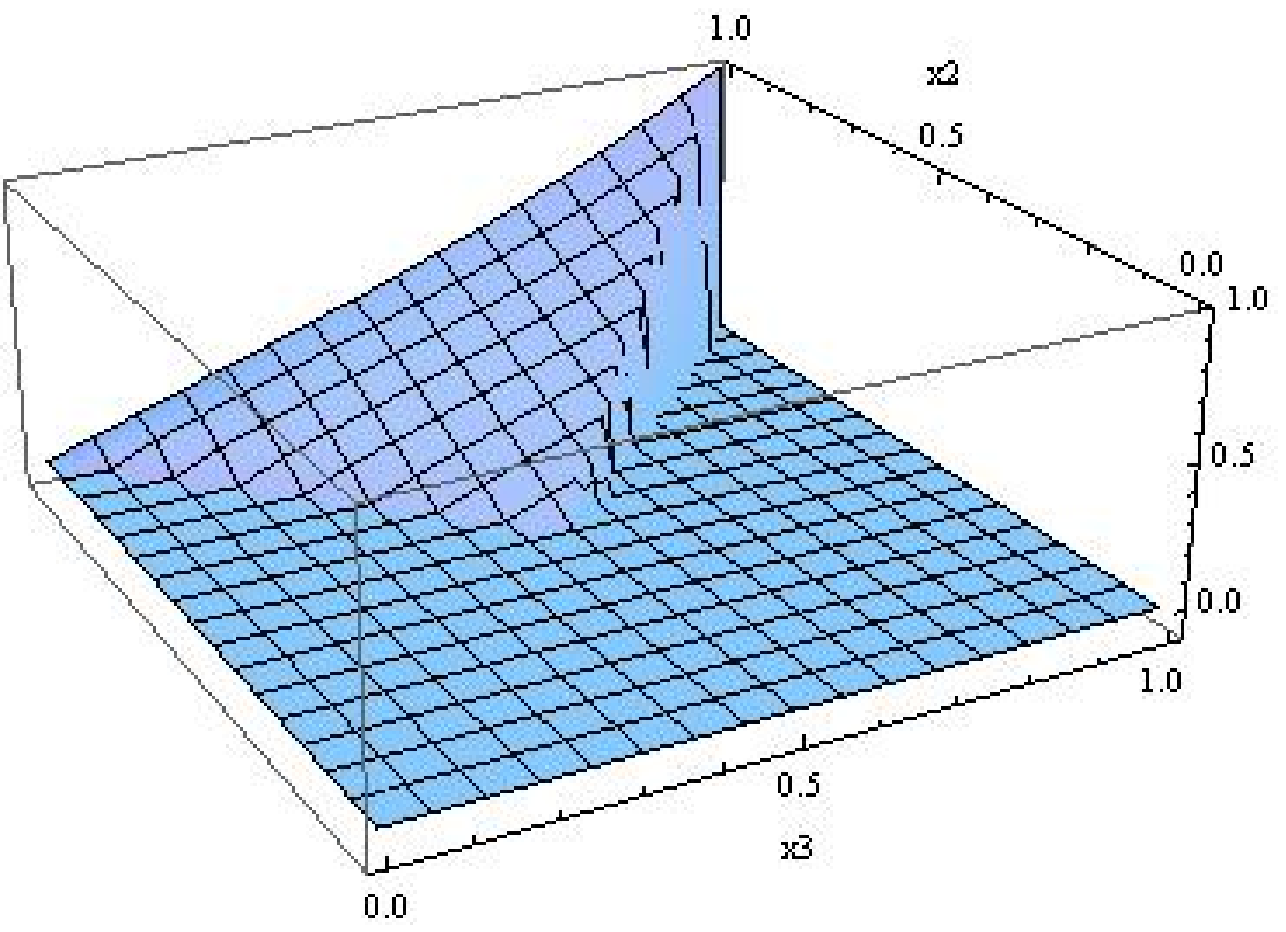}
	\label{Mb1-1} 
	\caption{exact DBI configuration on the left; approximated ghost shape on the right.}
\end{figure}

\begin{figure}[h]
	\includegraphics[scale=0.55]{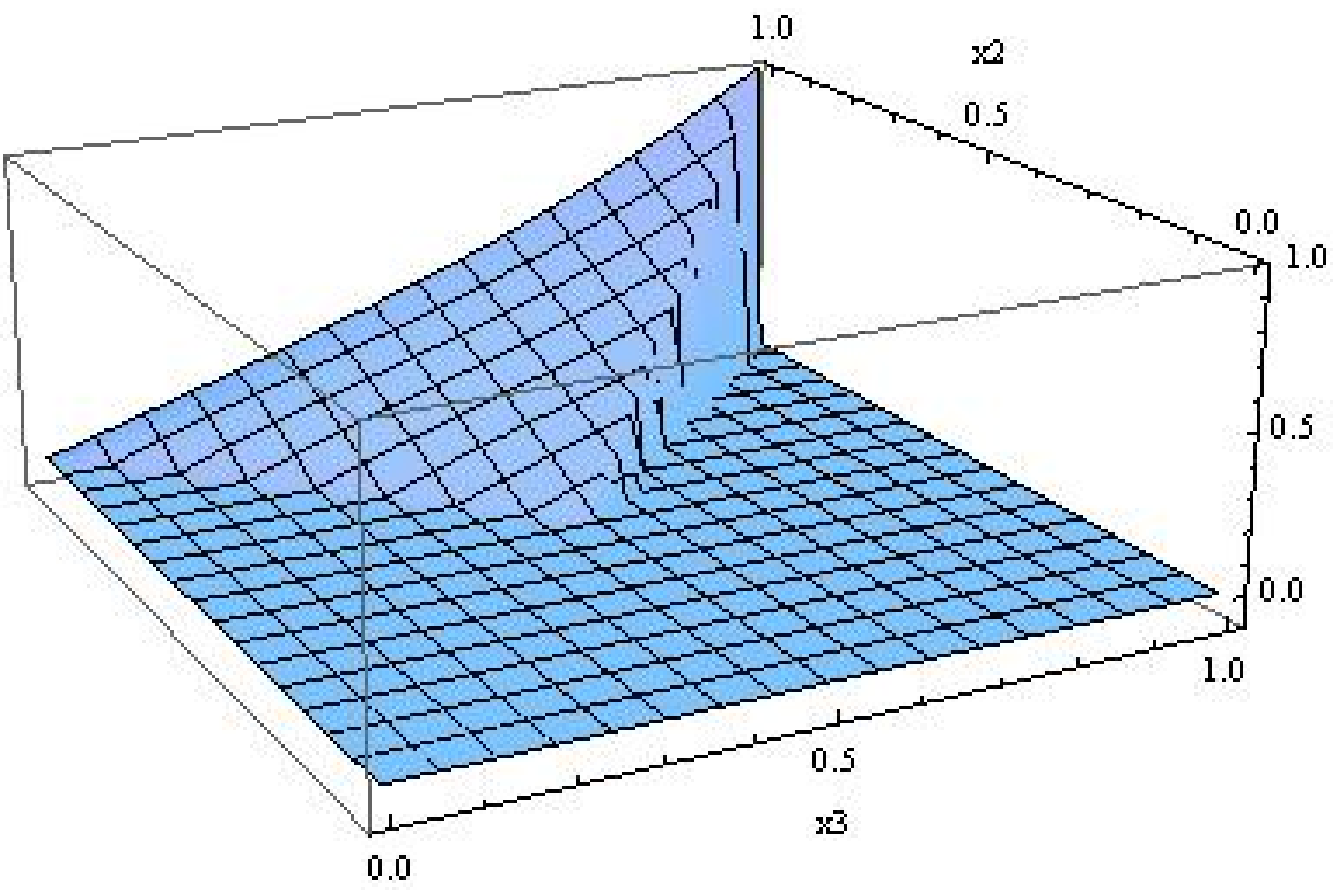}
\hspace{15mm}
	\includegraphics[scale=0.52]{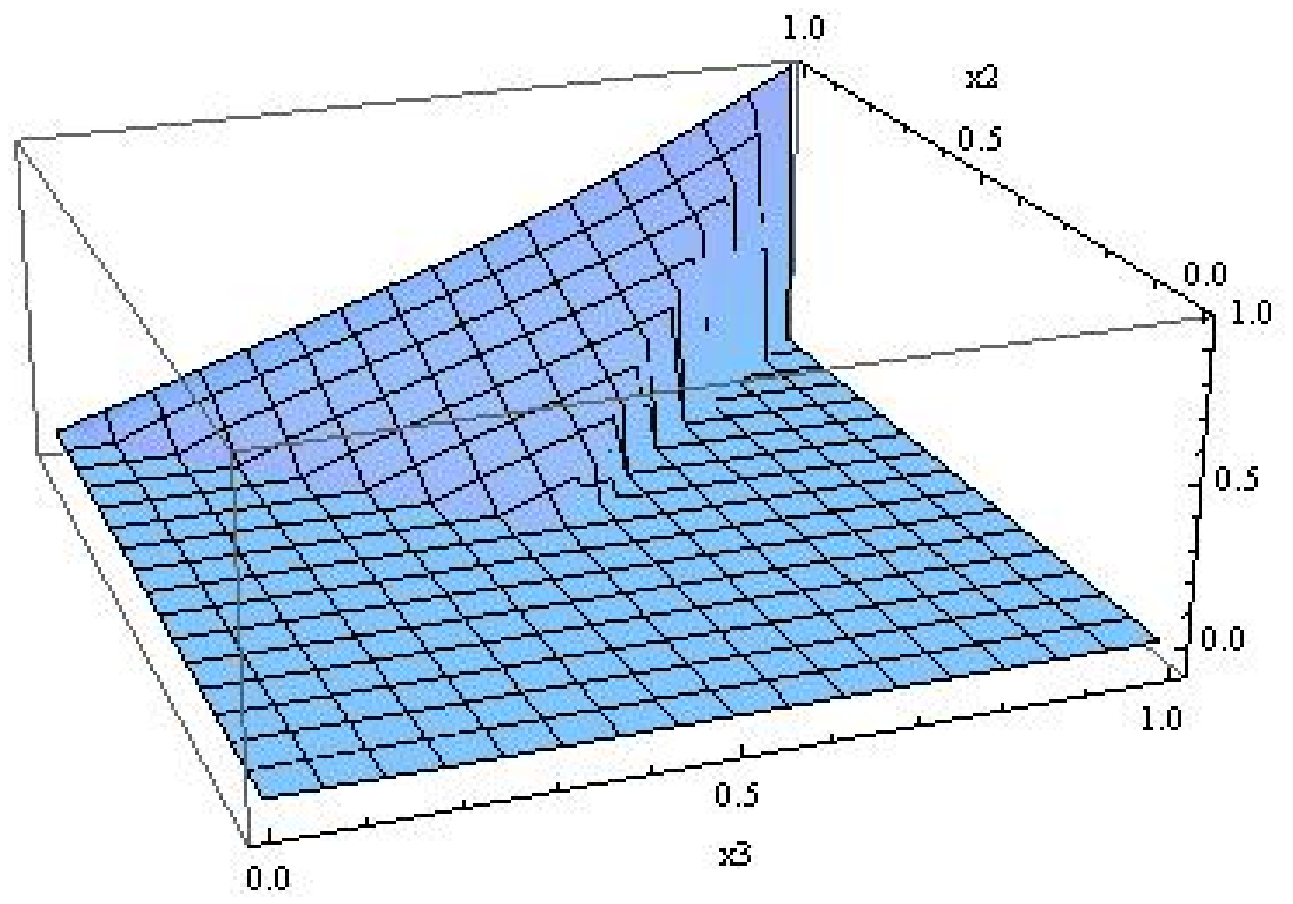}
	\label{Mb1-2} 
	\caption{$A$ on the left, $B$ configuration on the right for the $\bar M_1$-driven interaction term.}
\end{figure}
Also the bispectrum generated by this interaction term has an equilateral shape; the last two plots show, employing the general wavefunction, that also the interpolating models produce an equilateral shape.\\

$\bullet$  ${\cal O}_4= -1/2 \,\, \bar M_0^2/4\,\,  \left(5H (\partial_i^2\pi) (\partial_j \pi)^2  +4\dot\pi \partial_{i}^2 \partial_j \pi \partial_j \pi \right)/ a^4 $\\

\begin{figure}[h]
	\includegraphics[scale=0.55]{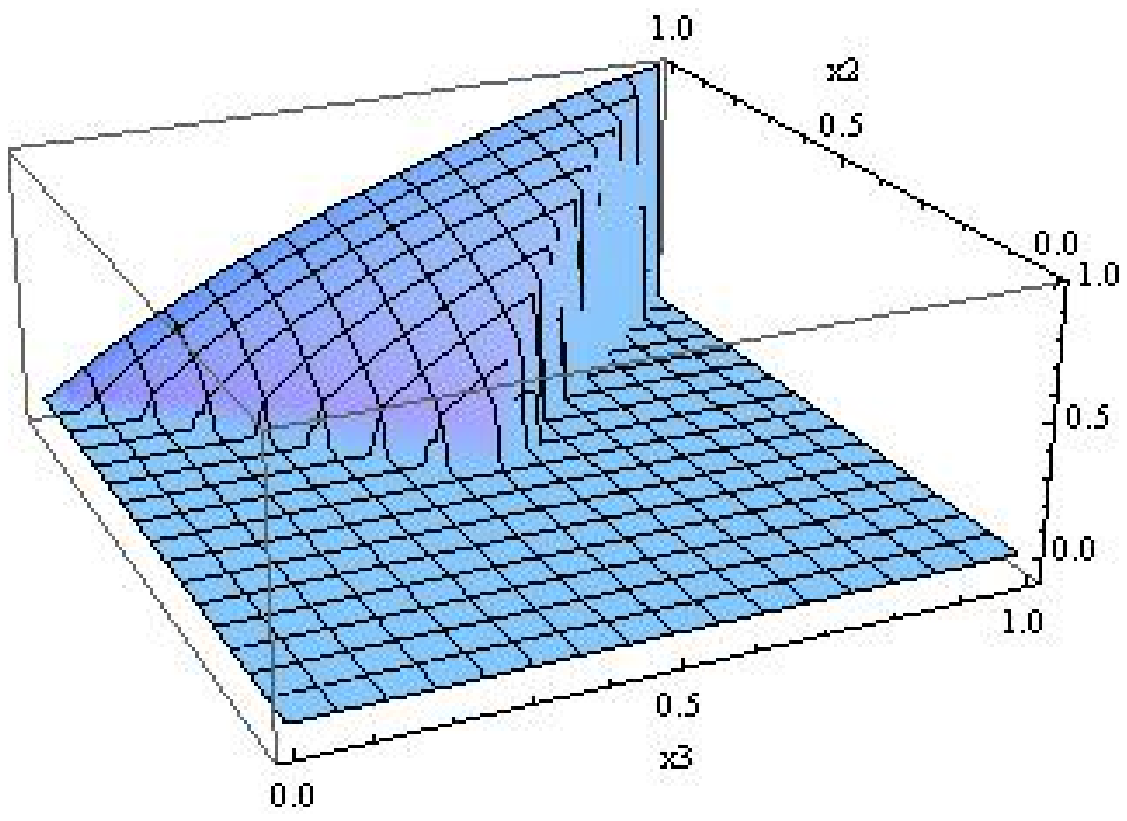}
\hspace{10mm}
	\includegraphics[scale=0.55]{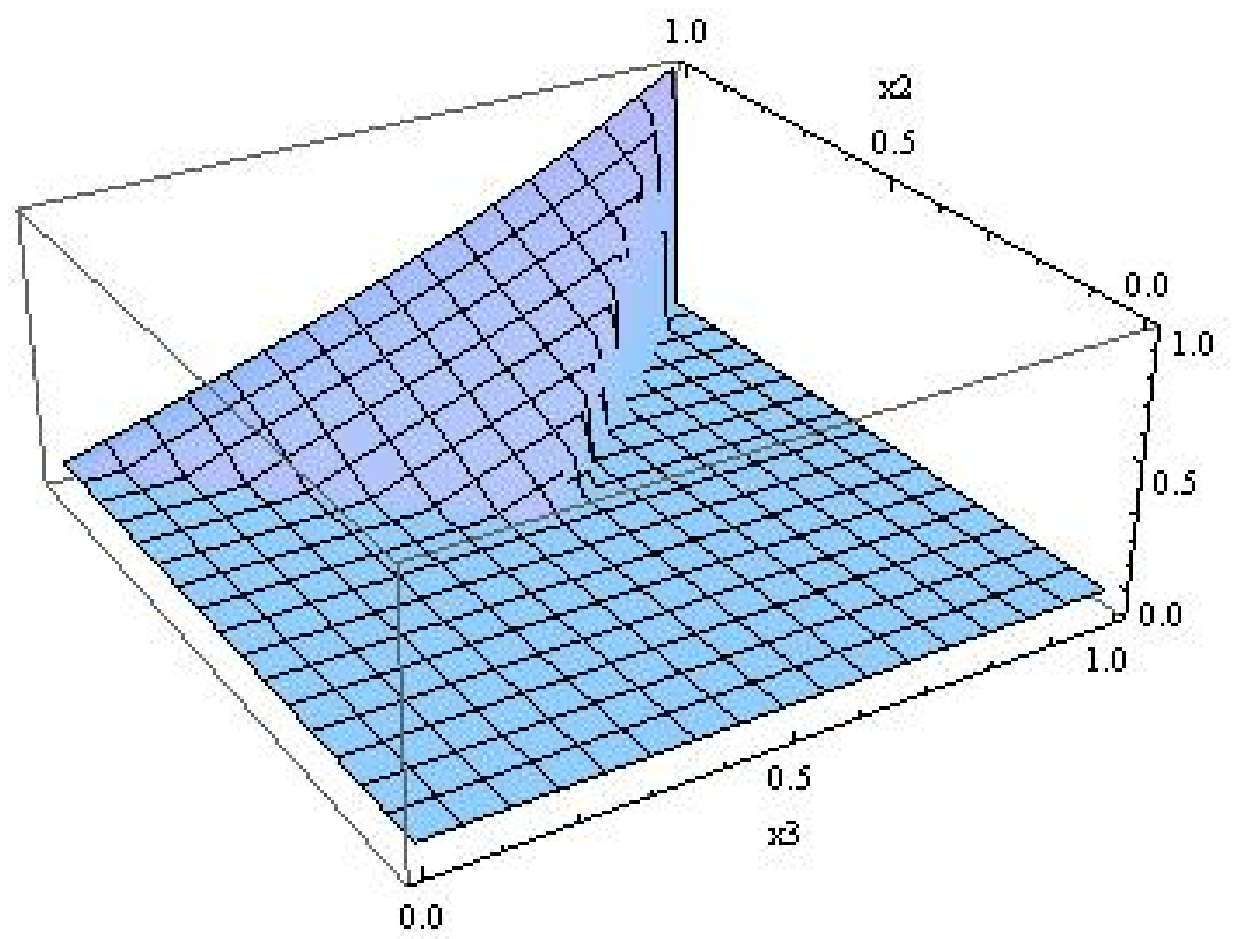}
	\label{fig:Mb2-1} 
	\caption{exact DBI configuration on the left; approx. ghost shape on the right.}
\end{figure}

\begin{figure}[h]
	\includegraphics[scale=0.55]{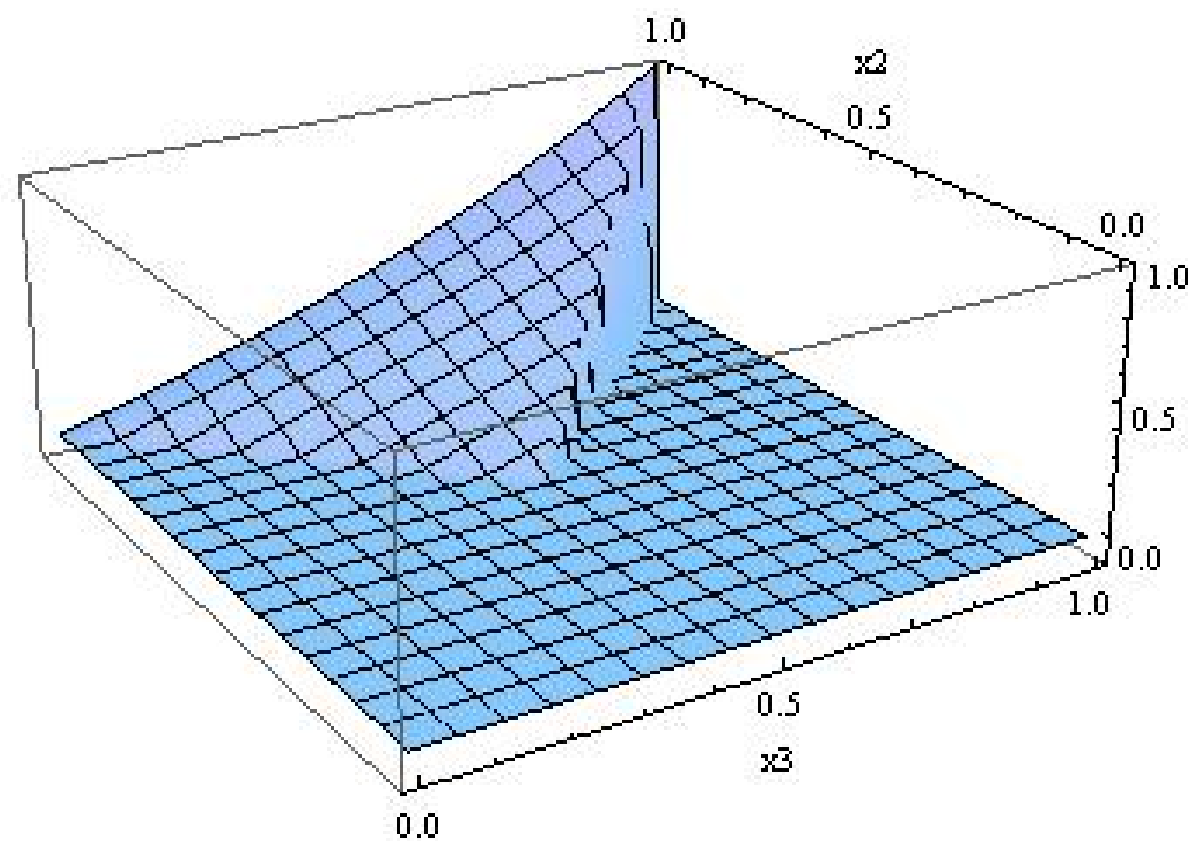}
\hspace{15mm}
	\includegraphics[scale=0.55]{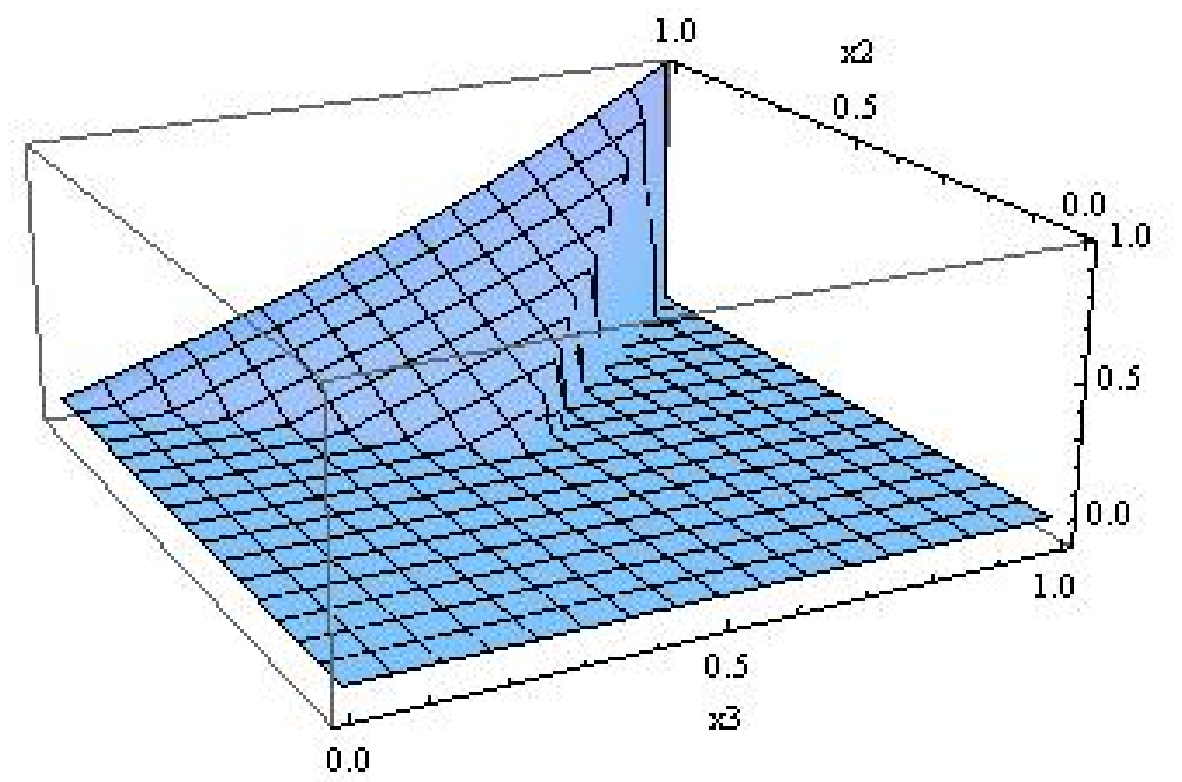}
	\label{Mb2-2} 
	\caption{$A$ on the left, $B$ configuration on the right for the $\bar M_0$ interaction term.}
\end{figure}
The various bispectra peak in the equilateral configuration. 
\newpage
\subsection{ Shapes from curvature-generated novel interaction terms.}
\vskip 0.5cm

$\bullet$  ${\cal O}_5= -2/3\,\,  \bar M_4^3 {\dot\pi}^2 \partial_i^2\pi \,\,/ a^2 $\\

With the term tuned by $\bar M_4$ we start including in our description the contributions that have so far been neglected in the literature.

\begin{figure}[h]
	\includegraphics[scale=0.45]{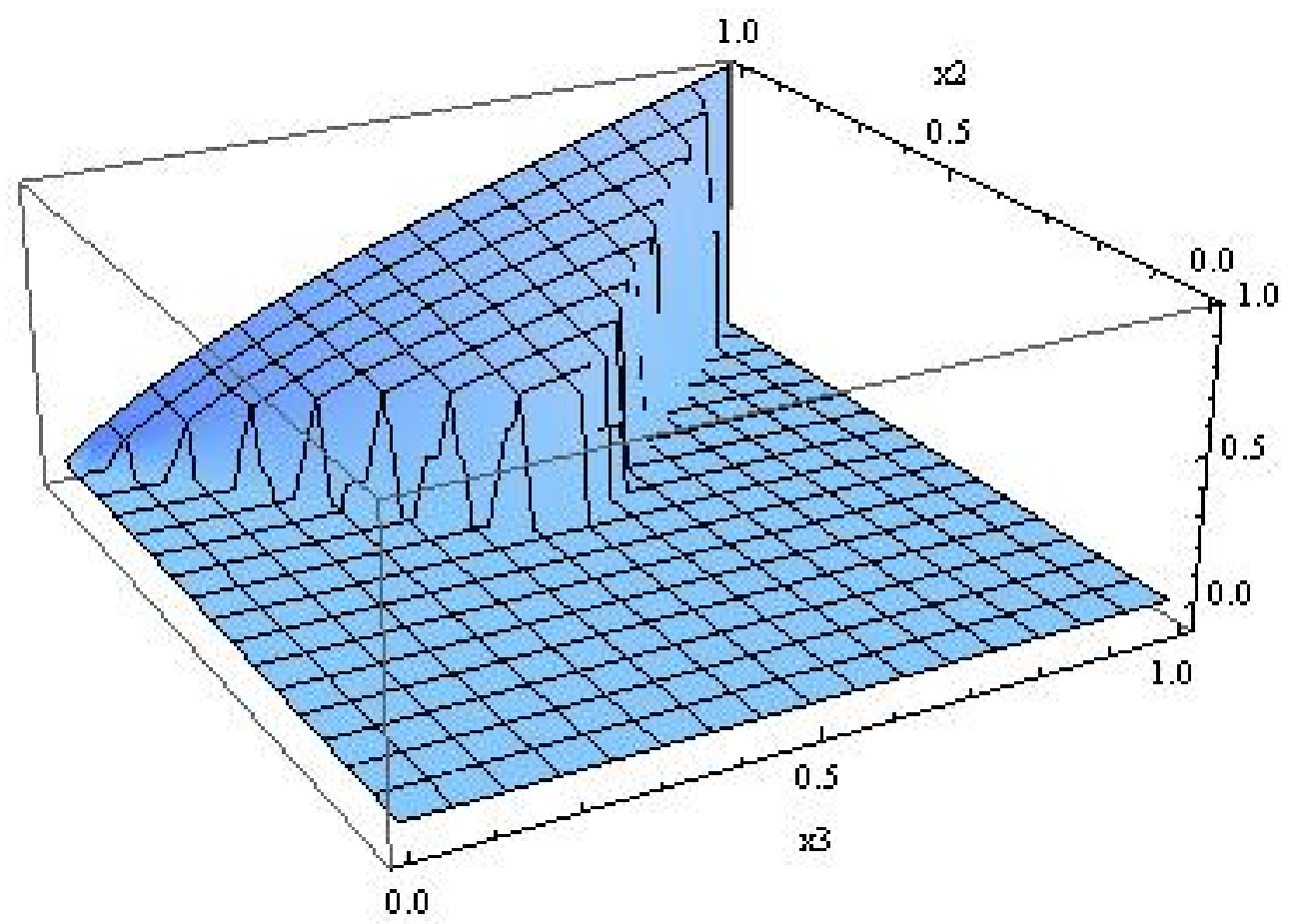}
\hspace{10mm}
	\includegraphics[scale=0.50]{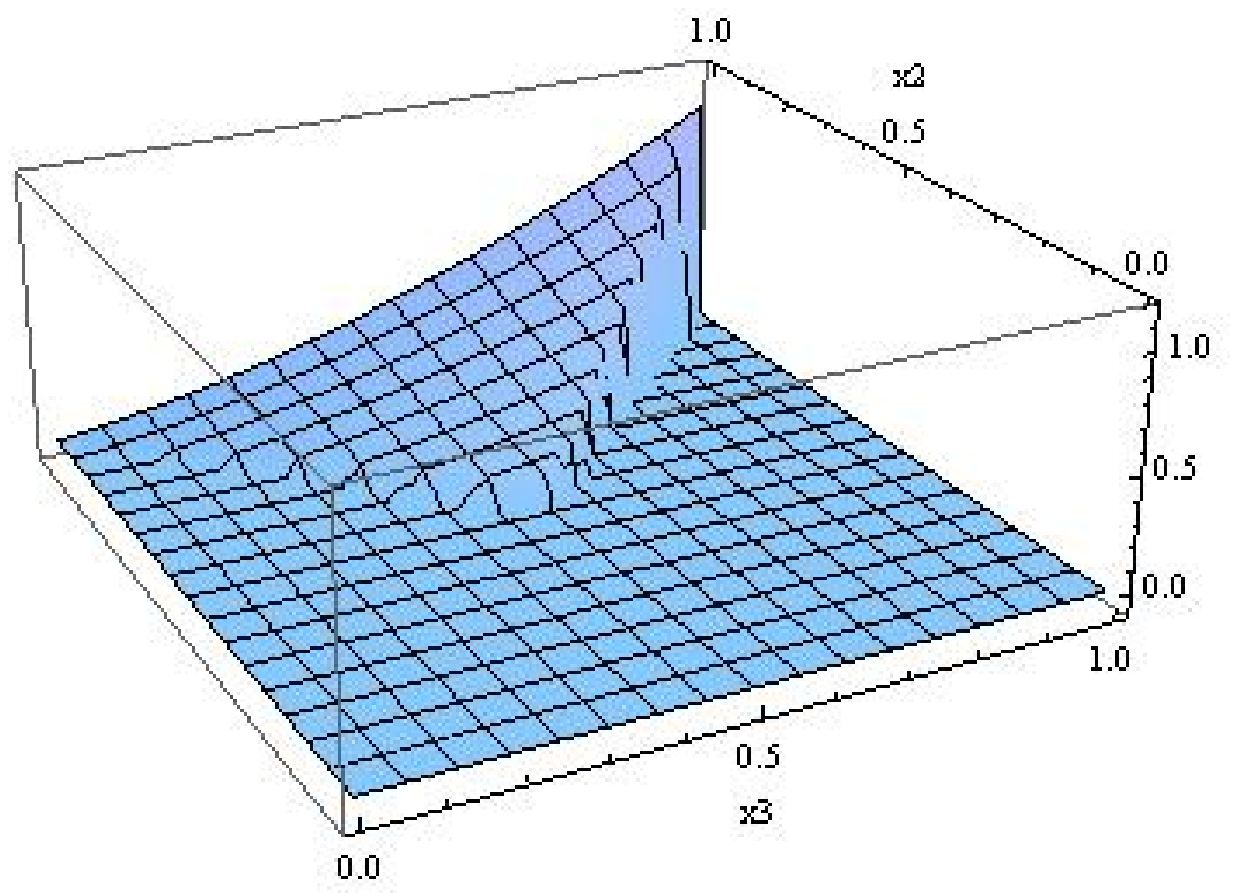}
	\label{fig:Mb4-1} 
	\caption{exact DBI configuration on the left; approximated ghost shape on the right.}
\end{figure}

\begin{figure}[h]
	\includegraphics[scale=0.53]{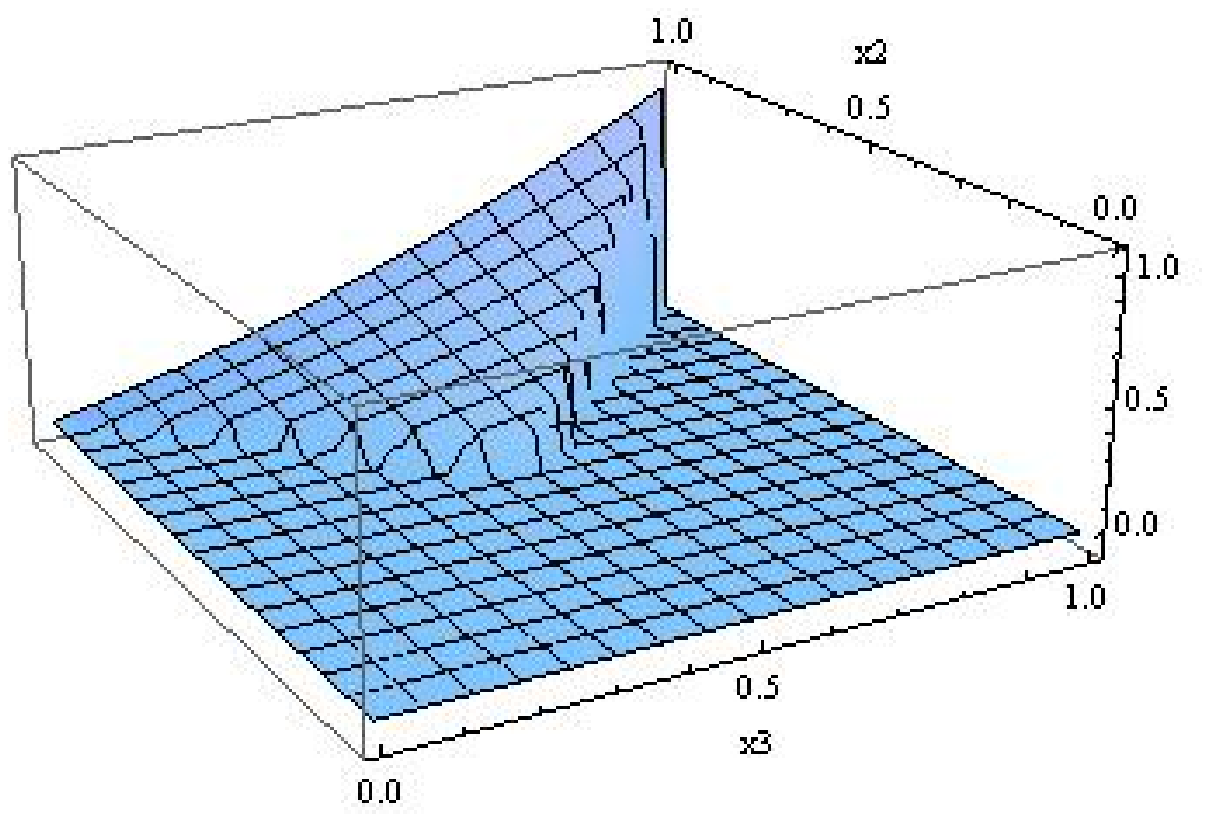}
\hspace{15mm}
	\includegraphics[scale=0.53]{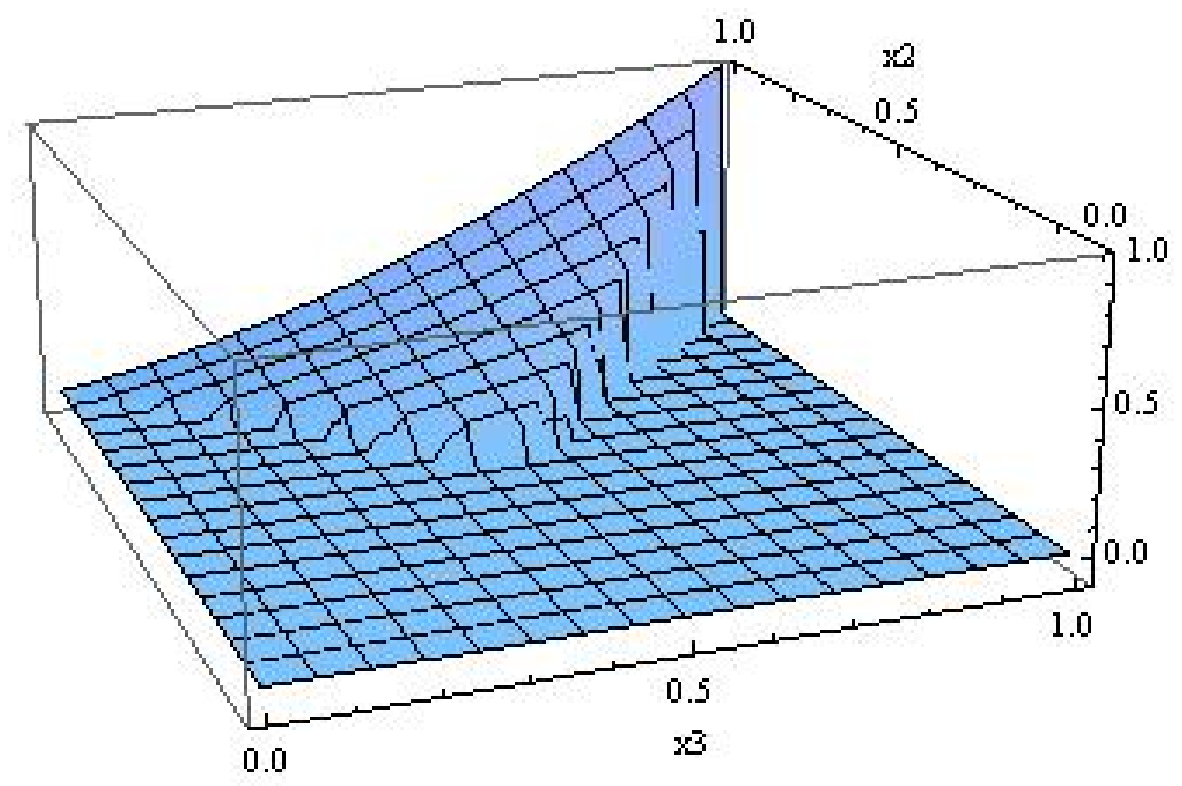}
	\label{fig:Mb4-2} 
	\caption{$A$ on the left, $B$ configuration on the right for the $\bar M_4$-driven interaction term.}
\end{figure}
Equilateral shapes for all configurations are obtained.\\

$\bullet$  ${\cal O}_6= 1/3\,\,  \bar M_5^2 \dot\pi (\partial_i^2\pi)^2 \,\,/ a^4 $\label{mb5}

\begin{figure}[h]
	\includegraphics[scale=0.45]{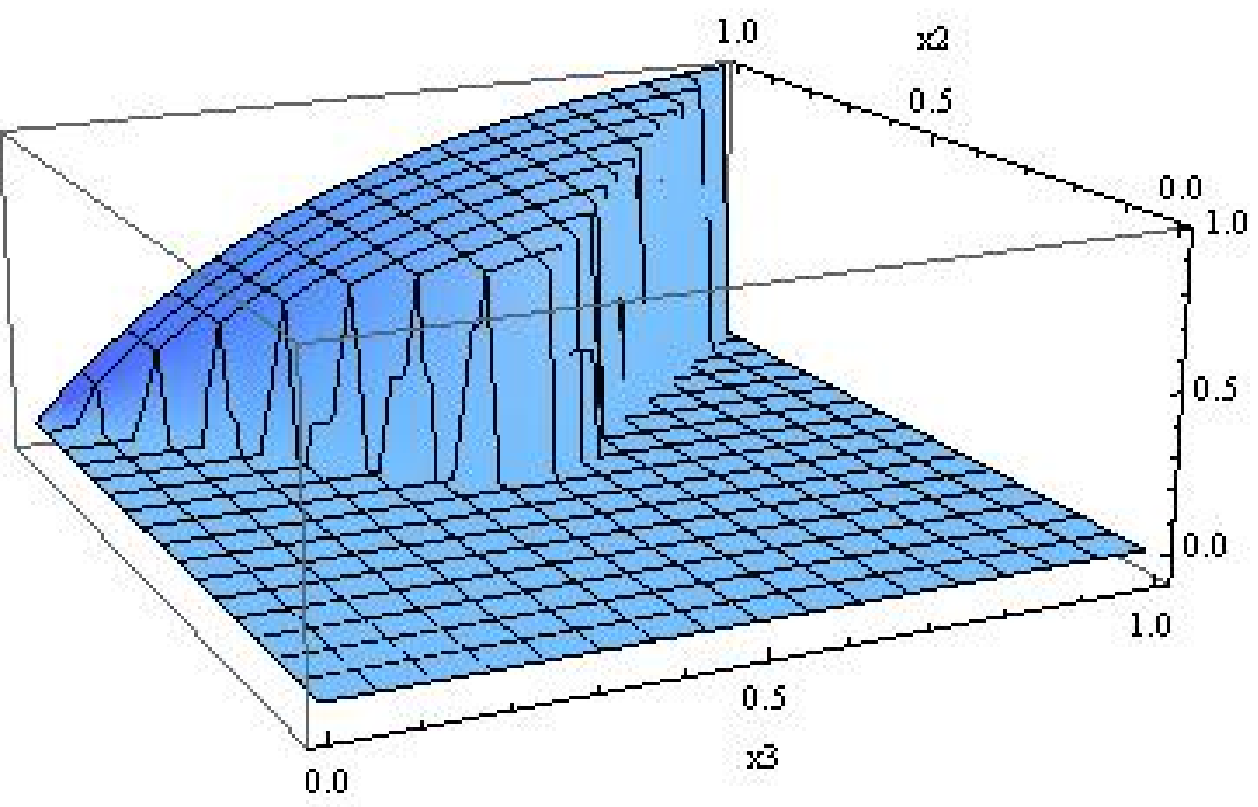}
\hspace{10mm}
	\includegraphics[scale=0.50]{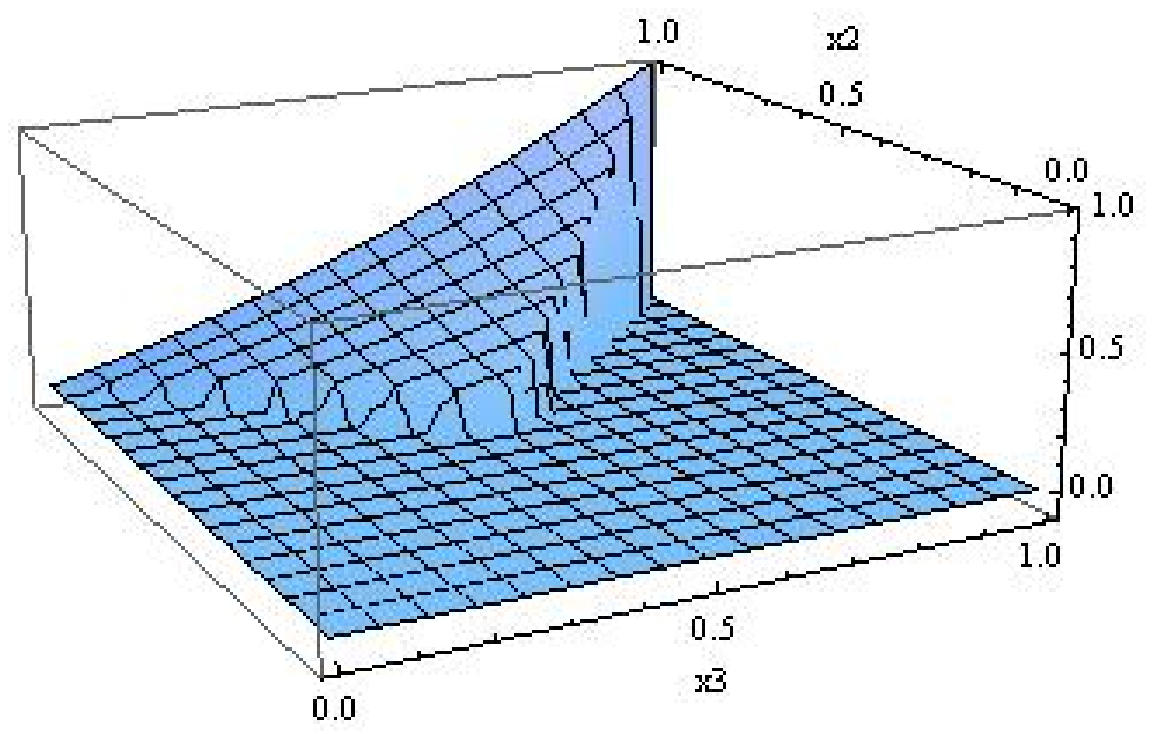}
	\label{fig:Mb5-1} 
	\caption{exact DBI configuration on the left; approximated ghost shape on the right.}
\end{figure}

\begin{figure}[h]
	\includegraphics[scale=0.60]{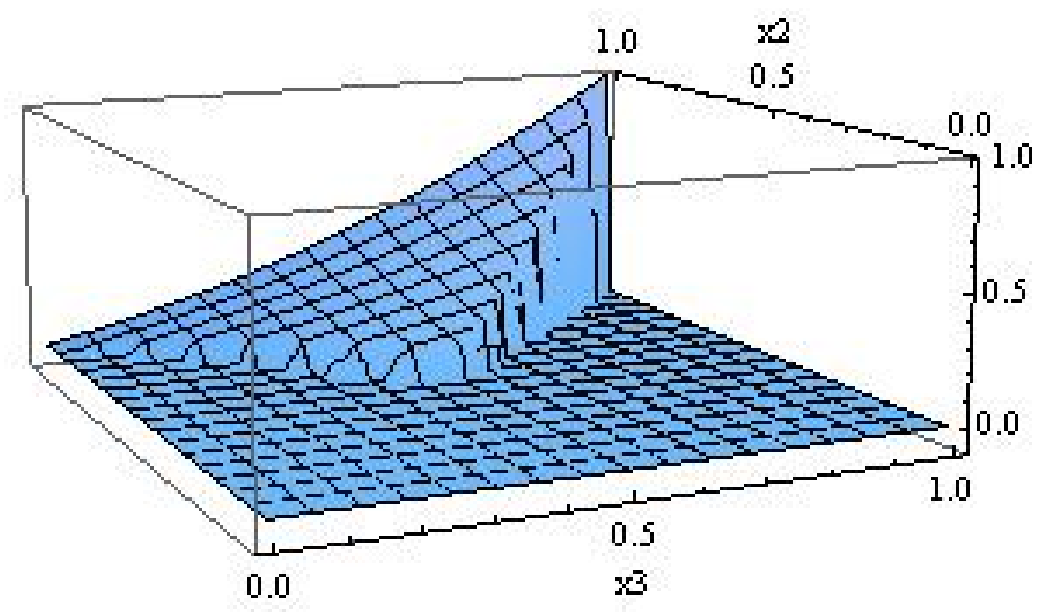}
\hspace{15mm}
	\includegraphics[scale=0.47]{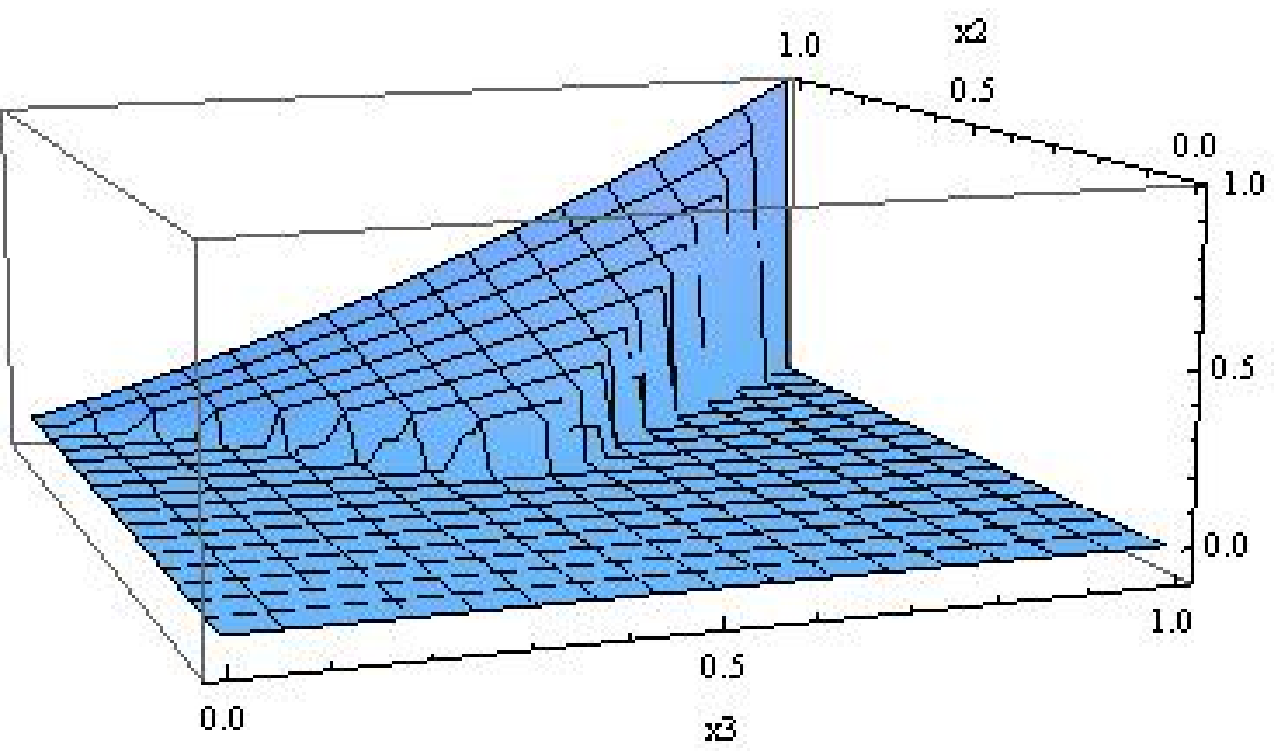}
	\label{fig:Mb5-2} 
	\caption{$A$ on the left, $B$ configuration on the right for the $\bar M_5$-driven interaction term.}
\end{figure}

Again, all the four bispectra peak in the equilateral configuration.\\
\newpage

$\bullet$  ${\cal O}_7= $ $1/3\,\,  \bar M_6^2 \dot\pi (\partial_{ij}\pi)^2 \,\,/ a^4 \label{mb6}$\\

\begin{figure}[h]
	\includegraphics[scale=0.53]{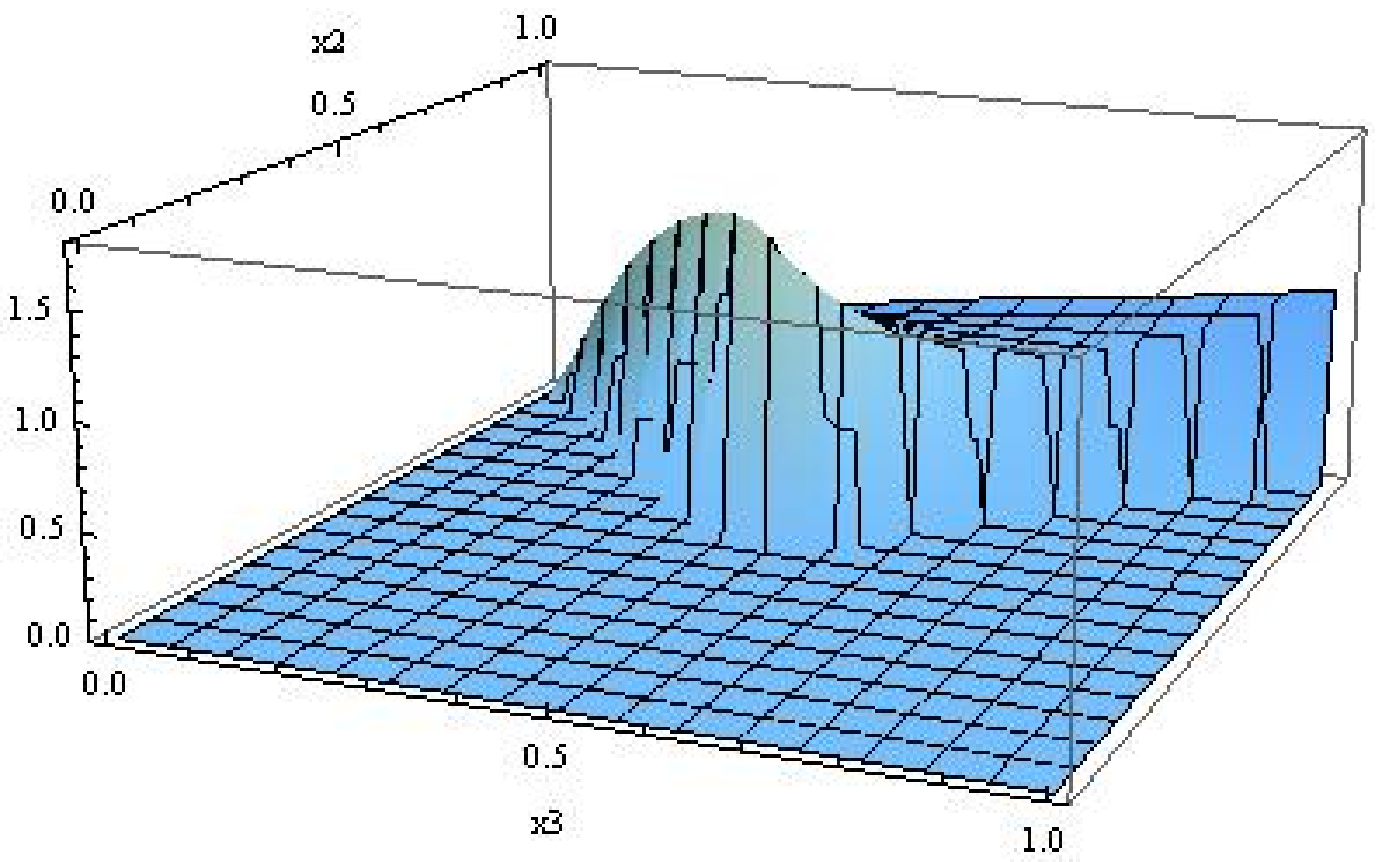}
\hspace{10mm}
	\includegraphics[scale=0.53]{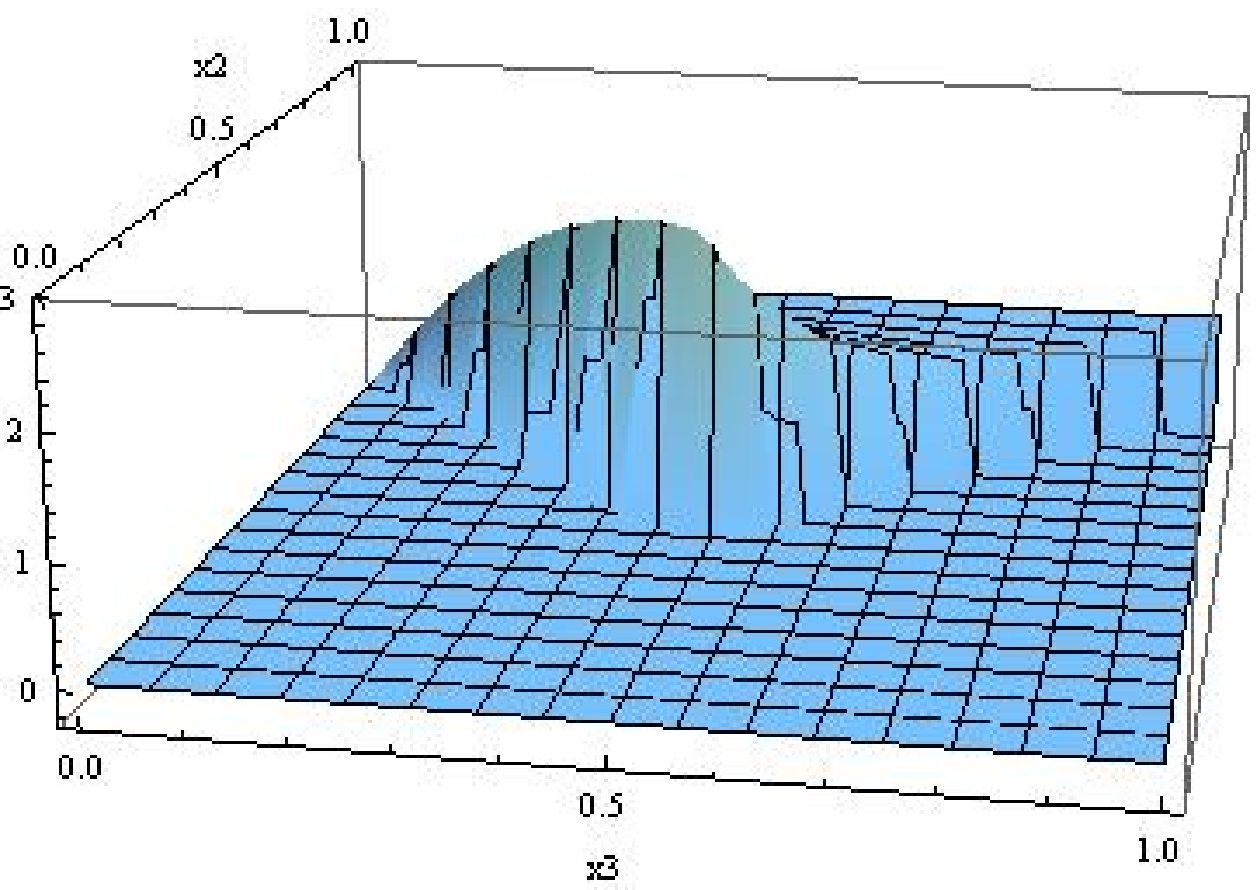}
	\label{fig:M_6_DBI} 
	\caption{exact DBI configuration on the left; approximated ghost shape on the right.}
\end{figure}

\begin{figure}[h]
	\includegraphics[scale=0.58]{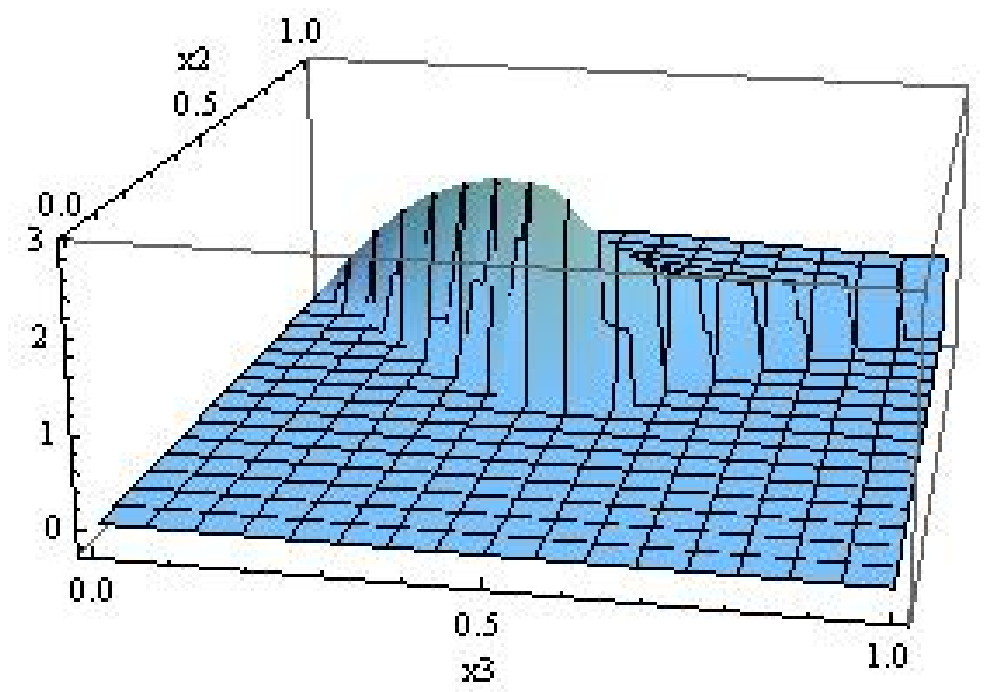}
\hspace{15mm}
	\includegraphics[scale=0.52]{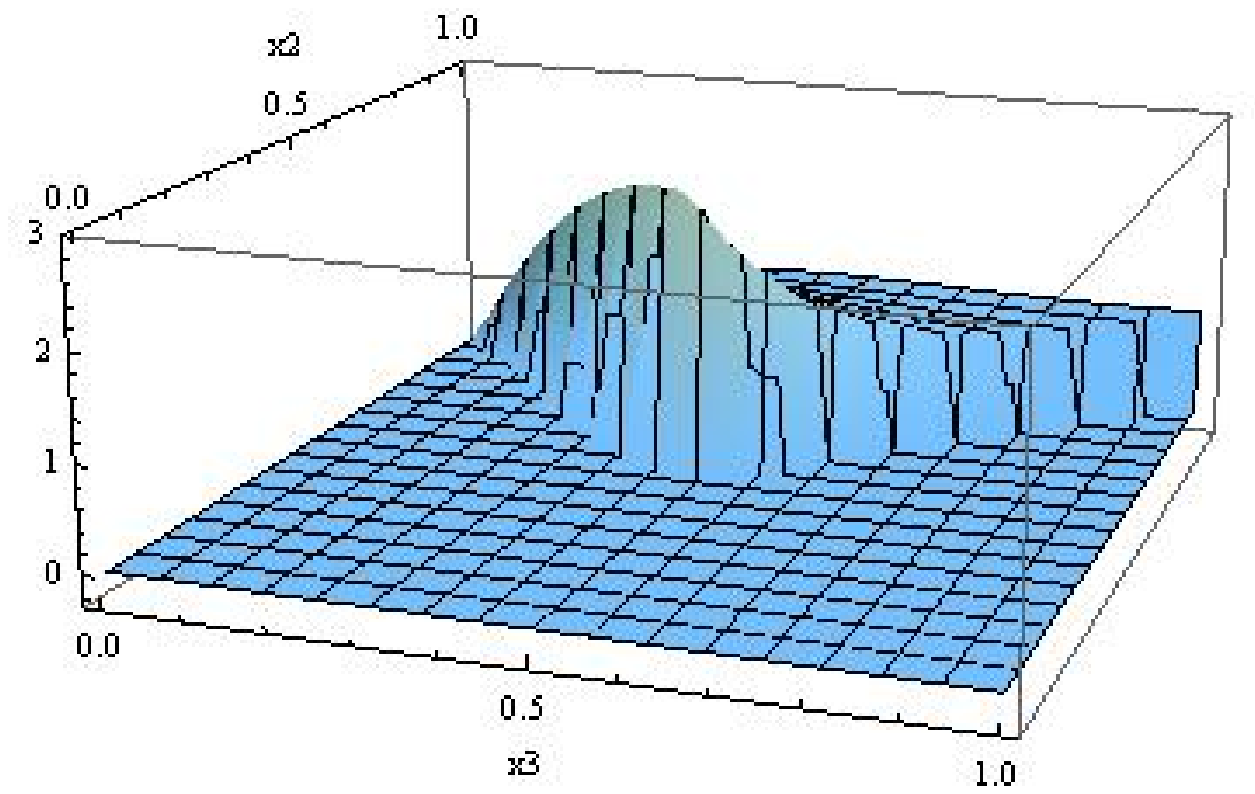}
	\label{fig:M_6b_DBI} 
	\caption{$A$ on the left, $B$ configuration on the right for the $\bar M_6$-driven interaction term.}
\end{figure}
This is one of the interesting novel curvature-generated terms that give rise to a flat shape (precisely the plot peaks at $k_1=1, k_2 \sim 1/2 \sim k_3$). 
Note that for a very similar interaction, namely the one generated by ${\cal O}_6$, we saw an equilateral plot. Here, derivatives combine to provide a different $k$-dependent factor outside the integral. Writing in Fourier space the interaction term ${\cal O}_6$  we obtain something proportional to ${k_2}^2 {k_3}^2$, while here we obtain something  like $(\vec k_2 \cdot \vec k_3)^2$.\\
\newpage
$\bullet$  ${\cal O}_8= -1/6\,\,  \bar M_7 (\partial_i^2 \pi)^3 \,\,/ a^6 $\label{mb7}\\

\begin{figure}[h]
	\includegraphics[scale=0.50]{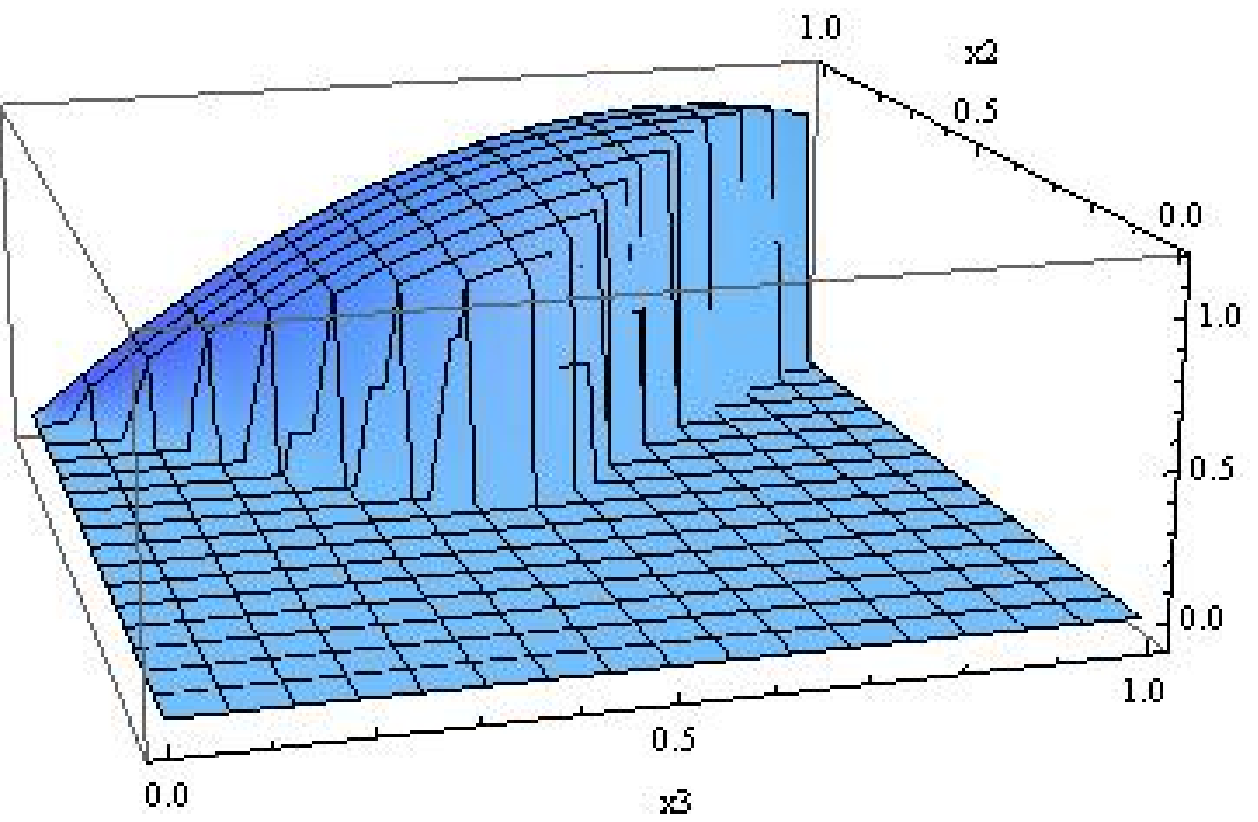}
\hspace{10mm}
	\includegraphics[scale=0.50]{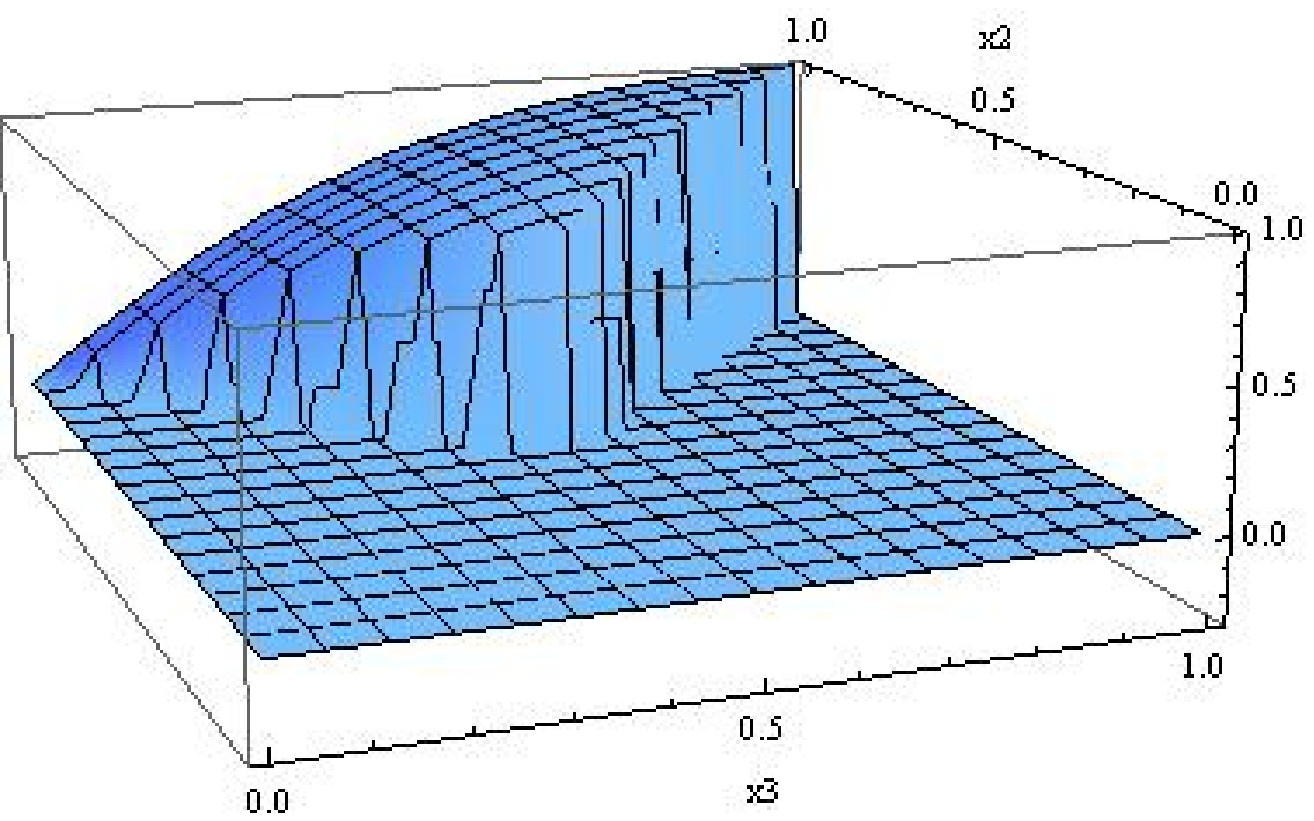}
	\label{fig:M_7_DBI} 
	\caption{exact DBI configuration on the left; approximated ghost shape on the right.}
\end{figure}

\begin{figure}[h]
	\includegraphics[scale=0.50]{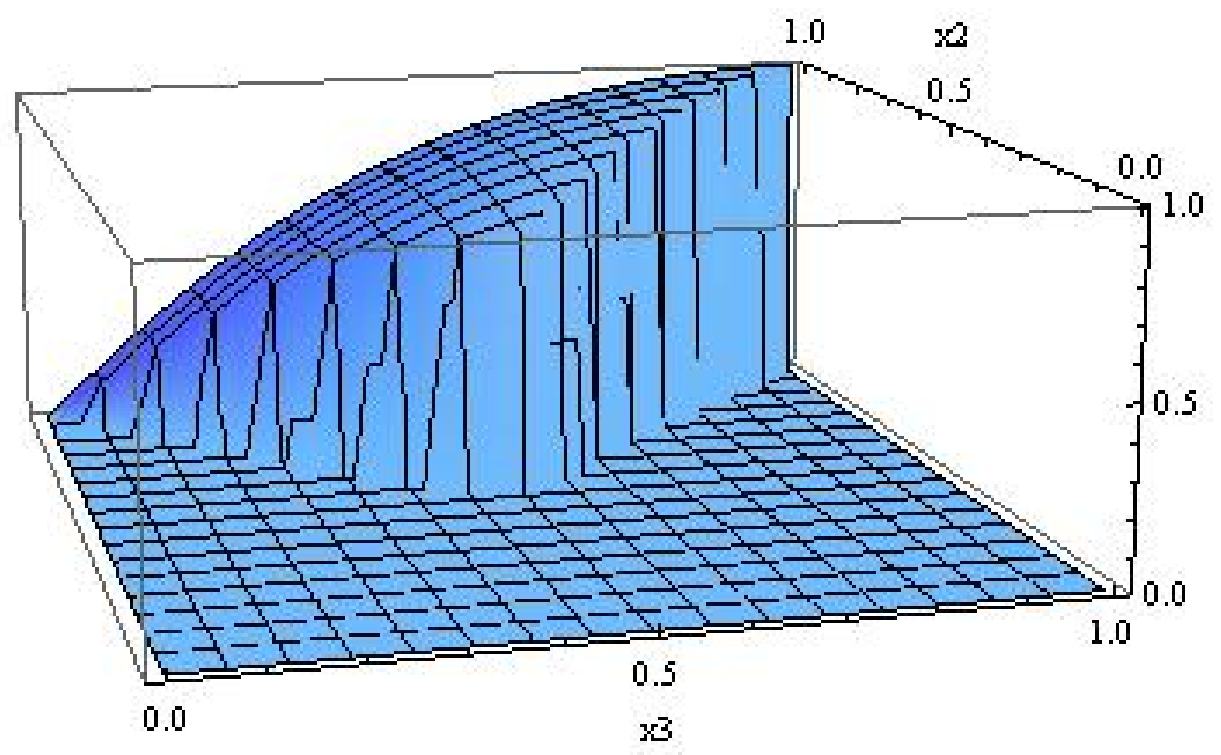}
\hspace{15mm}
	\includegraphics[scale=0.52]{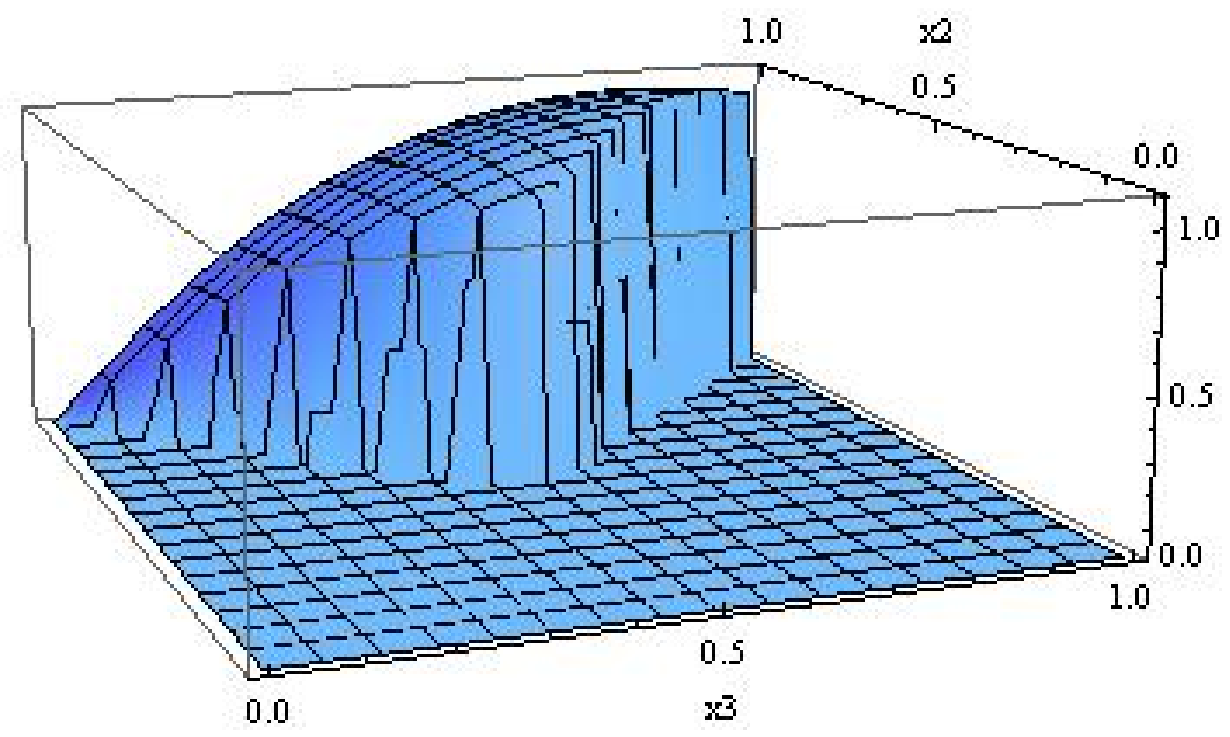}
	\label{fig:M_7b_DBI} 
	\caption{$A$ on the left, $B$ configuration on the right for the $\bar M_7$-driven interaction term.}
\end{figure}
All shapes peak in the equilateral configuration.\\
\newpage

$\bullet$  ${\cal O}_9= -1/6\,\,  \bar M_8 \, \partial_i^2 \pi (\partial_{jk} \pi)^2 \,\,/ a^6 $\label{mb8}\\
\begin{figure}[h]
	\includegraphics[scale=0.45]{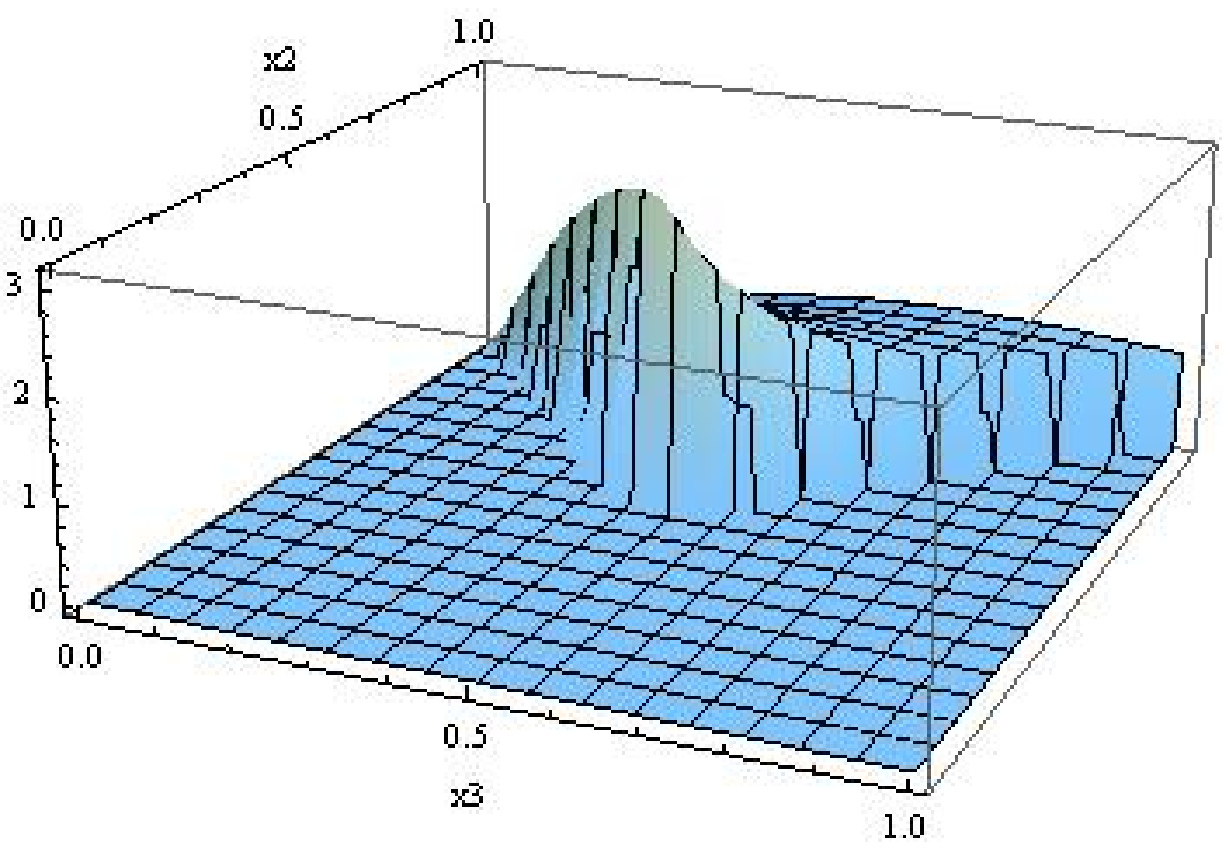}
\hspace{10mm}
	\includegraphics[scale=0.50]{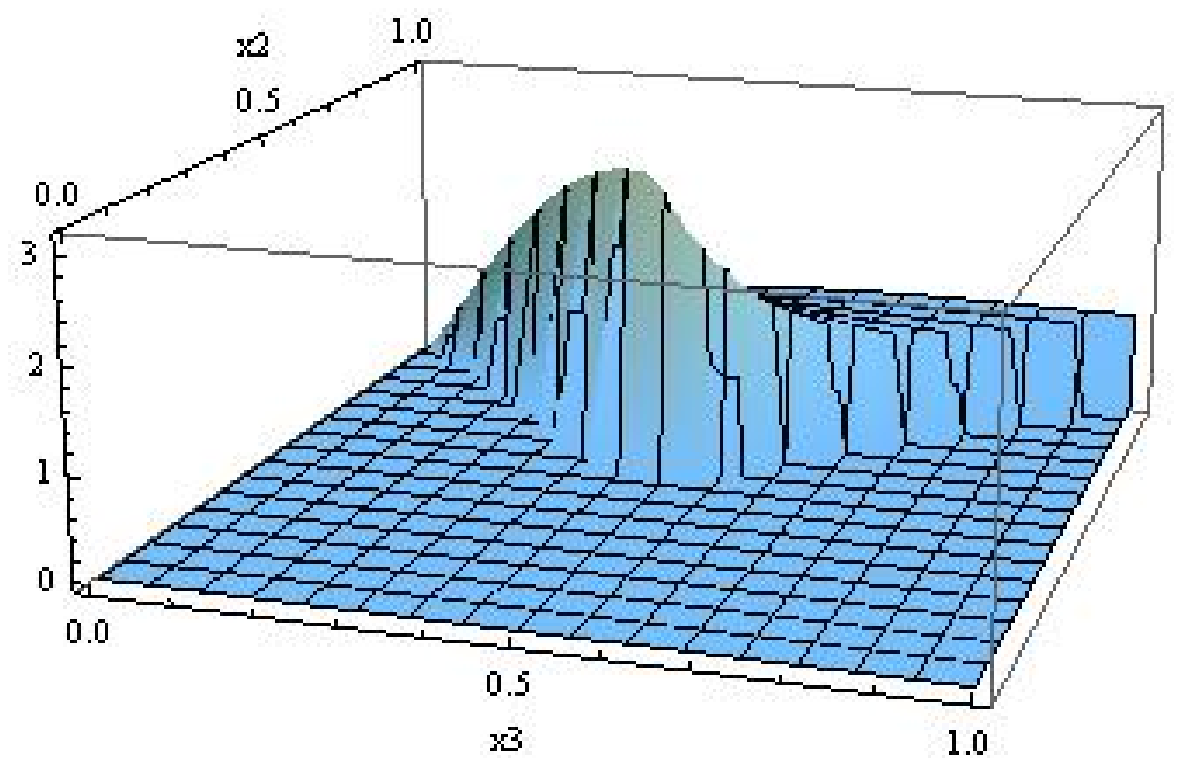}
	\label{fig:M_8_DBI} 
	\caption{exact DBI configuration on the left; approximated ghost shape on the right.}
\end{figure}

\begin{figure}[h]
	\includegraphics[scale=0.53]{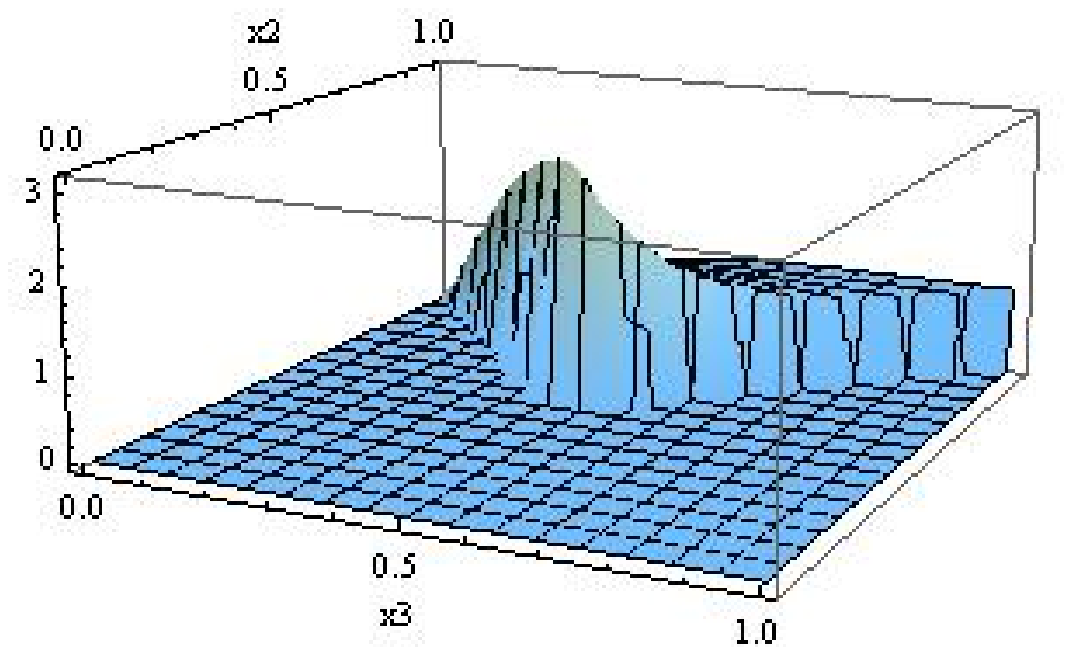}
\hspace{15mm}
	\includegraphics[scale=0.50]{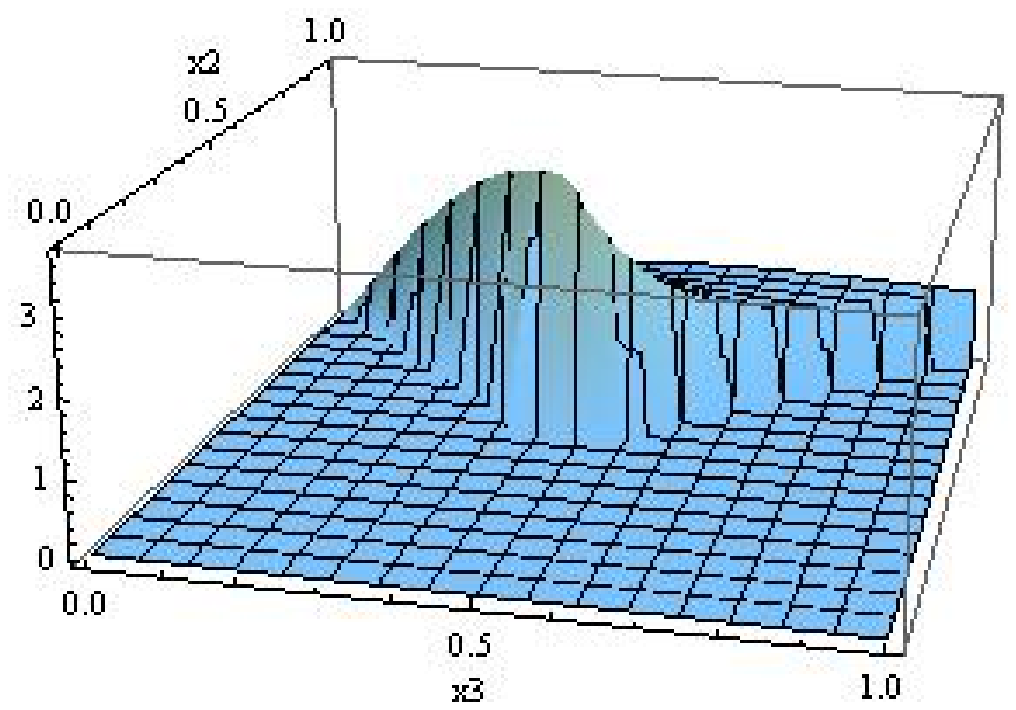}
	\label{fig:M_8b_DBI} 
	\caption{$A$ on the left, $B$ configuration on the right for the $\bar M_8$-driven interaction term}
\end{figure}
This is a second interaction term that produces, just as for  ${\cal O}_7$, a flat shape for the bispectra. Comparing it in Fourier space with our findings for ${\cal O}_8$, one can see that it is due to the way the spatial derivatives are combined. As shown in Figs. 19 and 20, also ${\cal O}_{10}$ gives rise to flat-shape bispectrum. We can see that 
the interactions ${\cal O}_8$, ${\cal O}_9$ and ${\cal O}_{10}$ have the same structure as far as the integral is concerned; on the other hand 
their $k$-dependence goes like $k_1^2 \, k_2^2 \, k_3^2 $, 
$k_1^2 (\vec{k_2}\cdot \vec{k_3})^2+ {\rm perm}$ and $(\vec{k_1}\cdot \vec{k_2})(\vec{k_2}\cdot \vec{k_3})(\vec{k_3}\cdot \vec{k_1})$, respectively. The last two produce a flat shape.\\
\newpage
$\bullet$ ${\cal O}_{10}=-1/6\,\,  \bar M_9 \, \partial_{ij} \pi \partial_{jk} \pi  \partial_{ki} \pi \,\,/ a^6 \label{mb9}$

\begin{figure}[h]
	\includegraphics[scale=0.50]{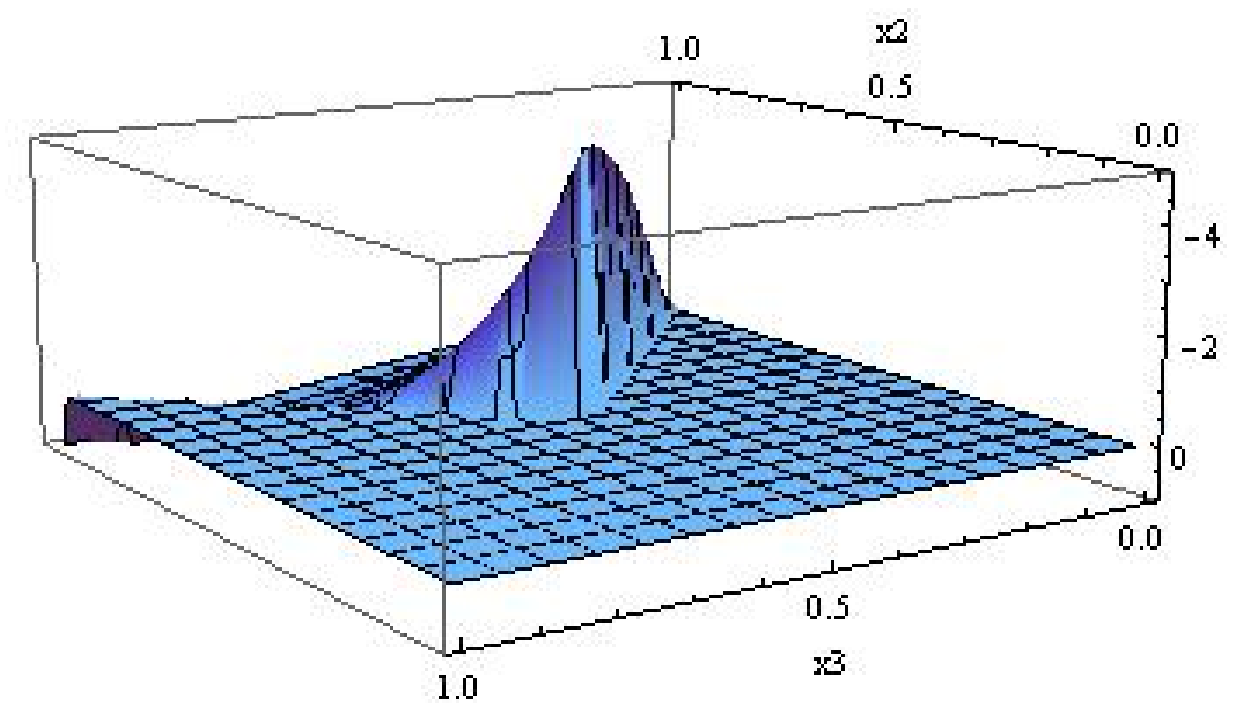}
\hspace{10mm}
	\includegraphics[scale=0.48]{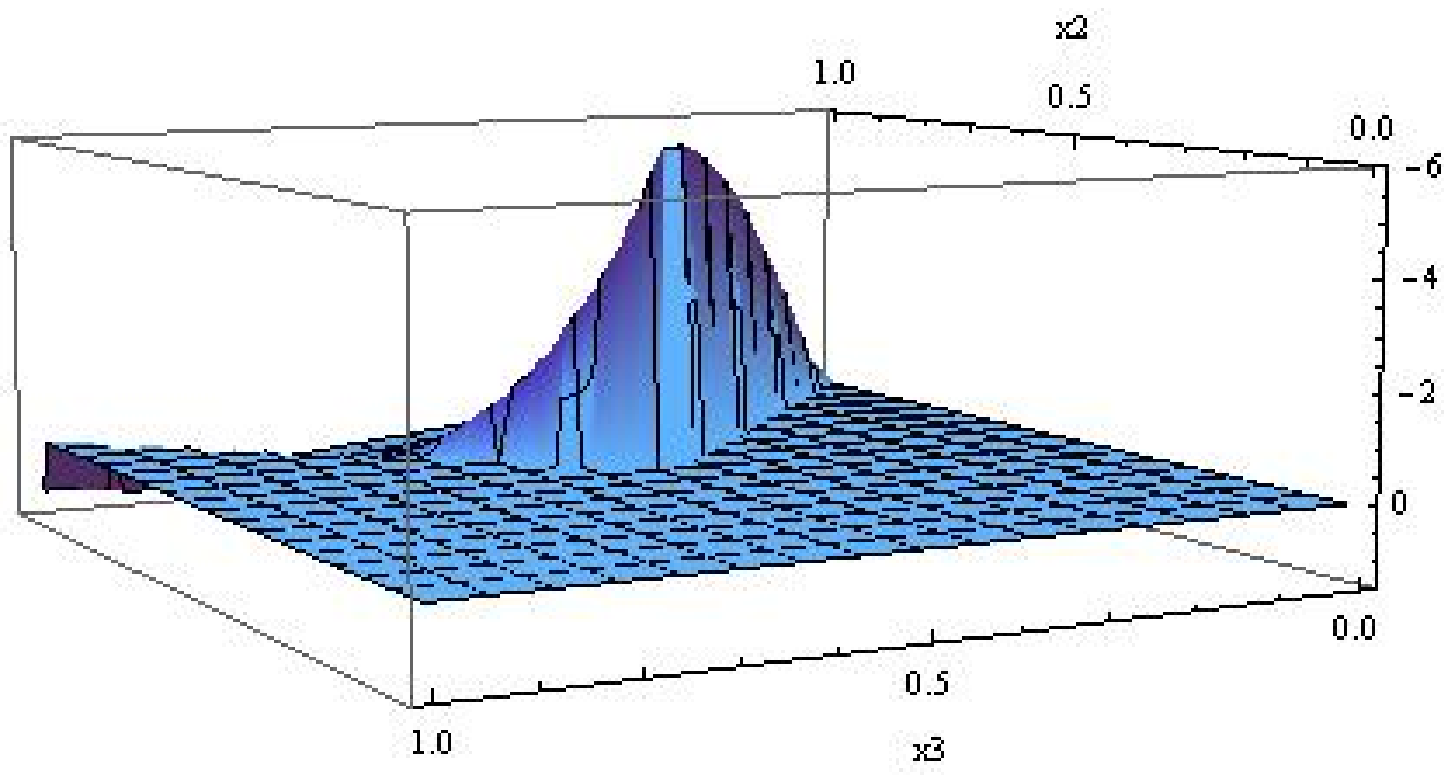}
	\label{fig:M_9_DBI} 
	\caption{exact DBI configuration on the left; approximated ghost shape on the right.}
\end{figure}

\begin{figure}[h]
	\includegraphics[scale=0.53]{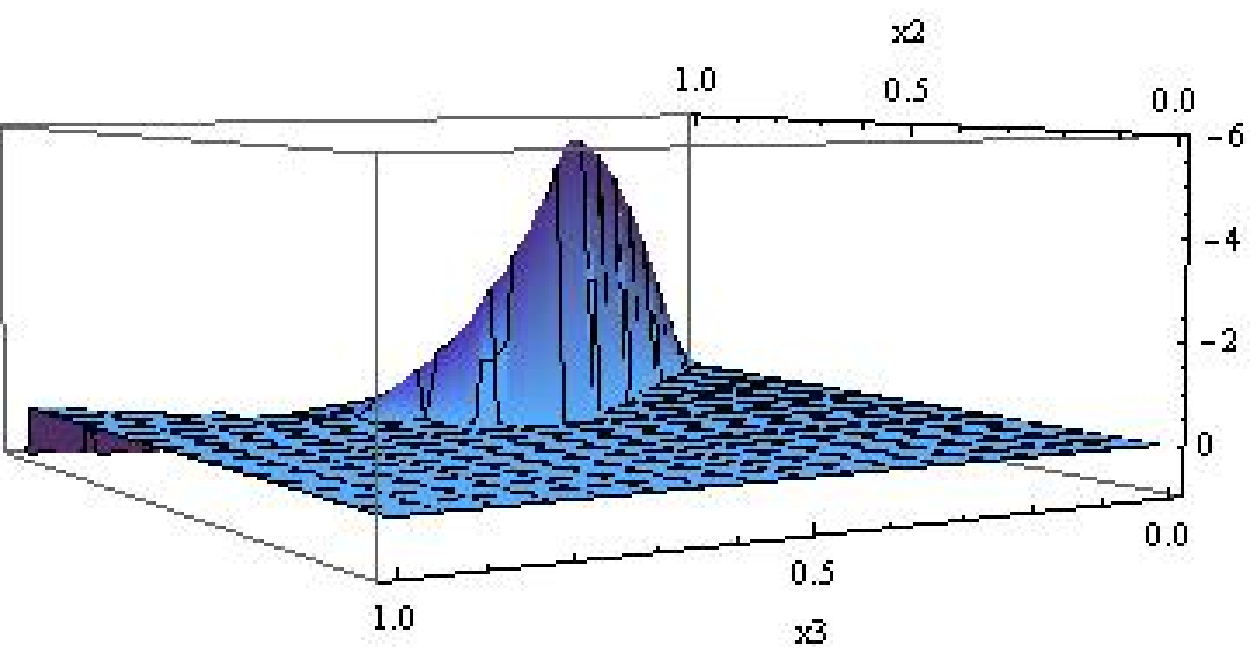}
\hspace{15mm}
	\includegraphics[scale=0.53]{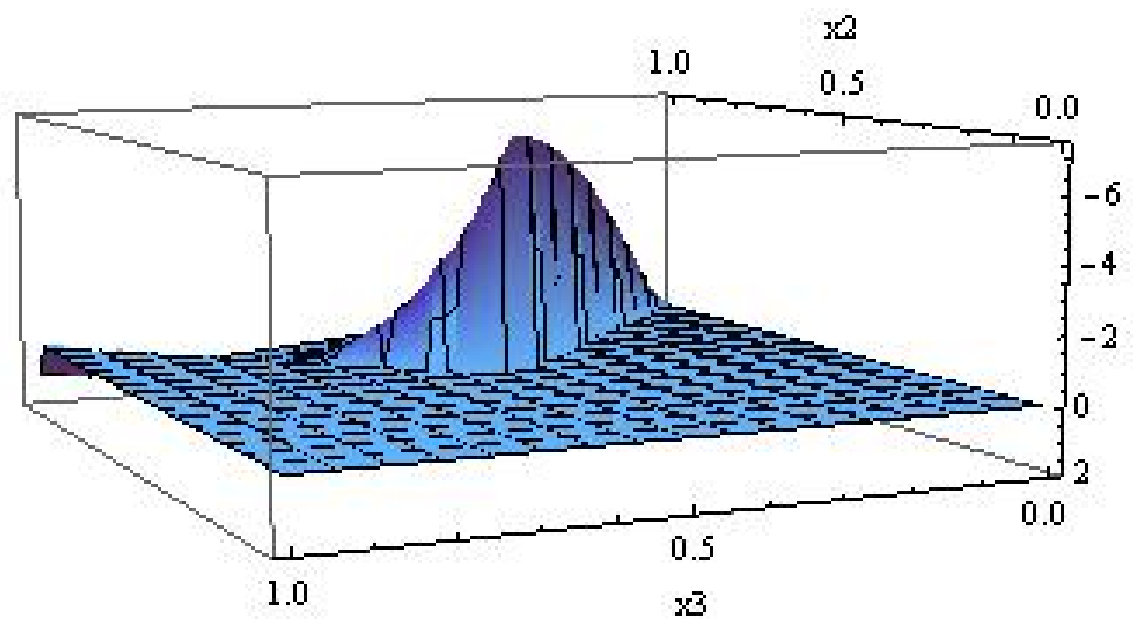}
	\label{fig:M_9b_DBI} 
	\caption{$A$ on the left, $B$ configuration on the right for the $\bar M_9$-driven interaction term.}
\end{figure}
The shapes are peaked in the flat configuration.

Obviously, having presented all the shapes due to each individual interaction, one might proceed with the study of the shape of linear combinations of them, much in the spirit of the orhtogonal shape recently introduced \cite{ssz05}. On the other hand, we are using approximated methods for three of the four configurations under scrutiny and it is therefore not a cautious step to infer new shapes from linear combinations of approximated ones, especially when delicate substractions are involved.

\subsection{Some general considerations on the shapes of non-Gaussianities}
From the above discussion we can read off some general qualitative features characterizing  the shapes of the bispectrum of a generic single-field model of inflation:
\begin{itemize}
\item Two qualitatively very different kinds of shapes appear: equilateral and flat.  As pointed out before, the next natural step would be to consider the shapes resulting from linear combination of the various interaction terms in the spirit of Ref. \cite{ssz05}.

\item In single field-models of inflation usually large non-Gaussianities are associated to equilateral shapes. 
In order to obtain a flat shape one needs to consider linear combinations of interaction operators such as what is done in \cite{ssz05} or models with an initial vacuum different from Bunch-Davies.
 Interestingly, in our case a flat shape emerges simply from individual operators generated by
curvature-related terms. When the flat shape appears, it does so  
 in all four configurations considered. We stress this point because it implies that this result does not depend on the type of wavefunction one employs in the calculation, be it the DBI inflation solution, the ghost inflationary one and in the intermediate cases. The results of the DBI configuration are exact and easily reproducible with analytical methods. In fact, the DBI wavefunction is the usual solution of the standard single-field slow-roll inflation.
 Adopting such a wavefunction we may  provide  analytic results for the flat bispectra
\[
\fl  \langle\zeta_{k1} \zeta_{k2}\zeta_{k3}\rangle_{\bar M_6}= \frac{\left(60 a^6 b^2+11 a^3 b^5+b^8+\left(48 a^6-4 a^3 b^3-3 b^6\right) c^2-4 b^4 c^4\right) H^4 \bar M_6^2}{4 a^9\, \alpha_0^2\, b^5\, \epsilon ^3\, M_{\rm Pl}^6},
\]
\[
\fl \langle\zeta_{k1} \zeta_{k2}\zeta_{k3}\rangle_{\bar M_8}=   \frac{\left(15 a^3+b^3+3 b c^2\right) \left(12 a^6+b^6-6 b^4 c^2+8 b^2 c^4+8 a^3 \left(b^3-2 b c^2\right)\right) H^5 \bar M_8}{ a^9\, \alpha_0^3\, b^6 \,\epsilon ^3\, M_{\rm Pl}^6},
\]
\bea
\fl \langle\zeta_{k1} \zeta_{k2}\zeta_{k3}\rangle_{\bar M_9}=   \frac{3 \left(15 a^3+b^3+3 b c^2\right) \left(8 a^6+b^6-6 b^4 c^2+8 b^2 c^4+8 a^3 \left(b^3-2 b c^2\right)\right) H^5 \bar M_9}{2 a^9 \,\alpha_0^3\, b^6\, \epsilon ^3 \,M_{\rm Pl}^6}\, ,\nonumber\\ \label{bisflat}
\eea
where the overall momentum conservation delta has been omitted, 
\[
a=(k_1 k_2 k_3)^{1/3};\quad b={k_1+k_2+k_3};\quad c=(k_1 k_2+ k_1 k_3+k_2 k_3)^{1/2}, 
\]
and $\alpha_0$ is now the usual speed of sound:
\[
 \alpha_0= -M_{\rm Pl}^2 \dot H/(-M_{\rm Pl}^2 \dot H +2M_2^4)=c_s^2\, .
\]
Note that all the three expression given above have a maximum precisely in the flat configuration ($k_1=1,k_2=1/2=k_3 $). What immediately stands out in Eq.~(\ref{bisflat}) is the presence in the numerator of factors consisting  of subtractions between generally positive $k$-symmetrized terms: it is this characteristic that selects a flat, rather than a  equilteral shape, as one can readily verify by checking the bispectrum of the ``equilateral'' interaction terms.\\ The expression in Eq.~(\ref{bisflat}) is exact in the DBI case and can be employed to get a shape qualitatively similar in the other three configurations.  For practical purposes,  we give below a very simple expression that very closely mimics the behaviour  of the typical bispectrum contribution that generates flat shape
\bea
\fl \langle\zeta_{k1} \zeta_{k2}\zeta_{k3}\rangle_{\bar M}\sim   \frac{\left(-k_1^2+k_2{}^2-k_3{}^2\right) \left(k_1^2+k_2{}^2-k_3{}^2\right) \left(-k_1^2+k_2{}^2+k_3{}^2\right)}{k_1^3 k_2{}^3 k_3{}^3 \left(k_1+k_2+k_3\right){}^6}   \label{bisflats}
\eea

\item As a general rule, the terms which are going to generate a flat shape can be read off already at  the Lagrangian level: indeed the flatness originates from the way the external momenta combine with each other and are summed over. Whenever mixed space derivatives act on a single $\pi$ term and the mixing is repeated on at least another $\pi$ field, the shape turns out to be flat (note that this criterium puts $\bar M_{6,8,9}$ contributions in the same, ``flat'' class, but correctly excludes apparently very similar ones such as $\bar M_{2,3,7}$). 
\end{itemize}

\section{Conclusions}
The purpose of this paper was to employ a powerful tool such as effective field theory to obtain as general as possible a knowledge of primordial non-Gaussianities generated in a very general set-up of single field models of inflation. We have performed a study of the corresponding bispectra, in terms of both amplitudes and shapes.  We extended the results existing in the literature in different ways. First, we have improved the treatment of the wavefunction describing the behaviour of  the cosmological  perturbations at second order in perturbation theory. Secondly, we computed the amplitude and the shape of bispectra for theories which interpolate between the most studied ones, {\it i.e.}, the DBI and the ghost models. Third, we have pointed out the importance of the
curvature-induced operators. Their study has revealed that large non-Gaussianities may be generated with a flat shape which is quite uncommon 
for single-field models of inflation so far analyzed. It would be interesting to identify in  which (class of) theories of inflation such operators arise. 
A natural way to extend our findings is to study with the same philosophy the four-point correlator, the trispectrum, and to identify, for instance, 
those classes of inflationary models where the  bispectra are suppressed, but large trispectra are generated.
These issues are currently under investigation \cite{future}.

\section*{Acknowledgments}
It is a pleasure to thank X. Chen, P. Creminelli and M. Pietroni for illuminating discussions.  This research has been partially supported by the ASI Contract No. I/016/07/0 COFIS, the ASI/INAF Agreement I/072/09/0 for the Planck LFI Activity of Phase E2. MF would like to thank Claudio Destri for insightful discussions and for kind encouragement during the completion of this work. MF is happy to thank the Physics Department of the University of Padua for warm hospitality.

\newpage
\noindent
\section*{References}
\bibliographystyle{JHEP}

\begin{thebibliography}{10}





\bibitem{smoot92} 
G.~F.~Smoot {\it et al.},
Astrophys.\ J.\  {\bf 396}, L1 (1992)

\bibitem{bennett96} 
C.~L.~Bennett {\it et al.},
Astrophys.\ J.\  {\bf 464}, L1 (1996). 

\bibitem{gorski96} 
K.~M.~Gorski, A.~J.~Banday, C.~L.~Bennett, G.~Hinshaw, A.~Kogut, 
G.~F.~Smoot and E.~L.~Wright,
Astrophys.\ J.\  {\bf 464} (1996) L11. 


\bibitem{wmap3}
 D.~N.~Spergel {\it et al.}  [WMAP Collaboration],
  Astrophys.\ J.\ Suppl.\  {\bf 170}, 377 (2007)
  [arXiv:astro-ph/0603449].



\bibitem{wmap5} 
  E.~Komatsu {\it et al.}  [WMAP Collaboration],
  arXiv:0803.0547 [astro-ph].

\bibitem{kom}
  E.~Komatsu {\it et al.},
  arXiv:1001.4538 [astro-ph.CO].
\bibitem{bisp}
  V.~Acquaviva, N.~Bartolo, S.~Matarrese and A.~Riotto,
  Nucl.\ Phys.\  B {\bf 667}, 119 (2003)
  [arXiv:astro-ph/0209156].
\\
  J.~M.~Maldacena,
  JHEP {\bf 0305} (2003) 013
  [arXiv:astro-ph/0210603].

\bibitem{bispectrum}
  N.~Bartolo, S.~Matarrese and A.~Riotto,
  Phys.\ Rev.\  D {\bf 65}, 103505 (2002)
  [arXiv:hep-ph/0112261].
\\
F.~Bernardeau and J.-P. Uzan,
 Phys. Rev. {\bf D66}, 103506 (2002), hep-ph/0207295.
\\
F.~Bernardeau and J.-P. Uzan,
 Phys. Rev. {\bf D67}, 121301 (2003), astro-ph/0209330.
\\
P.~Creminelli,
\newblock JCAP {\bf 0310}, 003 (2003), astro-ph/0306122.
\\
  N.~Bartolo, S.~Matarrese and A.~Riotto,
  JCAP {\bf 0401}, 003 (2004)
  [arXiv:astro-ph/0309692].
\\
  N.~Bartolo, S.~Matarrese and A.~Riotto,
  Phys.\ Rev.\  D {\bf 69}, 043503 (2004)
  [arXiv:hep-ph/0309033].
\\
  S.~Matarrese and A.~Riotto,
  JCAP {\bf 0308}, 007 (2003)
  [arXiv:astro-ph/0306416].
\\
  D.~Seery and J.~E.~Lidsey,
  JCAP {\bf 0506} (2005) 003
  [arXiv:astro-ph/0503692].
\\
  D.~Seery and J.~E.~Lidsey,
  JCAP {\bf 0509} (2005) 011
  [arXiv:astro-ph/0506056].
\\
  G.~I.~Rigopoulos, E.~P.~S.~Shellard and B.~J.~W.~van Tent,
  Phys.\ Rev.\  D {\bf 73}, 083522 (2006)
  [arXiv:astro-ph/0506704].
\\
\bibitem{Chen:2008wn}
  X.~Chen, R.~Easther and E.~A.~Lim,
  JCAP {\bf 0804}, 010 (2008)
  [arXiv:0801.3295 [astro-ph]].
\\
  D.~Langlois, S.~Renaux-Petel, D.~A.~Steer and T.~Tanaka,
  Phys.\ Rev.\  D {\bf 78}, 063523 (2008)
  [arXiv:0806.0336 [hep-th]].
\\
  F.~Arroja, S.~Mizuno and K.~Koyama,
  JCAP {\bf 0808}, 015 (2008)
  [arXiv:0806.0619 [astro-ph]].


\bibitem{BabichCZ}
  D.~Babich, P.~Creminelli and M.~Zaldarriaga,
  JCAP {\bf 0408}, 009 (2004)
  [arXiv:astro-ph/0405356].


\bibitem{chen-bis}
  X.~Chen, M.~x.~Huang, S.~Kachru and G.~Shiu,
  JCAP {\bf 0701}, 002 (2007)
  [arXiv:hep-th/0605045].



\bibitem{trispectrum}
  D.~Seery, J.~E.~Lidsey and M.~S.~Sloth,
  JCAP {\bf 0701}, 027 (2007)
  [arXiv:astro-ph/0610210].
\\
 X.~Chen, M.~x.~Huang and G.~Shiu,
  Phys.\ Rev.\  D {\bf 74} (2006) 121301
  [arXiv:hep-th/0610235].
\\
D.~Seery and J.~E. Lidsey,
\newblock JCAP {\bf 0701}, 008 (2007), astro-ph/0611034.
\\
X.~Chen, R.~Easther, and E.~A. Lim,
\newblock JCAP {\bf 0706}, 023 (2007), astro-ph/0611645.
\\
  F.~Arroja and K.~Koyama,
  Phys.\ Rev.\  D {\bf 77}, 083517 (2008)
  [arXiv:0802.1167 [hep-th]].
\\
  D.~Seery, M.~S.~Sloth and F.~Vernizzi,
  JCAP {\bf 0903}, 018 (2009)
  [arXiv:0811.3934 [astro-ph]].
\bibitem{loop}
\newblock  
  D.~Boyanovsky, H.~J.~de Vega and N.~G.~Sanchez,
  Nucl.\ Phys.\  B {\bf 747}, 25 (2006)
  [arXiv:astro-ph/0503669].
  \\
  S.~Weinberg,
  Phys.\ Rev.\  D {\bf 72}, 043514 (2005)
  [arXiv:hep-th/0506236].
\\
  D.~Boyanovsky, H.~J.~de Vega and N.~G.~Sanchez,
  Phys.\ Rev.\  D {\bf 72}, 103006 (2005)
  [arXiv:astro-ph/0507596].
\\
  M.~S.~Sloth,
  Nucl.\ Phys.\  B {\bf 748}, 149 (2006)
  [arXiv:astro-ph/0604488].
\\
  S.~Weinberg,
  Phys.\ Rev.\  D {\bf 74}, 023508 (2006)
  [arXiv:hep-th/0605244].
\\
  M.~S.~Sloth,
  Nucl.\ Phys.\  B {\bf 775}, 78 (2007)
  [arXiv:hep-th/0612138].
  \\
  D.~Seery,
  JCAP {\bf 0711}, 025 (2007)
  [arXiv:0707.3377 [astro-ph]].
\\
  D.~Seery,
  JCAP {\bf 0802}, 006 (2008)
  [arXiv:0707.3378 [astro-ph]].
\\
  A.~Riotto and M.~S.~Sloth,
  JCAP {\bf 0804}, 030 (2008)
  [arXiv:0801.1845 [hep-ph]].
\\
  L.~Leblond and S.~Shandera,
  JCAP {\bf 0808}, 007 (2008)
  [arXiv:0802.2290 [hep-th]].
  \\
  E.~Dimastrogiovanni and N.~Bartolo,
  JCAP {\bf 0811}, 016 (2008)
  [arXiv:0807.2790 [astro-ph]].
  \\
  P.~Adshead, R.~Easther and E.~A.~Lim,
  Phys.\ Rev.\  D {\bf 79}, 063504 (2009)
  [arXiv:0809.4008 [hep-th]].
  \\
  D.~Boyanovsky, C.~Destri, H.~J.~de Vega and N.~G.~Sanchez,
  Int.\ J.\ Mod.\ Phys.\  A {\bf 24}, 3669 (2009)
  [arXiv:0901.0549 [astro-ph.CO]].
  \\
L.~Senatore and M.~Zaldarriaga,
  arXiv:0912.2734 [hep-th].


\bibitem{reviewk}
For a review on non-Gaussianity, see,   
N.~Bartolo, E.~Komatsu, S.~Matarrese and A.~Riotto,
  Phys.\ Rept.\  {\bf 402}, 103 (2004)
  [arXiv:astro-ph/0406398].

\bibitem{Pl}
see http://planck.esa.int/.


\bibitem{Mand}
  N.~Mandolesi {\it et al.},
  arXiv:1001.2657 [astro-ph.CO].




\bibitem{Liguori1}
  M.~Liguori, E.~Sefusatti, J.~R.~Fergusson and E.~P.~S.~Shellard,
  arXiv:1001.4707 [astro-ph.CO].


\bibitem{VerdeM}
  L.~Verde and S.~Matarrese,
  Astrophys.\ J.\  {\bf 706}, L91 (2009)
  [arXiv:0909.3224 [astro-ph.CO]].




\bibitem{koyamarev}
  K.~Koyama,
  arXiv:1002.0600 [hep-th].

\bibitem{ourrev}
  N.~Bartolo, S.~Matarrese and A.~Riotto,
  arXiv:1001.3957 [astro-ph.CO].




\bibitem{chenrev}
  X.~Chen,
  arXiv:1002.1416 [astro-ph.CO].
 





\bibitem{luty}
  P.~Creminelli, M.~A.~Luty, A.~Nicolis and L.~Senatore,
  JHEP {\bf 0612}, 080 (2006)
  [arXiv:hep-th/0606090].



\bibitem{eft08}
  C.~Cheung, P.~Creminelli, A.~L.~Fitzpatrick, J.~Kaplan and L.~Senatore,
  JHEP {\bf 0803}, 014 (2008)
  [arXiv:0709.0293 [hep-th]].

\bibitem{w-e}
  S.~Weinberg,
  Phys.\ Rev.\  D {\bf 77}, 123541 (2008)
  [arXiv:0804.4291 [hep-th]].

\bibitem{mukh1}
  C.~Armendariz-Picon, T.~Damour and V.~F.~Mukhanov,
  Phys.\ Lett.\  B {\bf 458}, 209 (1999)
  [arXiv:hep-th/9904075].

\bibitem{mukh2}
  J.~Garriga and V.~F.~Mukhanov,
  Phys.\ Lett.\  B {\bf 458}, 219 (1999)
  [arXiv:hep-th/9904176].

\bibitem{DBI}
  M.~Alishahiha, E.~Silverstein and D.~Tong,
  Phys.\ Rev.\  D {\bf 70}, 123505 (2004)
  [arXiv:hep-th/0404084].


\bibitem{HT}
  R.~Holman and A.~J.~Tolley,
  JCAP {\bf 0805}, 001 (2008)
  [arXiv:0710.1302 [hep-th]].


\bibitem{Pier}
  P.~D.~Meerburg, J.~P.~van der Schaar and P.~S.~Corasaniti,
  JCAP {\bf 0905}, 018 (2009)
  [arXiv:0901.4044 [hep-th]].



\bibitem{FShellard}
  J.~R.~Fergusson and E.~P.~S.~Shellard,
  Phys.\ Rev.\  D {\bf 80}, 043510 (2009)
  [arXiv:0812.3413 [astro-ph]].


\bibitem{Liguori2}
  M.~Liguori, E.~Sefusatti, J.~R.~Fergusson and E.~P.~S.~Shellard,
  arXiv:1001.4707 [astro-ph.CO].






\bibitem{future}
 N.~Bartolo, M. Fasiello, S. Matarrese and A. Riotto,
  ``Large non-Gaussianities in the   the Effective  Field Theory  Approach to   Single-Field Inflation: the Trispectrum'', to appear.

\bibitem{tilted}
  L.~Senatore,
  Phys.\ Rev.\  D {\bf 71}, 043512 (2005)
  [arXiv:astro-ph/0406187].


\bibitem{3pt}
  C.~Cheung, A.~L.~Fitzpatrick, J.~Kaplan and L.~Senatore,
  JCAP {\bf 0802}, 021 (2008)
  [arXiv:0709.0295 [hep-th]].

\bibitem{ghost}
  N.~Arkani-Hamed, P.~Creminelli, S.~Mukohyama and M.~Zaldarriaga,
  JCAP {\bf 0404}, 001 (2004)
  [arXiv:hep-th/0312100].

\bibitem{b&d}
  N.~D.~Birrell and P.~C.~W.~Davies,
  ``Quantum Fields In Curved Space,''
{\it  Cambridge, Uk: Univ. Pr. (1982)}




\bibitem{chen-tris}
  X.~Chen, B.~Hu, M.~x.~Huang, G.~Shiu and Y.~Wang,
  JCAP {\bf 0908}, 008 (2009)
  [arXiv:0905.3494 [astro-ph.CO]].
\bibitem{ssz05}
  L.~Senatore, K.~M.~Smith and M.~Zaldarriaga,
  JCAP {\bf 1001}, 028 (2010)
  [arXiv:0905.3746 [astro-ph.CO]].

\bibitem{p.s.}
 N.~Bartolo, M. Fasiello, S. Matarrese and A. Riotto,  to appear.



  \bibitem{in-in1}
 J.~S.~Schwinger,
  J.\ Math.\ Phys.\  {\bf 2}, 407 (1961).

  \bibitem{in-in3}
R.~D.~Jordan,
  Phys.\ Rev.\  D {\bf 33}, 444 (1986).

\bibitem{in-in2}
E.~Calzetta and B.~L.~Hu,
  Phys.\ Rev.\  D {\bf 35}, 495 (1987).
  
\bibitem{w-qccc}
S.~Weinberg,
  Phys.\ Rev.\  D {\bf 72}, 043514 (2005)
  [arXiv:hep-th/0506236].
  








  
  
\end{thebibliography}

\end{document}